\pdfoutput=1
\documentclass[aps,prb,floatfix,amsmath,amssymb,preprint,longbibliography]{revtex4-1}

\usepackage{amsmath}
\usepackage{amssymb}
\usepackage{graphicx}
\usepackage{color}

\newcommand{\tc}        {$T_{c}$}
\newcommand{\lsco}      {La$\mathrm{_{2-x}}$Sr$\mathrm{_x}$CuO$\mathrm{_4}$}
\newcommand{\dxxyy}     {$d_{x^2-y^2}$}

\newcommand{\dzz}       {$d_{3z^2-r^2}$}
\newcommand{\psigma}    {$p_\sigma$}
\newcommand{\pz}        {$p_z$}

\begin{document}

\title{Latent Room-Temperature T$_{\rm c}$ in Cuprate Superconductors}
\author{Jamil \surname{Tahir-Kheli}}
\email{jamil@caltech.edu}
\affiliation{Applied Physics and Chemistry,\\ Beckman Institute (MC 139-74),
California Institute of Technology, Pasadena CA 91125}

\date{February 15, 2017}

\begin{abstract}
The ancient phrase, ``All roads lead to Rome" applies to Chemistry and Physics. Both are highly evolved sciences, with their own history, traditions, language, and approaches to problems. Despite all these differences, these two roads generally lead to the same place. For high temperature cuprate superconductors however, the Chemistry and Physics roads do not meet or even come close to each other. In this paper, we analyze the physics and chemistry approaches to the doped electronic structure of cuprates and find the chemistry doped hole (out-of-the-CuO$\mathrm{_2}$-planes) leads to explanations of a vast array of normal state cuprate phenomenology using simple counting arguments. The chemistry picture suggests that phonons are responsible for superconductivity in cuprates. We identify the important phonon modes, and show that the observed \tc$\ \sim 100$ K, the \tc-dome as a function of hole doping, the change in \tc\ as a function of the number of CuO$\mathrm{_2}$ layers per unit cell, the lack of an isotope effect at optimal \tc\ doping, and the D-wave symmetry of the superconducting Cooper pair wavefunction are all explained by the chemistry picture. Finally, we show that ``crowding" the dopants in cuprates leads to a pair wavefunction with S-wave symmetry and \tc\ $\approx280-390$ K. Hence, we believe there is enormous ``latent" \tc\ remaining in the cuprate class of superconductors.
\end{abstract}

\maketitle

The highest superconducting transition temperature, \tc, at ambient pressure is 138 K in the Mercury cuprate HgBa$_2$Ca$_2$Cu$_3$O$_{8+\delta}$ (Hg1223) with three CuO$_2$ layers per unit cell.\cite{Schilling1993,Mukuda2012} Hg1223 was discovered in 1993. The longest time period between record setting \tc\ discoveries is the 17 years between Pb (1913 with $T_c=7.2$ K) to Nb (1930 with $T_c=9.2$ K).  With the enormous increase in focus on superconductivity after the discovery of cuprates 30 years ago, the current 24 years without a new record at ambient pressure indicates we may be reaching the maximum attainable \tc.

In this paper, we show this conclusion to be wrong.  We demonstrate that \tc\ can be raised above room-temperature to $\approx400$K in cuprates by precise control of the spatial separation of dopants. Hence, there still remains substantial ``latent" \tc\ in cuprates. Our proposed doping strategy and superconducting mechanism is not restricted to cuprates and may be exploited in other materials.

Our room-temperature \tc\ result is based upon four observations:

\begin{itemize}
\item Cuprates are intrinsically inhomogeneous on the atomic-scale and are comprised of insulating and metallic regions. The metallic region is formed by doping the material.
\item A diverse set of normal state properties are explained solely from the topological properties of these two regions and their doping evolution.
\item Superconductivity results from phonons at or adjacent to the interface between the metallic and insulating regions. Transition temperatures $T_{c}\sim100$ K are possible because the electron-phonon coupling is of longer-range than metals (nearest neighbor).
\item These interface phonons explain the observed superconducting properties and lead to our prediction of room-temperature superconductivity.
\end{itemize}

How is our claim of room-temperature \tc\ possible?  The talent and funding invested into finding the mechanism of cuprate superconductivity and higher \tc\ materials has led to more than 200,000 refereed papers.\cite{Mann2011} After this mind-boggling quantity of literature, it seems unlikely that any unturned stones remain that could lead to our prediction.

Our claim does not come from locating an overlooked stone beneath the cuprate stampede. Instead, we believe the majority of the cuprate community settled upon the incorrect orbital nature of the doped hole.  This mistake led to Hamiltonians (Hubbard models) that ``threw the baby out with the bath water."

\begin{figure}[tbp]
\centering \includegraphics[width=0.6\linewidth]
{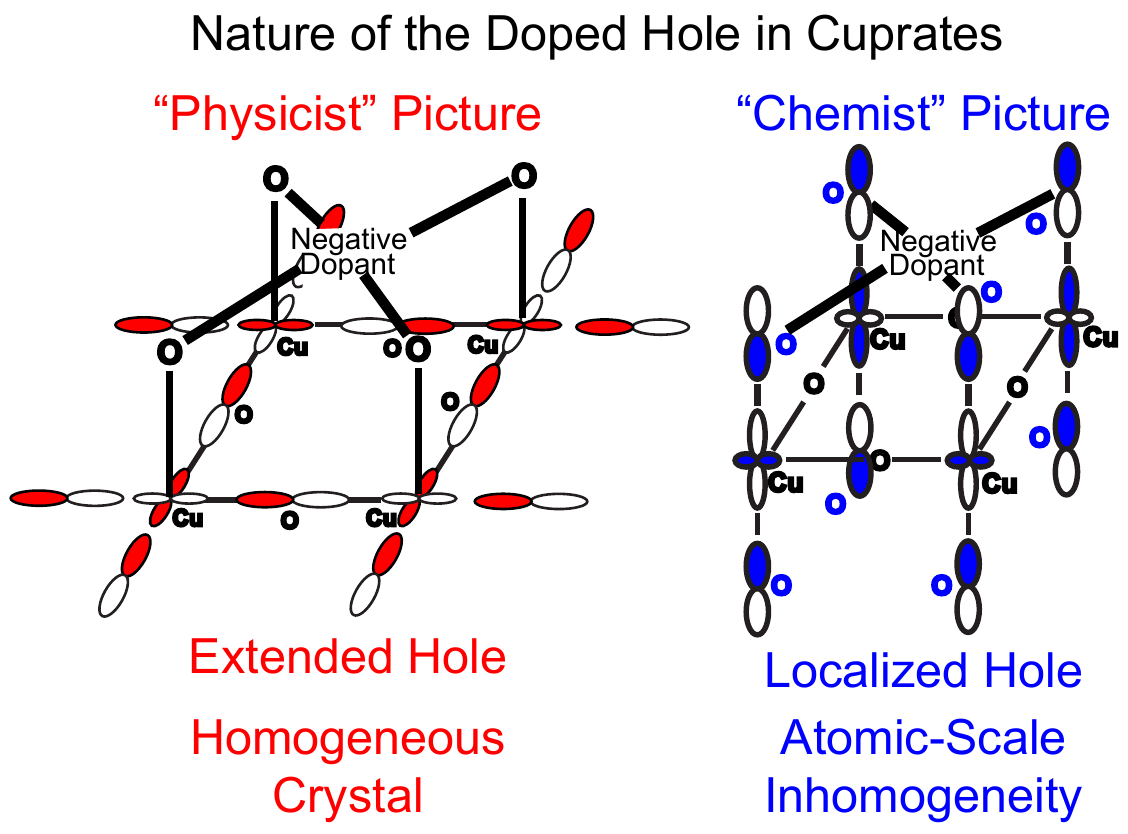}
\caption{
The difference in the character of the doped hole using the ``physicist's" density functionals, LDA and PBE, versus the ``chemist's" hybrid density functionals.  The dopant is negatively charged and resides out-of-the CuO$_2$ plane.  The physicist's hole state has density in the CuO$_2$ plane and is delocalized over the crystal. It is a peculiar hole state. Intuitively, one would expect the positive hole charge to point at the negatviely charged dopant. The chemist's hole state density is out-of-the-CuO$_2$ plane (points at dopant), and is localized around the four-Cu-site plaquette beneath the dopant.  In the chemist's view, the crystal has atomic-scale inhomogeneity that is not a small perturbation of translational symmetry. The physicist's view leads to an approximately homogeneous crystal.
}
\label{physchem}
\end{figure}

A major reason for the early adoption of these Hubbard models for cuprates was due to computational results using the ab initio local density approximation (LDA) in density functional theory (DFT).  While LDA is now deprecated, being replaced by the Perdew-Burke-Ernzerhof functional\cite{PBE} (PBE), both functionals lead to exactly the same doped hole wavefunction in cuprates.  These ``physicist" functionals find the doped hole to be a \emph{delocalized wavefunction} comprised of orbitals residing in the CuO$_2$ planes common to all cuprates.\cite{Yu1987,Mattheiss1987,Pickett1989} Unfortunately, LDA and PBE both contain unphysical Coulomb repulsion of an electron with itself.\cite{Perdew1981} The ``chemist" hybrid density functionals, invented in 1993 (seven years after the discovery of cuprate superconductivity), corrected for this self-Coulomb error, and thereby found the doped hole residing in a \emph{localized wavefunction} surrounding the dopant atom with orbital character pointing out of the CuO$_2$ planes.\cite{Perry2001,Perry2002} The physicist and chemist doped holes are shown in Figure \ref{physchem}. A discussion of the superiority of the chemist's DFT to the physicist's DFT is in Appendix \ref{DFT}.

The ``chemist's" ab initio doped hole leads to eight \emph{electronic structure concepts} that explain a vast array of normal and superconducting state phenomenlogy using simple counting. These eight structural concepts are described below.

\noindent\textsc{\underline{Structural Concept 1}:} \textbf{Cuprates are inhomogeneous on an atomic-scale.} The inhomogeneity is not a small perturbation to translational symmetry. It must be included at zeroth order.

\begin{figure}[h]
\centering \includegraphics[width=15cm]
{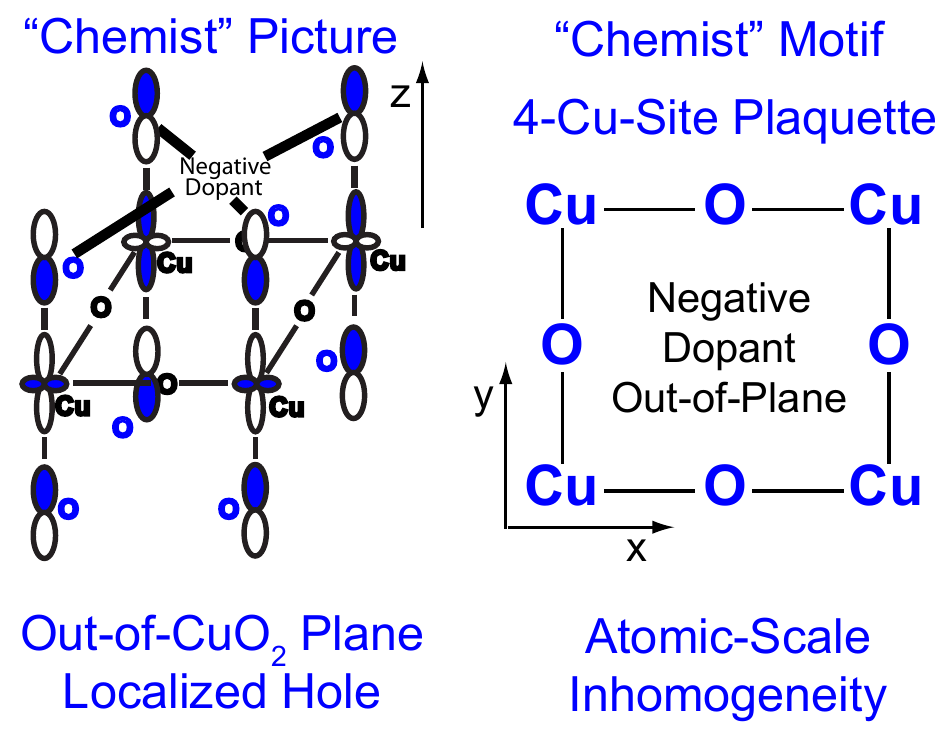}
\caption{
The doped four-Cu-site plaquette ``motif."
}
\label{motif}
\end{figure}

\noindent\textsc{\underline{Structural Concept 2}:} \textbf{The Cuprate motif is a four-Cu-site plaquette formed by each dopant.} See Figure \ref{motif}. The out-of-the-CuO$_2$ plane negative dopant is surrounded by an out-of-the-CuO$_2$ plane hole. The hole is comprised of apical Oxygen \pz\ and planar Cu \dzz\ character. There is also some planar O \psigma\ character that is not drawn.

\clearpage

\begin{figure}[h]
\centering \includegraphics[width=9cm]
{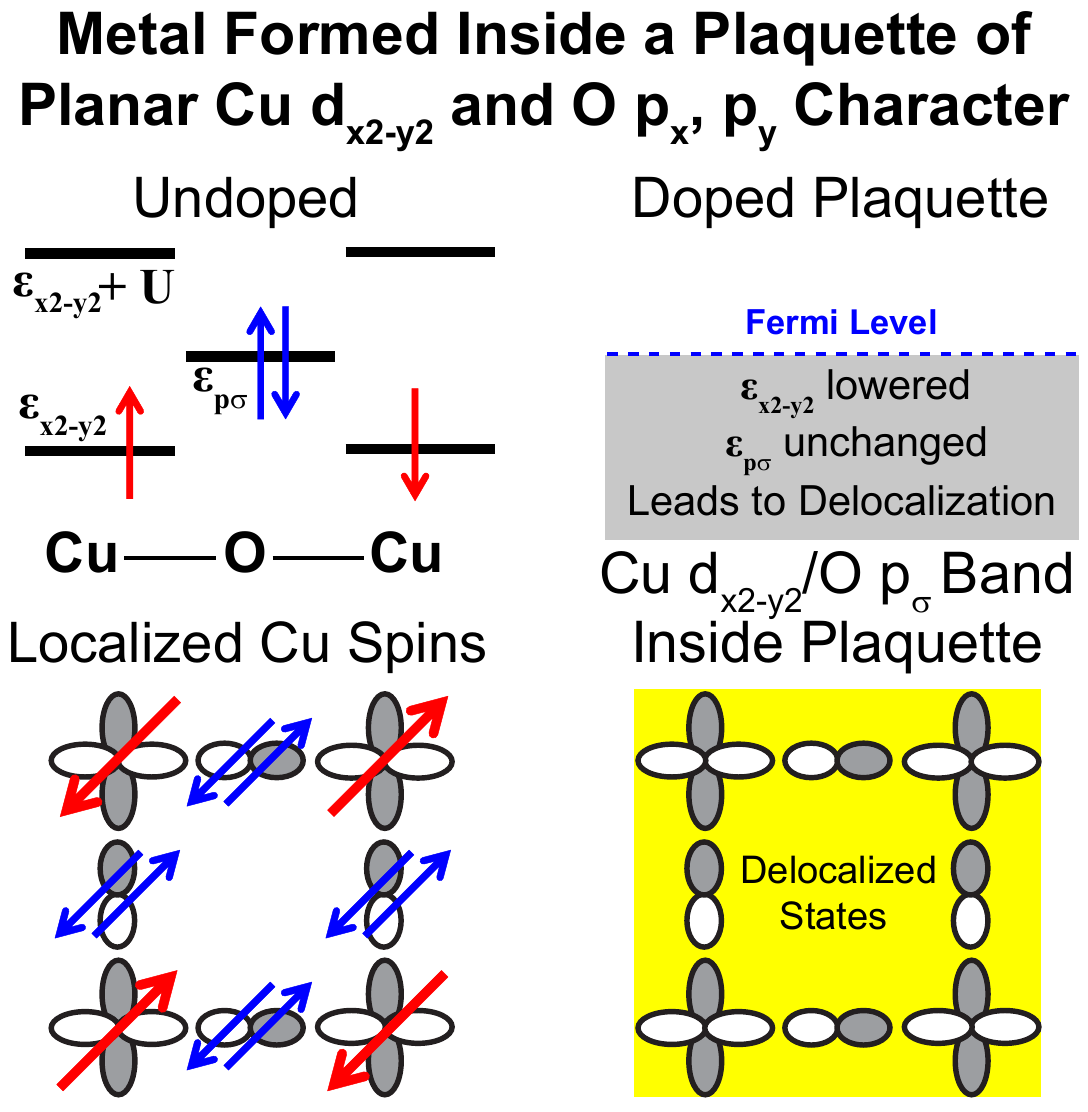}
\caption{
The mechanism for the creation of delocalized \emph{planar} metallic wavefunctions inside a doped plaquette. (a) The Cu \dxxyy\ and O \psigma\ orbital energies in an undoped plaquette. The energy ordering, $\epsilon_{x^2-y^2}<\epsilon_{p\sigma}<\epsilon_{x^2-y^2}+U$, where $U$ is the large on-site Cu \dxxyy\ Coulomb repulsion leads to localization of the spins on the Cu sites, and hence the undoped AF state. (b) The induced delocalization (yellow overlay) of the \emph{planar} Cu \dxxyy\ and O \psigma\ electrons in the plaquette from the decrease of the Cu \dxxyy\ orbital energy relative to the O \psigma\ orbital energy. The decrease occurs because the positive charge of the chemist's out-of-the-plane hole from Figures \ref{physchem} and \ref{motif} is closer to the planar Cu site. The out-of-plane hole is not shown here. There are a total of 4 O atom \psigma\ orbitals and 4 Cu \dxxyy\ orbitals inside the four-Cu-site plaquette. Since there are two spin states for each orbital, there is a total of $(4+4)\times2=16$ states in the yellow overlay in the bottom right figure. These 16 states are filled with the 12 electrons (4 red plus 8 blue) shown in the lower left figure.
}
\label{delocalization}
\end{figure}

\noindent\textsc{\underline{Structural Concept 3}:} \textbf{A tiny piece of metal is formed within each plaquette from electron delocalization in the \textit{planar} Cu $\mathbf{d_{x^2-y^2}}$ and O $\mathbf{p_\sigma}$ ($\mathbf{p_x}$ and $\mathbf{p_y}$) orbitals.} See Figure \ref{delocalization}. Delocalization occurs because the positive charge of the out-of-plane hole lowers the Cu \dxxyy\ orbital energy relative to the O \psigma\ orbital energy. In contrast, these electrons are localized in a spin-1/2 antiferromagnetic (AF) state in an undoped plaquette.

\clearpage

\begin{figure}[h]
\centering \includegraphics[width=8cm]
{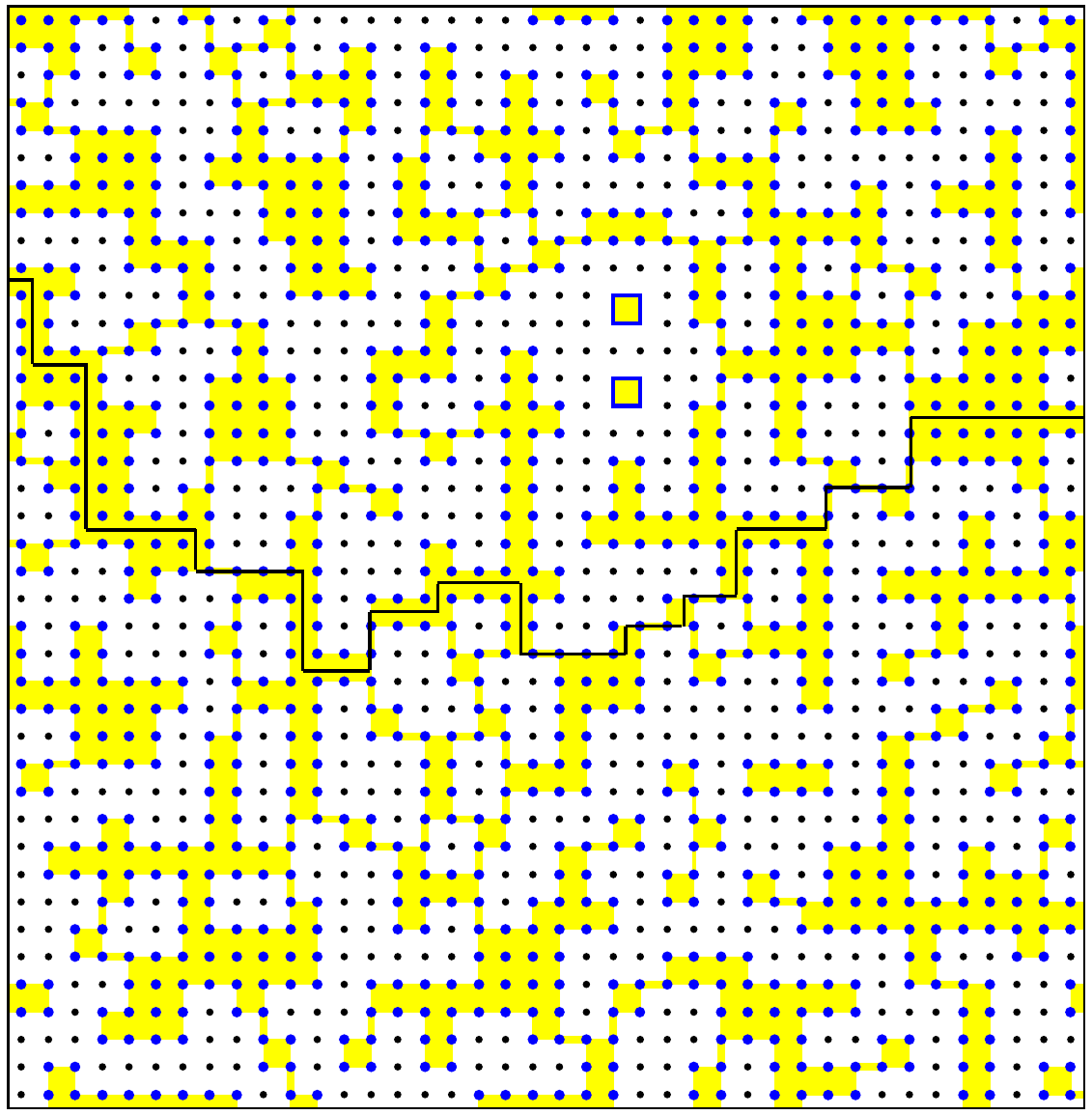}
\caption{
A 2D percolating pathway in an optimally doped CuO$_2$ plane ($x=0.16$ dopants per planar Cu). The black line shows a pathway from the left to the right side of this $40\times40$ CuO$_2$ lattice. The localized antiferromagnetic (AF) spins on the Cu sites are shown as black dots. The blue dots are Cu sites inside the doped metallic region. The O atoms are not shown. The yellow overlay represents the delocalized metallic band comprised of planar orbitals inside the doped region. The blue squares represent isolated plaquettes (not adjacent to another plaquette). There is a degeneracy at the Fermi level inside each isolated plaquette (see Appendix \ref{sec:PG}) that is split by its interaction with the crystal environment, and thereby leads to the pseudogap.\cite{Tahir-Kheli2011} The number of isolated plaquettes ``vanish" (become of measure zero) at hole doping $x\approx0.19$ where the cuprate pseudogap is known to disappear.\cite{Tahir-Kheli2011} The \emph{tenuous} percolating pathway has poor critical current.
}
\label{pathway}
\end{figure}

\noindent\textsc{\underline{Structural Concept 4}:} \textbf{A metal is formed when the doped plaquettes percolate through the crystal.} When a three-dimensional (3D) pathway of adjacent doped plaquettes is created through the crystal (percolation of the plaquettes), a metallic band comprised of \textit{planar} Cu \dxxyy\ and O \psigma\ orbitals is created inside the percolating region. These delocalized metallic wavefunctions do not have momentum, $\mathbf{k}$, as a good quantum number. Two-dimensional (2D) percolation occurs at a higher doping ($x\approx0.15$ holes per planar Cu) than the start of 3D percolation (at $x\approx0.05$ holes per planar Cu). See Figure \ref{pathway}. The undoped (non-metallic) region remains an insulating spin-1/2 AF. Thus cuprates have both insulating and metallic regions on an atomic-scale.

\clearpage

\noindent\textsc{\underline{Structural Concept 5}:} \textbf{The out-of-the-CuO$_2$ plane hole shown in Figures \ref{physchem} and \ref{motif} is a dynamic Jahn-Teller distortion that is a linear superposition of two ``frozen dumbbell" states.} See Figure \ref{dynamicJT}. Figure \ref{hole-in-dumbbell} in Appendix \ref{sec:JT} shows that the out-of-the-plane hole goes into the states in Figure \ref{dynamicJT}. We call the dynamic Jahn-Teller hole state a ``fluctuating dumbbell."

\begin{figure}[h]
\centering \includegraphics[width=14cm]
{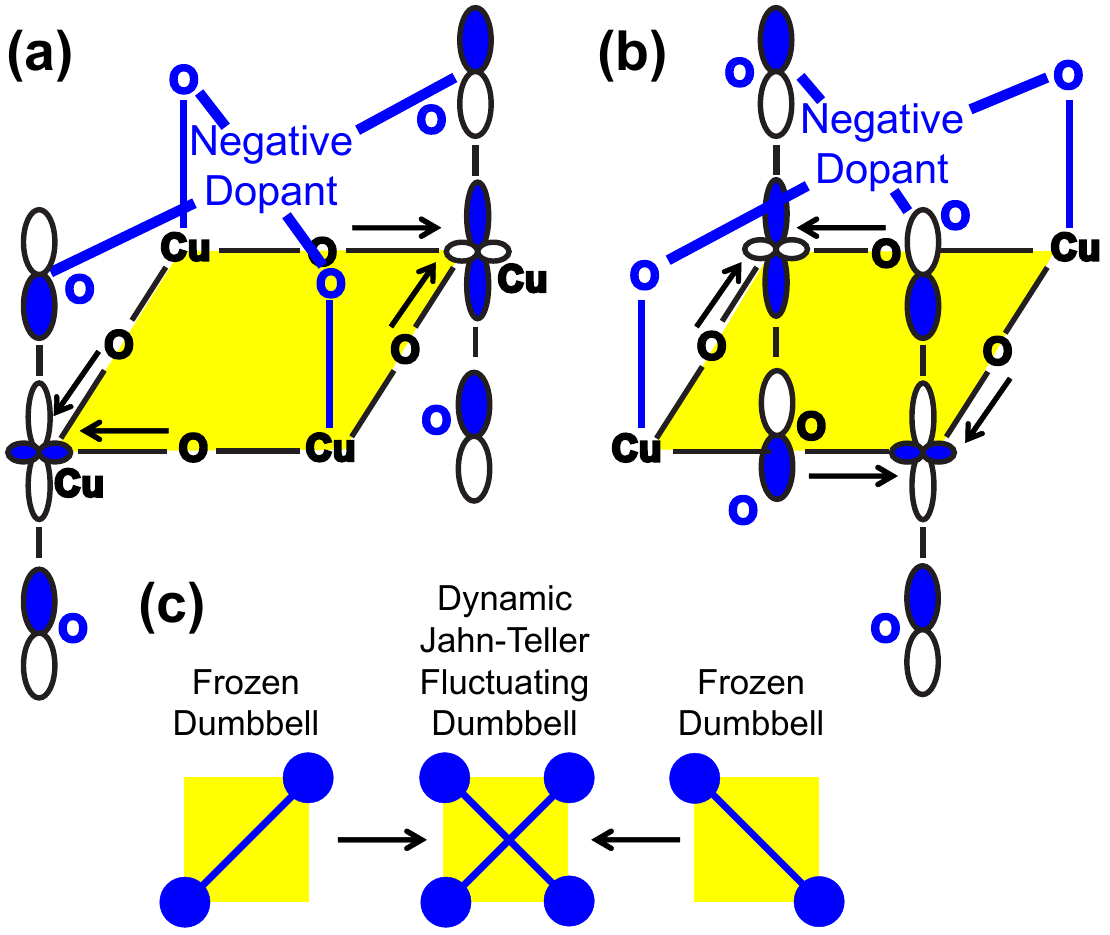}
\caption{
The character of the out-of-plane hole inside a doped plaquette. The arrows show the displacement of the O atoms in the CuO$_2$ plane with each ``frozen dumbbell" configuration.  We propose that the out-of-plane hole wavefunction is a dynamic Jahn-Teller state that is a linear superposition of the two dumbbells shown in (a) and (b) (called a ``fluctuating dumbbell"). We represent the two frozen dumbbells in (a) and (b) schematically by the left and right figures in the bottom row.  The fluctuating dumbbell is shown in the center figure at the bottom. The yellow overlay represents the {\it planar} Cu \dxxyy\ and O \psigma\ electrons (not shown) that are delocalized inside the plaquette.
}
\label{dynamicJT}
\end{figure}
\clearpage

\begin{figure}[h]
\centering \includegraphics[width=10cm]
{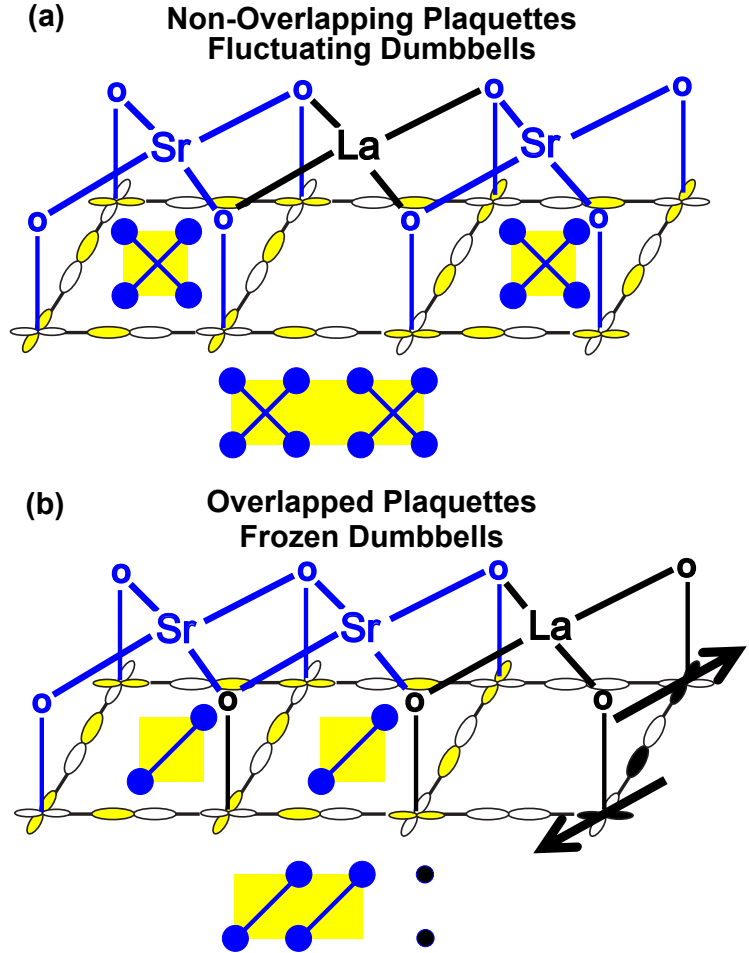}
\caption{
The effect of overlapping plaquettes on the fluctuating dumbbells for the case of Sr doping of \lsco. (a) Two non-overlapping plaquettes with fluctuating dumbbells. The top figure shows the CuO$_\mathrm{2}$ layer and the out-of-the-plane Sr dopants. The schematic figure below shows the metallic delocalization of the \textit{planar} Cu \dxxyy\ and O \psigma\ orbital electrons (yellow overlay) and the two fluctuating dumbbells. (b) Two overlapping plaquettes. The degeneracy of the two dumbbell states inside each plaquette is broken and the two dynamic Jahn-Teller fluctuating dumbbells become two frozen dumbbells. The bottom figure shows a schemtic of the metallic regions (yellow overlay) and the frozen dumbbells. The orientation of the dumbbells in this figure is arbitrary. The actual orientation in the crystal will depend on the environment. 
}
\label{staticJT}
\end{figure}

\noindent\textsc{\underline{Structural Concept 6}:} \textbf{A fluctuating dumbbell can be frozen by overlapping its plaquette with another plaquette.} See Figure \ref{staticJT}.

\clearpage

\noindent\textsc{\underline{Structural Concept 7}:} \textbf{If possible, plaquettes avoid overlapping.} Since the dopant atom is negatively charged, two plaquettes will repel each other. Their Coulomb repulsion is short-ranged because of screening from the planar metallic electrons. Hence, plaquettes do not overlap, but are otherwise distributed randomly. Plaquettes can avoid overlap up to a hole doping of $x=0.187$. For dopings greater than $x=0.187$, plaquettes must overlap. Plaquettes overlap as little as possible to minimize their mutual repulsion. Up to $x=0.187$ doping, there always exists a four-site square of AF spins where the next plaquette can be placed. For the doping range $0.187<x<0.226$, added plaquettes can cover three AF spins. In the range $0.226<x<0.271$, plaquettes cover two AF spins, and from $0.271<x<0.316$, a single localized spin. At $x=0.316$, the crystal is fully metallic with no localized spins. Further doping cannot increase the number of metallic sites. Figure \ref{pathway} shows that plaquettes can avoid overlap at $x=0.16$ doping. Figure \ref{dop23} below shows $x=0.23$ doping, where plaquettes must overlap.

\begin{figure}[h]
\centering \includegraphics[width=9cm]
{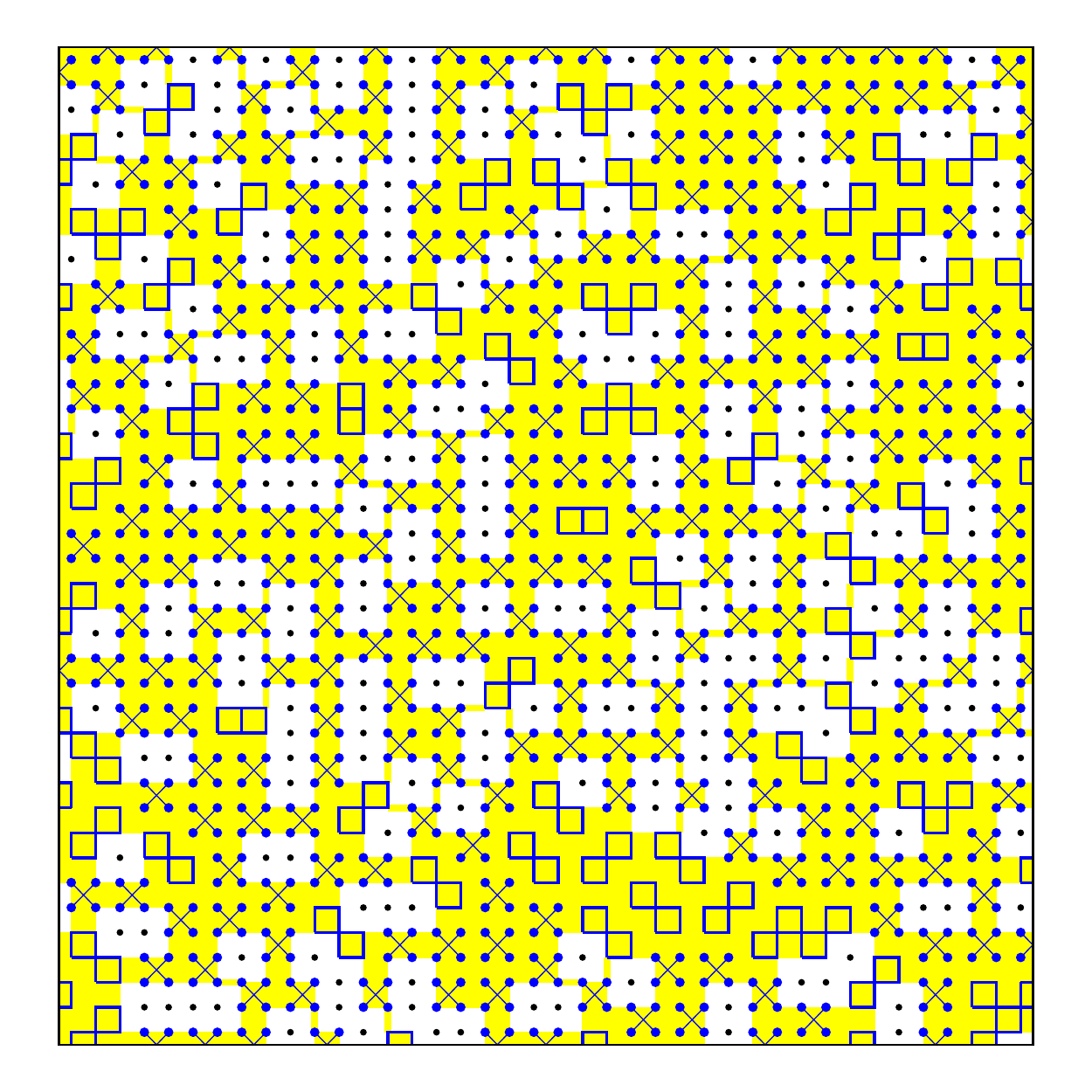}
\caption{Plaquette doping at $x=0.23$ on a $40\times40$ CuO$\mathrm{_2}$ lattice. The black dots are undoped AF Cu sites. The O atoms are not shown. The yellow overlay represents the delocalized metal comprised of planar Cu \dxxyy\ and planar O \psigma\ character. Fluctuating dumbbells (blue crosses) are seen in non-overlapped plaquettes. The overlapped plaquettes are shown as blue squares. There are frozen dumbbells inside each blue square. The frozen dumbbells are not shown in the figure.
}
\label{dop23}
\end{figure}
\clearpage

\begin{figure}[h]
\centering \includegraphics[width=9cm]
{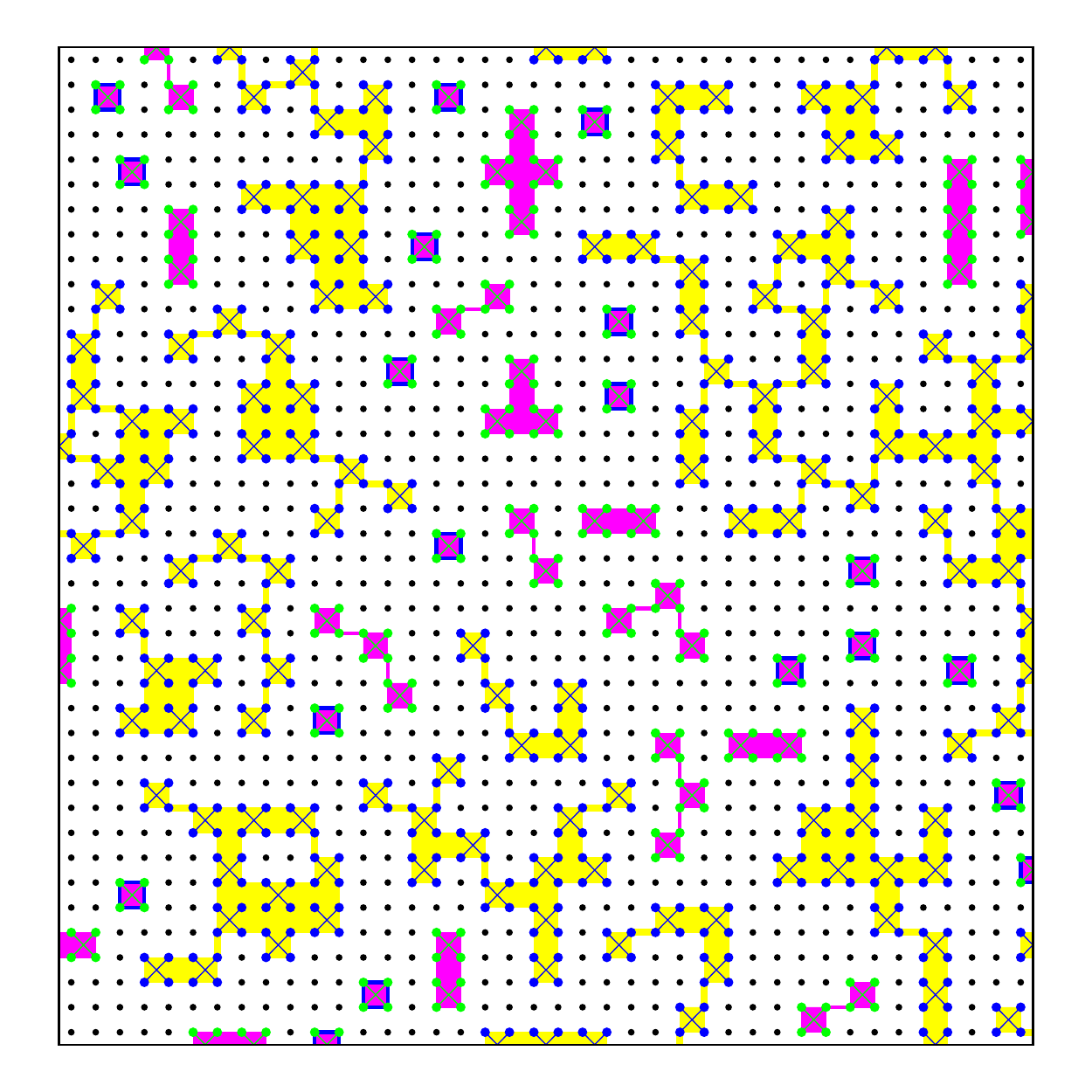}
\caption{Plaquette doping of $x=0.12$ on a $40\times40$ CuO$\mathrm{_2}$ lattice. The black dots are undoped AF Cu sites. No O atoms are shown. The blue squares are isolated plaquettes (no neighboring plaquette). The yellow overlay represents the metallic region comprised of planar Cu \dxxyy\ and planar O \psigma\ character. Fluctuating dumbbells (blue and green crosses) are seen in non-overlapped plaquettes. The magenta clusters are smaller than the Cooper pair coherence length and do not contribute to superconducting pairing. Since the size of a single plaquette (the Cu$-$Cu distance) is $\approx3.8$ \AA\ and the superconducting coherence length is $\sim2$ nm, we have chosen plaquette clusters fewer than or equal to 4 plaquettes in size to be fluctuating. The magenta overlay means there is metallic delocalization of the planar Cu and O orbitals in these clusters. The isolated plaquettes do not contribute to the superconducting pairing and the superconducting fluctuations above \tc\ because they contribute to the pseudogap, as shown in Appendix \ref{sec:PG}.
}
\label{dop12}
\end{figure}

\noindent\textsc{\underline{Structural Concept 8}:} \textbf{Plaquette Clusters smaller than the superconducting coherence length ($\sim2$ nm) thermally fluctuate and do not contribute to the superconducting pairing.} At low dopings, the plaquettes have not yet merged into a single connected region. There exist plaquette clusters smaller than the coherence length, as shown in magenta in Figure \ref{dop12}. They cannot contribute to the superconducting \tc. These fluctuating clusters lead to superconducting fluctuations above \tc.

\clearpage

\begin{figure}[h]
\centering \includegraphics[width=13cm]
{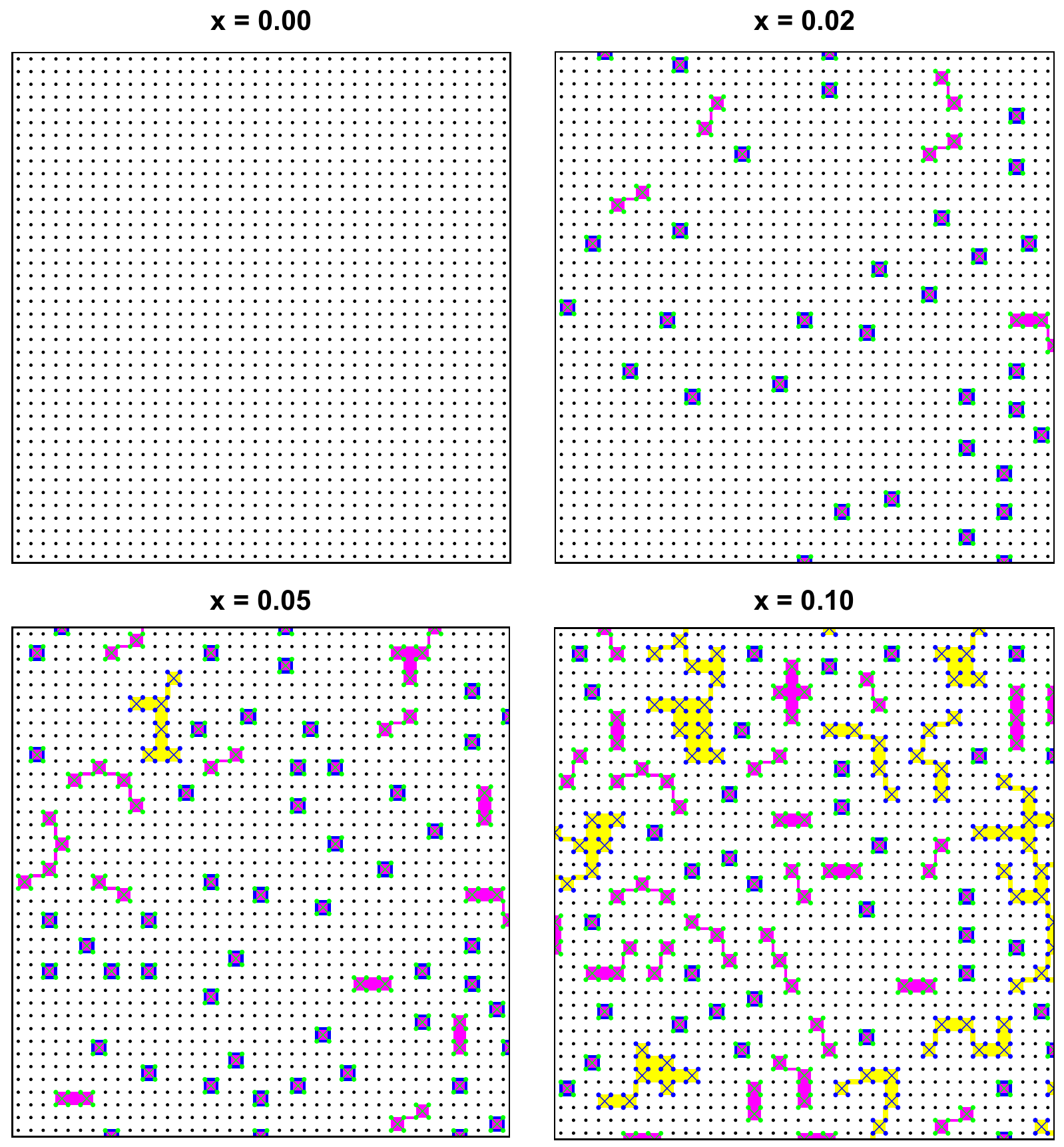}
\caption{Plaquette doping of a $40\times40$ square CuO$\mathrm{_2}$ lattice for dopings $x=0.00,0.02,0.05,$ and $0.10$.
Only the Cu sites are shown. The black dots are undoped AF Cu sites. The blue squares are isolated plaquettes (no neighboring plaquette). The yellow plaquette clusters are larger than 4 plaquettes in size (larger than the coherence length), and thereby contribute to the superconducting pairing. The magenta clusters are metallic clusters that are smaller than the coherence length. The blue and green crosses are the fluctuating dumbbells. There is no plaquette overlap in this doping range.
}
\label{dop00-10}
\end{figure}

\begin{figure}[t]
\centering \includegraphics[width=14cm]
{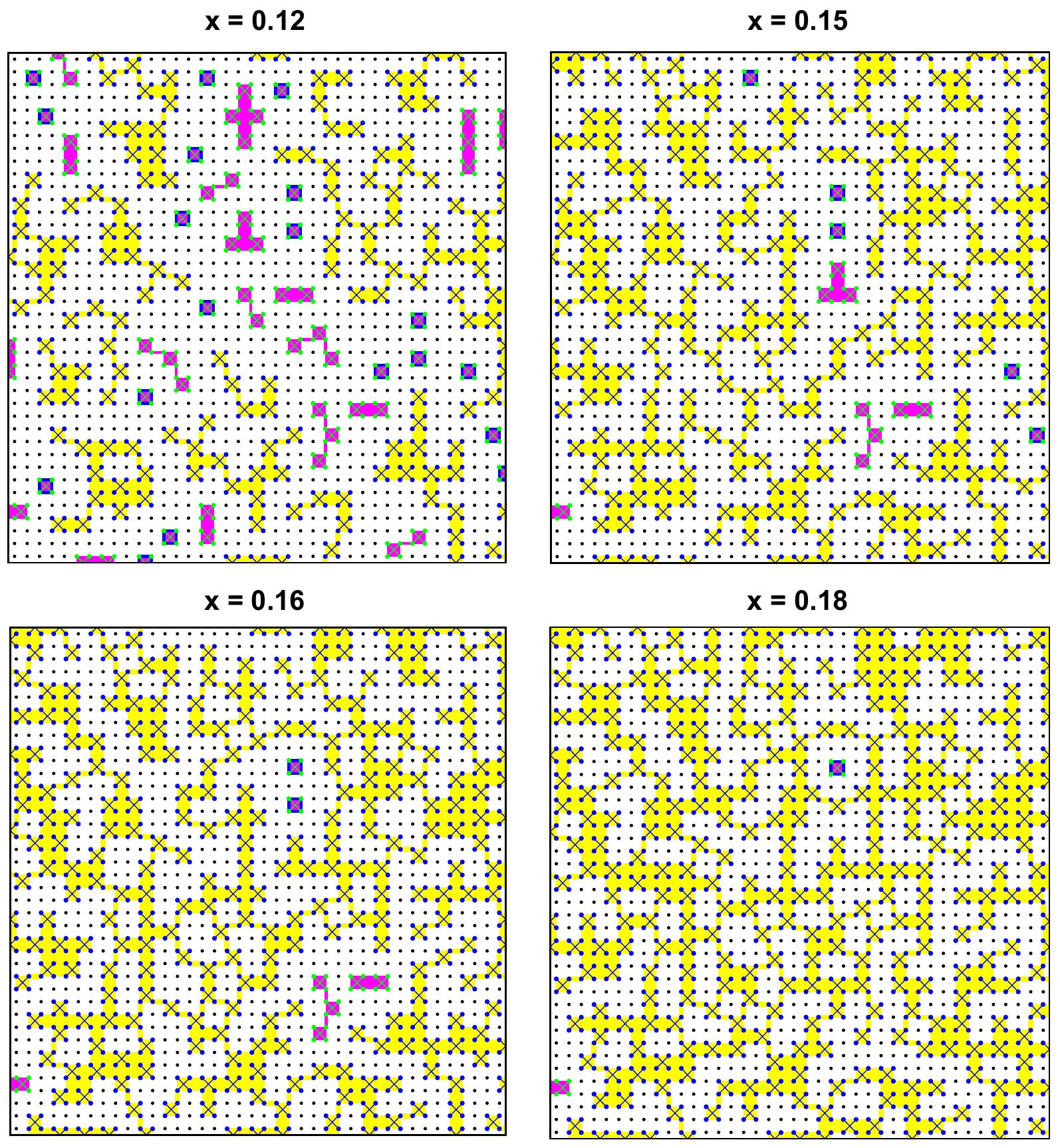}
\caption{Plaquette doping at $x=0.12,0.15,0.16,$ and $0.18$. The caption of Figure \ref{dop00-10} explains the symbols in this figure. Optimal \tc\ occurs for $x\approx0.16$ because the \emph{product} of the size of the metallic region (number of electrons that can participate in superconducting pairing) and the size of the interface (the number of pairing phonon modes) is maximized. Percolation in 2D occurs at $x\approx0.15$. In the finite lattice shown here, there is no 2D percolating metallic pathway for $x=0.15$. A 2D percolating pathway appears at $x=0.16$, as shown in Figure \ref{pathway}. The number of isolated plaquettes rapidly decreases over this doping range. By $x=0.18$ doping, there is only one isolated plaquette in its $40\times40$ lattice. At optimal doping, the percolation pathway is \emph{very} tenuous. The maximum critical current density, $J_c$, will be much less than the maximum due to Cooper pair depairing. Using current fabrication methods, cuprates have a low $J_c$ at the highest \tc. Crossing continuous metallic pathways one plaquette in width would have both large \tc\ and $J_c$.
}
\label{dop12-18}
\end{figure}

\begin{figure}[t]
\centering \includegraphics[width=14cm]
{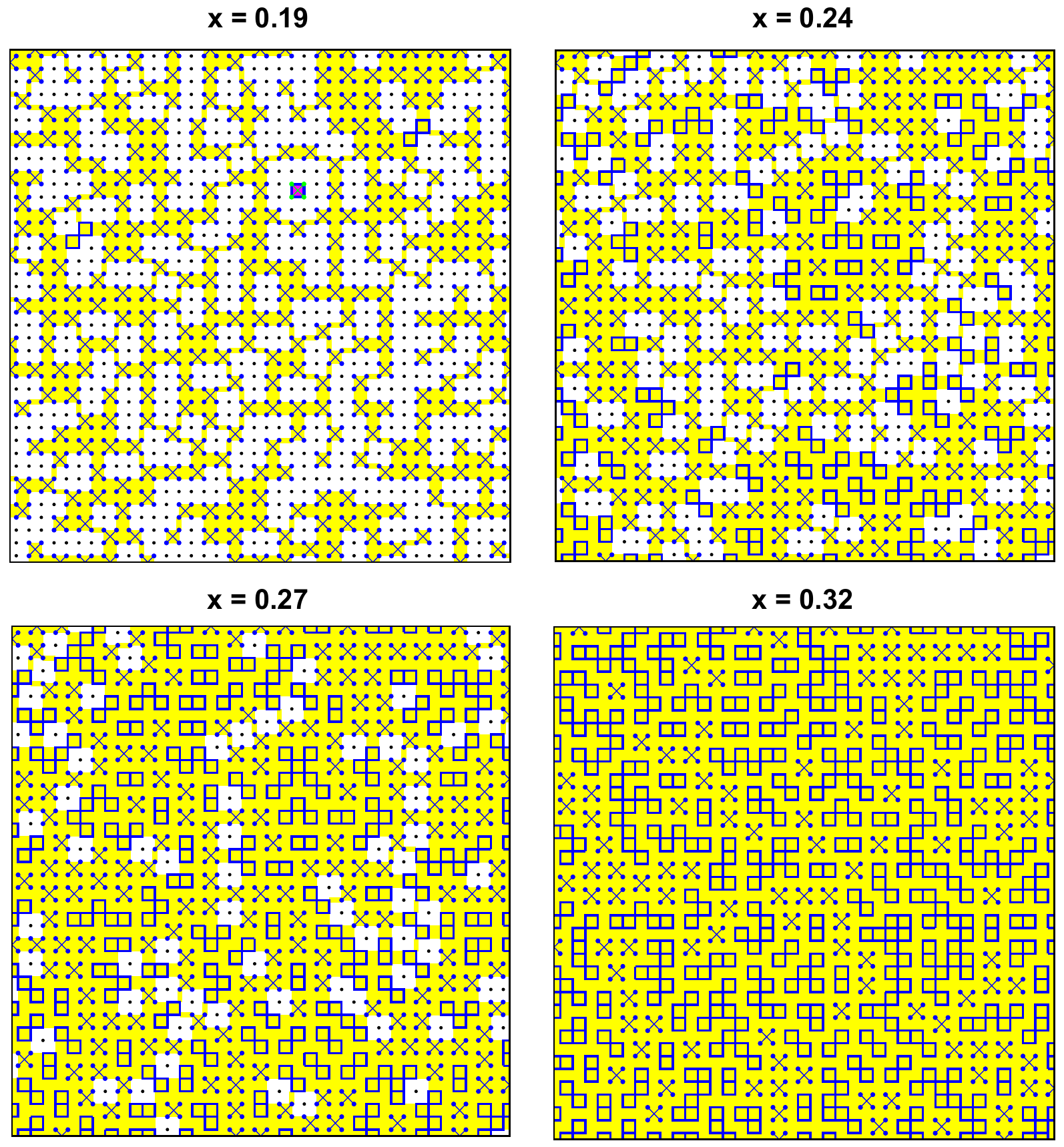}
\caption{Plaquette doping of a $40\times40$ square CuO$\mathrm{_2}$ lattice for dopings $x=0.19,0.24,0.27,$ and $0.32$. See the caption of Figure \ref{dop00-10} for an explanation of the symbols in this figure. The first plaquette overlap can be seen at $0.19$ doping (blue squares). The degeneracy of the two dumbbells states inside each overlapping plaquette has been broken. A frozen dumbbell configuration exists inside each overlapped plaquette. It is not drawn. At $0.27$ doping, only isolated spins remain (black dots). At $0.32$ doping, there are no localized spins remaining. The crystal is purely metallic. Only one isolated plaquette (blue square with magenta interior) remains at $x=0.19$ doping. The number of isolated plaquettes never vanishes entirely. Instead, their number becomes of measure zero above $x\approx0.19$ doping and leads to the vanishing of the pseudogap at $x\approx0.19$. At $x=0.32$, the critical current density, $J_c$, will be close to the Cooper pairing depairing limit since the metallic pathways through the crystal are \emph{not} tenuous. However, there is no interface pairing, leading to $T_c=0$.
}
\label{dop19-32}
\end{figure}

Figures \ref{dop00-10}, \ref{dop12-18}, and \ref{dop19-32} show the evolution of the plaquettes as a function of doping. Only twelve dopings are shown here from the range $x=0.00$ to $x=0.32$. The Appendix has similar figures for all dopings in this range in $0.01$ increments (Figures S0$-$S32). Only one CuO$\mathrm{_2}$ plane is shown in each of these figures.

\clearpage

\clearpage

\clearpage

The above eight electronic structural concepts explain a diverse set of normal state cuprate phenomenology as a function of doping by simple \textit{counting} arguments,\cite{Tahir-Kheli2015,Tahir-Kheli2013,Tahir-Kheli2011,Tahir-Kheli2010} as we have shown previously. These include (See reference \onlinecite{Tahir-Kheli2015} for a videotaped seminar summarizing all of these results.):

\begin{itemize}
\item{the low and high-temperature normal state resistivity by \textit{counting} the number of overlapped plaquettes and the size of the metallic region.\cite{Tahir-Kheli2015,Tahir-Kheli2013}}

\noindent For \lsco, the fluctuating dumbbells in adjacent CuO$\mathrm{_2}$ layers become decorrelated above $\sim 1$ K. Phonon modes with character predominantly inside these pla\-quettes become 2D, leading to the low-temperature linear resistivity term. For the double-chain cuprate, YBa$\mathrm{_2}$Cu$\mathrm{_4}$O$\mathrm{_8}$, if the fluctuating dumbbells between adjacent CuO$\mathrm{_2}$ layers are correlated, then these phonons remain 3D, leading to a low-temperature resistivity that is quadratic in temperature, as observed.\cite{Proust2016}

\item{the pseudogap and its vanishing at $x\approx0.19$ doping from \textit{counting} isolated plaquettes (not adjacent to another doped plaquette in the same CuO$_2$ plane) and their spatial distribution.\cite{Tahir-Kheli2015,Tahir-Kheli2011}}

\noindent As discussed in Appendix \ref{sec:PG}, there is a degeneracy near the Fermi level of the planar states inside an isolated plaquette. The degeneracy is broken by interaction with the environment. A nearby isolated plaquette strongly splits the degeneracy and leads to the pseudogap.

\item{the ``universal" room-temperature thermopower by \textit{counting} the sizes of the insulating AF and metallic regions and taking the weighted average of the thermopower of each region.\cite{Tahir-Kheli2015,Tahir-Kheli2010}}

\noindent Since the room-temperature thermopower of the AF region is $\sim 100\ \mu V/K$ and the metallic region thermopower is $\sim -10\ \mu V/K$, there is a rapid decrease in the thermopower as the size of the metallic region increases with doping.

\item{the STM doping incommensurability by \textit{counting} the size of the metallic regions.\cite{Tahir-Kheli2015,Tahir-Kheli2010}}

\item{the energy of the $(\pi,\pi)$ neutron spin scattering resonance peak by \textit{counting} the size of the AF regions.\cite{Tahir-Kheli2015,Tahir-Kheli2010}} The resonance peak arises from the finite spin correlation length of the AF regions.
\end{itemize}
\clearpage

In this paper, we use exactly the same doped electronic structure described above to explain the superconducting \tc\ and its evolution with doping. We show that Oxygen atom phonon modes at and adjacent to the interface between the insulating and metallic regions lead to superconductivity. We estimate the magnitude of the electron-phonon coupling and obtain the following:

\begin{itemize}
\item{a large $T_{c}\sim100$ K from phonons (because the range of the electron-phonon coupling near the metal-insulator interface increases from poor metallic screening),}
\item{the observed \tc-dome as a function of hole doping (since the total pairing is the product of the size of the metallic region times the interface size),}
\item{the large \tc\ changes as a function of the number of CuO$_2$ layers per unit cell (from inter-layer phonon coupling of the interface O atoms plus inhomogeneous hole doping of the layers),}
\item{the D-wave symmetry of the superconducting Cooper pair wavefunction (also known as the D-wave superconducting gap).}

\noindent In general, an isotropic S-wave superconducting pair wavefunction is energetically favored over a D-wave pair wavefunction for phonon induced superconductivity. However, the fluctuating dumbbells reduce the S-wave \tc\ below the D-wave \tc\ by drastically increasing the Cooper pair electron repulsion.
\item{the lack of a superconducting \tc\ isotope effect at optimal doping (due to the random anharmonic potentials of each pairing O atom).}
\item{by overlapping plaquettes, the fluctuating dumbbells become frozen and the S-wave pair wavefunction \tc\ rises above the D-wave \tc. (Figure \ref{hightc}).}

\noindent{\textbf{While maintaining the same metallic ``footprint" of optimal doping ($\mathbf{x=0.16}$), completely frozen dumbbells lead to an S-wave $\mathbf{T_c}$ of $\mathbf{\approx400}$ K when the D-wave $\mathbf{{T_c}=100}$ K (Figure \ref{hightc}).}}
\end{itemize}

All the \tc's in this paper are computed using the strong coupling Eliashberg equations\cite{Allen1982} as detailed in Appendix \ref{eliashberg1}. These equations include the electron ``lifetime" effects that substantially decrease \tc\ from the simple BCS \tc\ expression.

These results are shown in the following set of $\mathit{T_c}$ \textsc{Concepts}.
\clearpage

\begin{figure}[h]
\centering \includegraphics[width=14cm]
{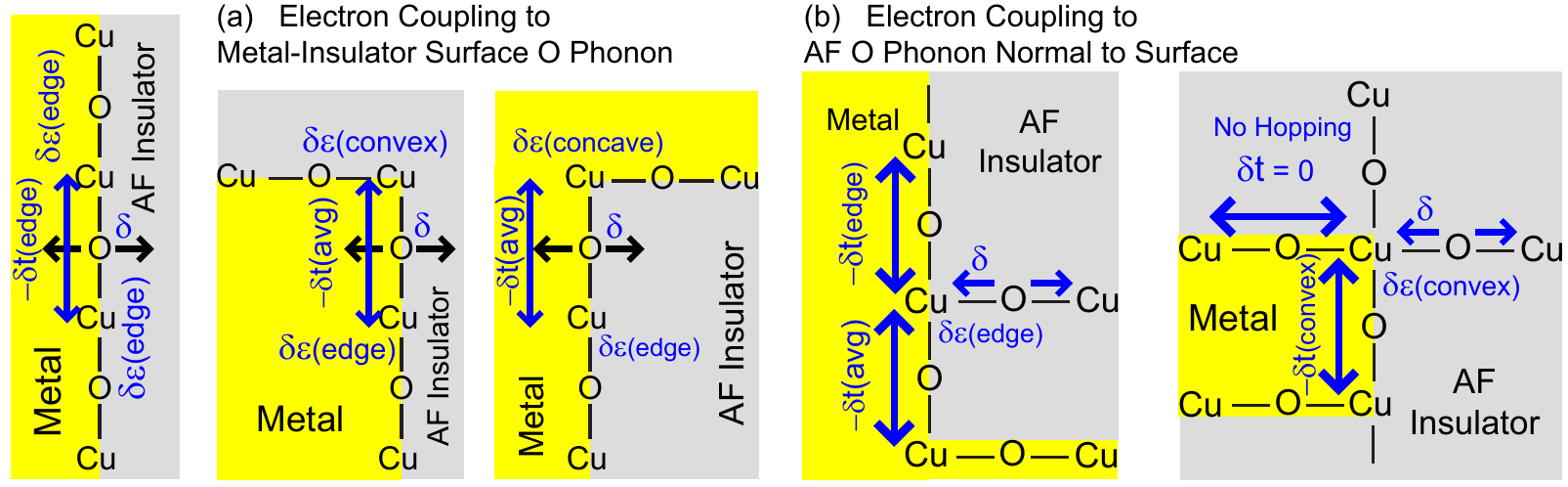}
\caption{
(a) The planar O atom phonon mode at the metal-insulator interface. Its displacement is planar and normal to the $\mathrm{Cu-O-Cu}$ along the metal-insulator interface. Displacement by $\delta$ leads to changes to the neighboring interface Cu \dxxyy\ orbital energy by $\delta\epsilon$ ($\delta\epsilon>0$). There are three kinds of Cu atoms, edge, convex, and concave. Similarly, there is a change in the Cu to Cu hopping matrix element $-\delta t$ ($\delta t>0$). We choose $\delta t$ to equal the average of the $\delta t$ at each Cu. (b) The planar O atom phonon mode in the insulating region adjacent to the metal-insulator interface. The displacement is planar and normal to the metal-insulator interface. The $\delta\epsilon$ and $\delta t$ are defined in the same way as in (a). There is a $\delta\epsilon$ for the metallic Cu closest to the O atom and two $-\delta t$ for hopping to neighboring metallic Cu sites. The metallic screening of the O atom charge is not strong because it resides in the insulating AF region. Hence, the two $\delta t$ are large. In this paper, the vibrational energy of these two phonon modes is set to 60 meV.\cite{Pintschovius2005} See Table \ref{fixedpars}.
}
\label{phonons}
\end{figure}

\noindent\textsc{\underline{\tc\ Concept 1}:} \textbf{There are two planar O atom phonon modes (one at the metal-AF insulator interface and the other adjacent to the interface on the insulating side) that have longer-range electron coupling due to poor electron screening from the metallic region.} See Figure \ref{phonons}. For the remainder of the paper, we use the ``effective" single band model for the metallic band.\cite{Hashimoto2008} In this model, the planar O atoms are eliminated. The model has a single effective Cu \dxxyy\ orbital per Cu in the CuO$\mathrm{_2}$ plane with an effective hopping to neighboring metallic Cu atoms. The parameters of the band structure are the Cu \dxxyy\ orbital energy and the hopping terms (Table \ref{band}, Appendix \ref{parameters}).

\noindent\textsc{\underline{\tc\ Concept 2}:} \textbf{The typical magnitude of the electron-phonon coupling matrix element, $\mathbf{g}$, is the geometric mean\cite{Allen1982} of the Debye energy, $\mathbf{\omega_D}$, and the Fermi energy, $\mathbf{E_F}$, or $\mathbf{g=\sqrt{\omega_D E_F}}$.} The derivation is given in Appendix \ref{sec:e-ph}. For $0.02$ eV $<\omega_D<0.1$ eV and $E_F=1$ eV, we find $0.14$ eV $<g<0.32$ eV. All \tc\ results in this paper use electron-phonon coupling parameters in this range.
\clearpage

\begin{figure}[h]
\centering \includegraphics[width=5cm]
{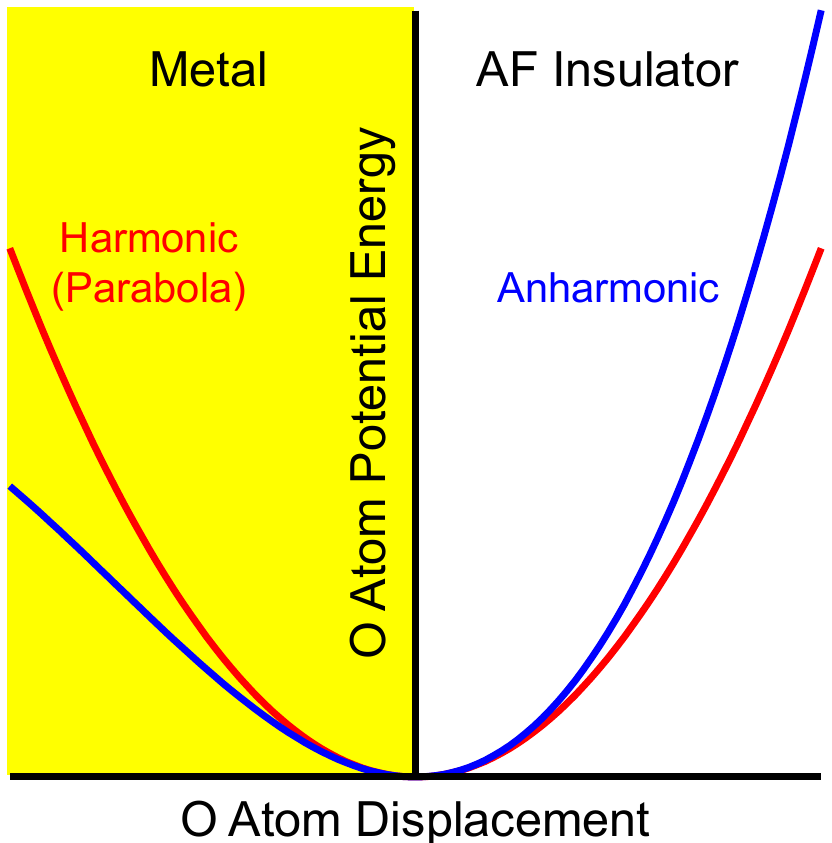}
\caption{Anharmonicity of the O atom phonon mode in Figure \ref{phonons}a due to the difference of the electron screening in the metallic and insulating regions. For Figure \ref{phonons}b, the yellow overlay is shifted to the left.
}
\label{anharmonic}
\end{figure}

\noindent\textsc{\underline{\tc\ Concept 3}:} \textbf{The potential energy of each O atom in Figure \ref{phonons} is strongly anharmonic due to the difference of the electron screening in the metallic and insulating regions.} See Figure \ref{anharmonic}. In fact, the phonon mode shown in Figure \ref{phonons}a is anharmonic even without a nearby metal-insulator boundary. The ``floppiness" of the bond-bending of a linear chain (here, the planar Cu$-$O$-$Cu chain) has been emphasized by Phillips,\cite{Phillips2007} and seen by neutron scattering (the F atom\cite{Li2011} in ScF$\mathrm{_3}$ and the Ag atom\cite{Lan2014} in Ag$\mathrm{_2}$O). However, without the metal-insulator boundary, reflection symmetry would force the electron-phonon coupling for this mode to be zero.

\begin{figure}[h]
\centering \includegraphics[width=14cm]
{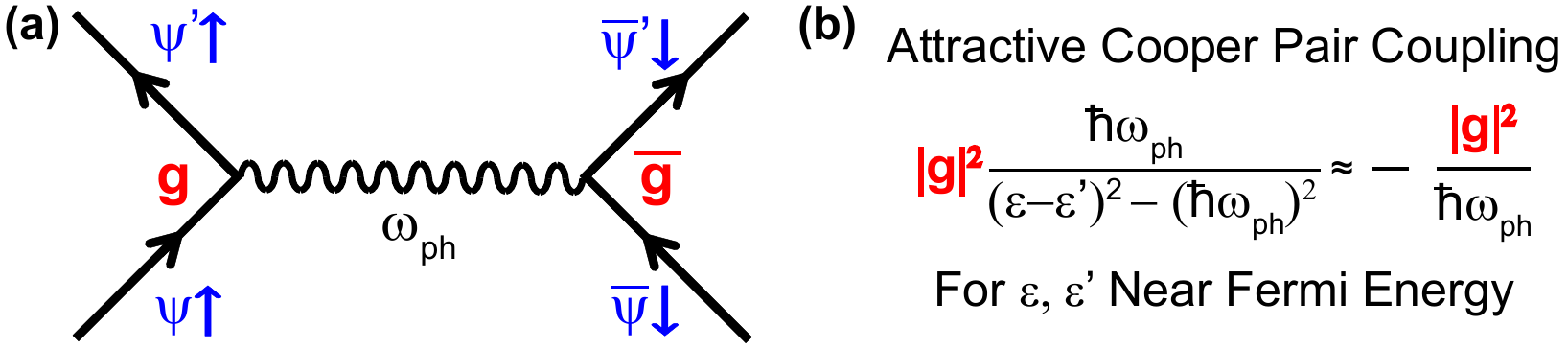}
\caption{The scattering matrix element between two Cooper pairs $(\psi\uparrow,\overline\psi\downarrow)$ and $(\psi'\uparrow,\overline{\psi'}\downarrow)$ with the exchange of a phonon. The two electrons in each pair are time-reversed partners. The symbol $\overline{x}$ means the complex conjugate of $x$. For scatterings in the vicinity of the Fermi level, the matrix element is always negative, leading to a superconducting pair wavefunction that is a linear superposition of Cooper pairs with the same sign. Such a pair wavefunction is called isotropic S-wave. Unfortunately, it is known that the cuprate pair wavefunction changes sign and is of D-wave form (more specifically, of \dxxyy\ form). See \tc\ \textsc{Concept} 6 for a resolution to the problem.
}
\label{pairing}
\end{figure}

\noindent\textsc{\underline{\tc\ Concept 4}:} \textbf{Near optimal doping, $\mathbf{x\approx0.16}$, there is no T$\mathbf{_c}$ isotope effect.} Harmonic potentials have no isotope variation of the superconducting pairing strength because the pairing is inversely proportional to $M\omega^2$ where $M$ is the O atom mass and $\omega$ is the angular frequency of the phonon mode. For a derivation of this result, substitute $g=\sqrt{(\hbar/M\omega)}\nabla V$ into the pairing coupling in Figure \ref{pairing}b, where $V$ is the electron potential. Since $M\omega^2=K$, where $K$ is the spring constant, there is no pairing isotope effect. For anharmonic potentials, the phonon pairing strength becomes dependent on the isotope mass.\cite{Hui1974} Anharmonic potentials can decrease or increase the \tc\ isotope effect depending on the details of the anharmonicity.\cite{Crespi1991,Crespi1991a,Crespi1993} Near optimal doping, the metallic and insulating environments for each O atom phonon is random, leading to an average isotope effect of zero, as observed.\cite{Keller2005,Keller2008} The O atom environment becomes less random at lower dopings, as seen in Figure \ref{dop00-10}. Hence, the isotope effect appears at low dopings.\cite{Keller2005,Keller2008}

\clearpage

\noindent\textsc{\underline{\tc\ Concept 5}:} \textbf{Cooper pairing from phonons is maximally phase coherent for an isotropic S-wave pair wavefunction because the sign of the pairing matrix element in Figure \ref{pairing} is always negative. However, a D-wave pair wavefunction is observed for cuprates. It appears \textit{prima facie} that phonons cannot be responsible for superconductivity in cuprates.} Since Cooper pairs are comprised of two electrons in time-reversed states, the sign of the Cooper pair scattering is always \emph{negative} and of the form $\sim(-)|g|^2/\hbar\omega_{ph}$,\cite{deGennes-book,Schrieffer-book} where $g$ is the matrix element to emit a phonon and $\hbar\omega_{ph}$ is the energy of the phonon mode. See Figure \ref{pairing}. Hence, the lowest energy superconducting pairing wavefunction is a linear superposition of Cooper pairs with the \emph{same} sign. It is called the isotropic ``S-wave" state. In theory, the pair Coulomb repulsion, $\mu$, could suppress the S-wave state and lead to a D-wave state because $\mu$ cancels out of \tc\ when performing the angular integral around the D-wave pair wavefunction. However, the electrons in a pair can couple via a phonon while avoiding each other (due to retardation of phonons). The ``effective" repulsion, $\mu^*$, known as the Morel-Anderson pseudopotential\cite{Bogoliubov-book,Morel1962,deGennes-book,Schrieffer-book,Cohen1969,Allen1982} is too small to raise the D-wave \tc\ higher than the S-wave \tc. Unless there is a mechanism for drastically increasing $\mu^*$, any phonon model for cuprate superconductivity is bound to fail to obtain the correct superconducting pair wavefunction. We show in \tc\ \textsc{Concept} 6 that the fluctuating dumbbells in Figure \ref{dynamicJT} increase $\mu^*$ to $\mu^* \sim \mu$, leading to a D-wave pairing wavefunction.

\clearpage

\begin{figure}[h]
\centering \includegraphics[width=14cm]
{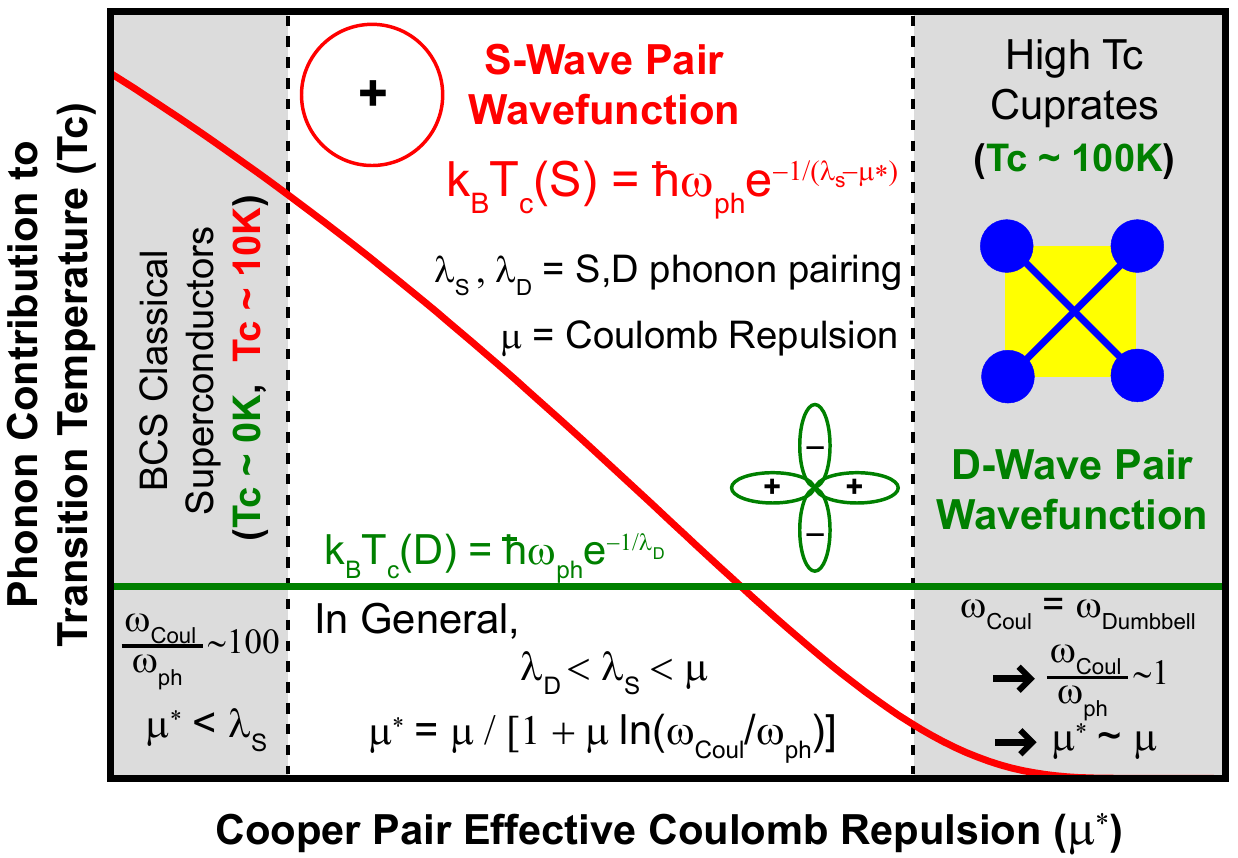}
\caption{Evolution of the S-wave and D-wave superconducting pairing wavefunctions as a function of the Coulomb pseudopotential, $\mu^*$. Typically, $\mu^*$ is small, leading to S-wave pairing as seen on the left-side of the figure. The fluctuating dumbbells raise $\mu^*$ to $\mu^*\sim\mu$ by disrupting the metallic electrons from screening the bare Coulomb repulsion, $\mu$. As $\mu^*$ increases, the S-wave \tc\ decreases while the D-wave \tc\ remains unchanged as shown in the red and green \tc\ expressions for S and D-wave, respectively. As $\mu^*$ approaches the ``bare" Coulomb repulsion, $\mu$, the D-wave \tc\ becomes the favored pair symmetry for the superconductor. The right-side of the figure applies to cuprates.
}
\label{S-D-wave}
\end{figure}

\noindent\textsc{\underline{\tc\ Concept 6}:} \textbf{The fluctuating dumbbells suppress the S-wave pairing wavefunction and lead to a D-wave pairing wavefunction.} See Figure \ref{S-D-wave}. The expression for the Morel-Anderson Coulomb pseudopotential,\cite{Bogoliubov-book,Morel1962,deGennes-book,Schrieffer-book,Cohen1969,Allen1982} $\mu^*$, is shown in Figure \ref{S-D-wave}. It depends on the ratio of the Coulomb and phonon energy scales, $\omega_{Coul}/\omega_{phonon}$. Since this ratio is large, $\mu^*$ is small, leading to an S-wave pair wavefunction rather than the experimentally observed D-wave pair wavefunction.\cite{Tsuei2000} The fluctuating dumbbell frequency, $\omega_{Dumbbell}$, is of the same order as $\omega_{phonon}$ because of the dynamic Jahn-Teller distortion of the planar O atoms in Figure \ref{dynamicJT}. The O atom distortion disrupts the metallic screening of the Coulomb repulsion, and thereby increases $\mu^*$ as shown in Figure \ref{S-D-wave}. In essence, $\omega_{Dumbbell}$ substitutes for $\omega_{Coul}$ in the expression for $\mu^*$. When $\mu^*\sim\mu$, a D-wave pair wavefunction is formed.

\clearpage

\begin{figure}[h]
\centering \includegraphics[width=9.5cm]
{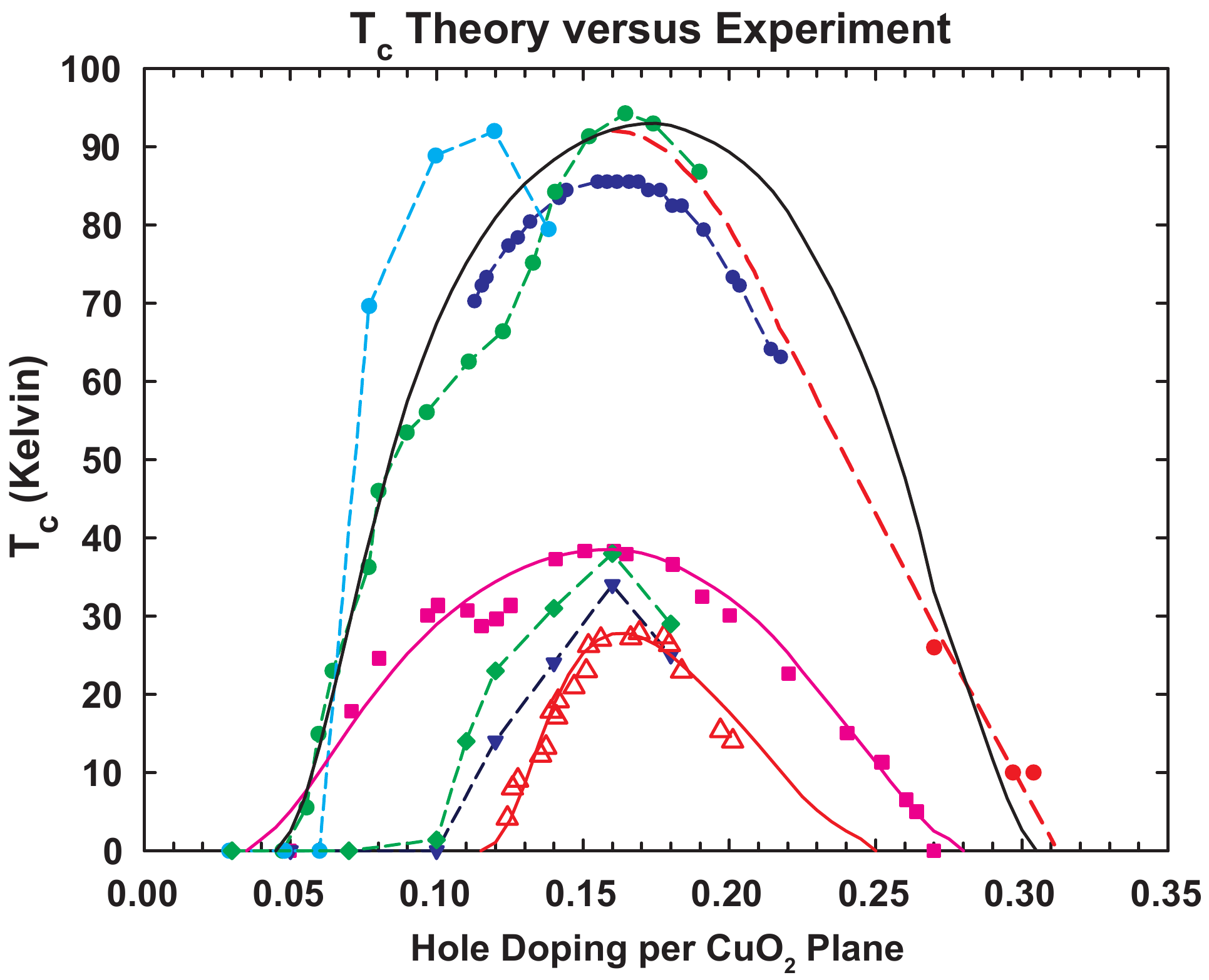}
\caption{Comparison of Experimental and Computed \tc-domes. Bi$_2$Sr$_2$CaCu$_2$O$_{8+\delta}$ (light blue solid circles\cite{Karppinen2003}), YBa$_2$Cu$_3$O$_{7-\delta}$ (green circles\cite{Liang2006}), Y$_{1-y}$Ca$_y$Ba$_2$Cu$_3$O$_{7-\delta}$ (dark blue circles\cite{Naqib2003}), \lsco\ (magenta squares\cite{Yoshida2007}), Bi$_2$Sr$_2$CuO$_{6+\delta}$ (green diamonds\cite{Ono2003} and solid blue triangles,\cite{Hashimoto2008})  6\% Zn doped YBa$_2$(Cu$_{0.94}$Zn$_{0.06}$)$_3$O$_{7-\delta}$ (open red triangles\cite{Naqib2003}), and Tl$_2$Ba$_2$CuO$_{6+\delta}$ (solid red circles\cite{Bangura2010,Rourke2010}). The dashed red line is the proposed Tl$_2$Ba$_2$CuO$_{6+\delta}$ curve by Hussey et al.\cite{Bangura2010, Rourke2010} The solid black, magenta, and red lines are computed. All parameters are described in Appendix \ref{parameters}.
}
\label{tcdome}
\end{figure}

\noindent\textsc{\underline{\tc\ Concept 7}:} \textbf{Interface O atom phonon pairing explains the experimental $\mathbf{T_c}$ domes.} Figure \ref{tcdome} shows the calculated \tc-domes versus experiment as a function of doping for different cuprates using the phonon modes from Figure \ref{phonons} and the electron-phonon couplings estimated in \tc\ \textsc{Concept} 2. All three computed D-wave \tc\ domes were obtained from the strong-coupling Eliashberg equations for \tc.\cite{Allen1982,Schrieffer1963,Scalapino1966} Other phonon modes also contribute to \tc. These phonons primarily reduce the magnitude of \tc\ due to their contribution to electron pair ``lifetime effects" (strictly speaking, the `` wavefunction renormalization effects"). The effect of all the phonon modes on \tc\ are included in our computations. All the details of the band structure, the interface O phonon coupling parameters, and the inclusion of the remaining phonons into the Eliashberg calculations are described in appendices \ref{parameters} and \ref{eliashberg1}. We intentionally chose our parameters to be simple and conceptual. We did not attempt to fit the experimental points exactly. Our goal is to demonstrate that reasonable electron-phonon couplings and our proposed inhomogeneous cuprate electronic structure are sufficient to understand the experimental \tc-domes.

\clearpage

\begin{figure}[h]
\centering \includegraphics[width=13cm]
{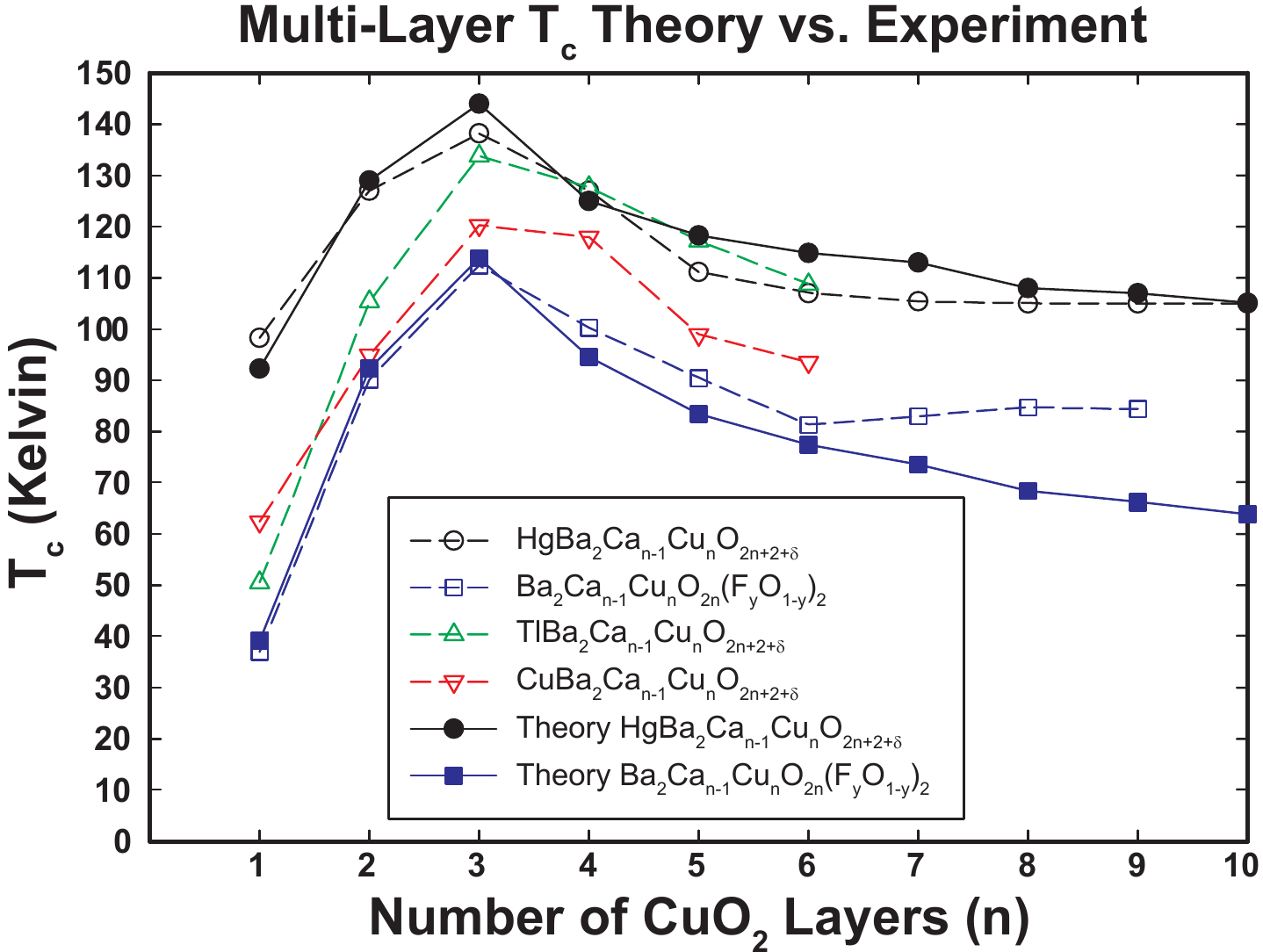}
\caption{Comparison of the computed and experimental \tc\ as a function of CuO$\mathrm{_2}$ layers per unit cell. The experimental data points (open symbols and dashed lines) are from Mukuda et al.\cite{Mukuda2012} Two theoretical \tc\ curves are shown (solid symbols and solid lines). All parameters, including the hole dopings in each CuO$\mathrm{_2}$ layer, are in Appendix \ref{parameters}. The generalization of the Eliashberg equations for one CuO$\mathrm{_2}$ plane to multi-layers is described in Appendix \ref{eliashberg-multi}. If the hole doping in each layer was the same, then the computed \tc\ curve would monotonically increase with the number of layers, $n$, and saturate for large $n$. Since the hole doping for the inner layers is less than the hole doping on the outermost layers (see Table \ref{layer-doping} in Appendix \ref{parameters}), there are fewer interface O atom phonons that contribute to the high \tc\ in the inner layers. The maximum \tc\ occurs at three layers.
}
\label{tclayer}
\end{figure}

\noindent\textsc{\underline{\tc\ Concept 8}:} \textbf{The experimental variation of $\mathbf{T_c}$ with the number of CuO$\mathbf{_2}$ layers per unit cell is due to interlayer coupling of the interface O atom phonons and the nonuniform hole doping between layers.} Since the O atom phonons near the metal-insulator interface are longer-ranged, they couple to adjacent CuO$\mathrm{_2}$ planes. Hence, there is a strong dependence of \tc\ on the number of CuO$\mathrm{_2}$ layers per unit cell. In addition, the Cu Knight shift measurements of Mukuda et al.\cite{Mukuda2012} have shown that the hole doping is not the same in each CuO$\mathrm{_2}$ layer. The computed \tc\ as a function of the number of CuO$\mathrm{_2}$ layers is shown in Figure \ref{tclayer}.

\clearpage

\noindent\textsc{\underline{\tc\ Concept 9}:} \textbf{The D-wave T$\mathbf{_c}$ values computed in Figures \ref{tcdome} and \ref{tclayer} are weakly dependent on the orbital energy change, $\mathbf{\delta\epsilon}$, and strongly dependent on the hopping energy change, $\mathbf{\delta t}$.} See Table \ref{deltatc} for the change in \tc\ at optimal doping of $x=0.16$ for the computed black, red, and magenta curves in Figure \ref{tcdome}.

\begin{table}[h]
\caption{The change in the \tc\ at optimal doping ($x=0.16$) for the three computed curves in Figure \ref{tcdome}. The orbital energy parameter, $\delta\epsilon$, and the hopping energy parameter, $\delta t$, are each changed by 0\% and $\pm10$\% from their initial values found in Appendix \ref{parameters}. In appendices \ref{lam0s} and \ref{lam0perp}, the $\delta\epsilon$ terms lead to a more isotropic electron-phonon pairing, and the $\delta t$ terms are more anisotropic. For a D-wave \tc, an isotropic electron-phonon pairing does not contribute to \tc. In the fourth column (red curve), $\delta\epsilon=0$ (see Appendix \ref{parameters}). Hence, changes to $\delta\epsilon$ do not affect \tc.}
\label{deltatc}
\begin{tabular}{c|ccc}
 Change in & Black Curve & Magenta Curve & Red Curve \\
$(\delta\epsilon,\delta t)$ &\ \ YBa$\mathrm{_2}$Cu$\mathrm{_3}$O$\mathrm{_{7-\delta}}$\ \  &\ \ \lsco\ \ \ &\ \ YBa$\mathrm{_2}$(Cu$\mathrm{_{0.94}}$Zn$\mathrm{_{0.06}}$)$\mathrm{_3}$O$\mathrm{_{7-\delta}}$\ \  \\
\hline
 $(0\%,0\%)$       & 92.2 K  & 38.5 K & 27.7 K \\
 $(0\%,-10\%)$     & 83.1 K  & 30.7 K & 19.9 K \\
 $(0\%,+10\%)$     & 100.0 K & 46.3 K & 35.6 K \\
 $(-10\%,0\%)$     & 93.6 K  & 39.4 K & 27.7 K \\
 $(-10\%,-10\%)$   & 84.7 K  & 31.3 K & 19.9 K \\
 $(-10\%,+10\%)$   & 101.1 K & 47.5 K & 35.6 K \\
 $(+10\%,0\%)$     & 90.5 K  & 37.5 K & 27.7 K \\
 $(+10\%,-10\%)$   & 81.4 K  & 30.1 K & 19.9 K \\
 $(+10\%,+10\%)$   & 98.5 K  & 45.0 K & 35.6 K \\
\hline
\end{tabular}
\end{table}

From Table \ref{deltatc}, a 10\% increase in $\delta\epsilon$ always decreases the D-wave \tc\ by $\approx 2-3\%$. A $\pm10\%$ change in $\delta t$ leads to $\approx\pm10-30\%$ change in the D-wave \tc. In appendices \ref{lam0s} and \ref{lam0perp}, the exact dependence of the electron-phonon pairing parameter, $\lambda$, is derived. The contribution of $\delta\epsilon$ to $\lambda$ is approximately isotropic around the Fermi surface leading to a weak dependence of the D-wave \tc\ on changes in $\delta\epsilon$. In contrast, an S-wave pairing symmetry \tc\ depends strongly on both $\delta\epsilon$ and $\delta t$. The weak dependence of the D-wave \tc\ on $\delta\epsilon$ implies our choices for the $\delta\epsilon$ parameters for the \tc\ curves in Figures \ref{tcdome} and \ref{tclayer} are not accurately fitted by the experimental \tc\ data. The uncertainty in the magnitude of $\delta\epsilon$ leads to an S-wave \tc\ range from $\approx270-400$ K due to dopant ``crowding," as shown next.
\clearpage

\begin{figure}[h]
\centering \includegraphics[width=16cm]
{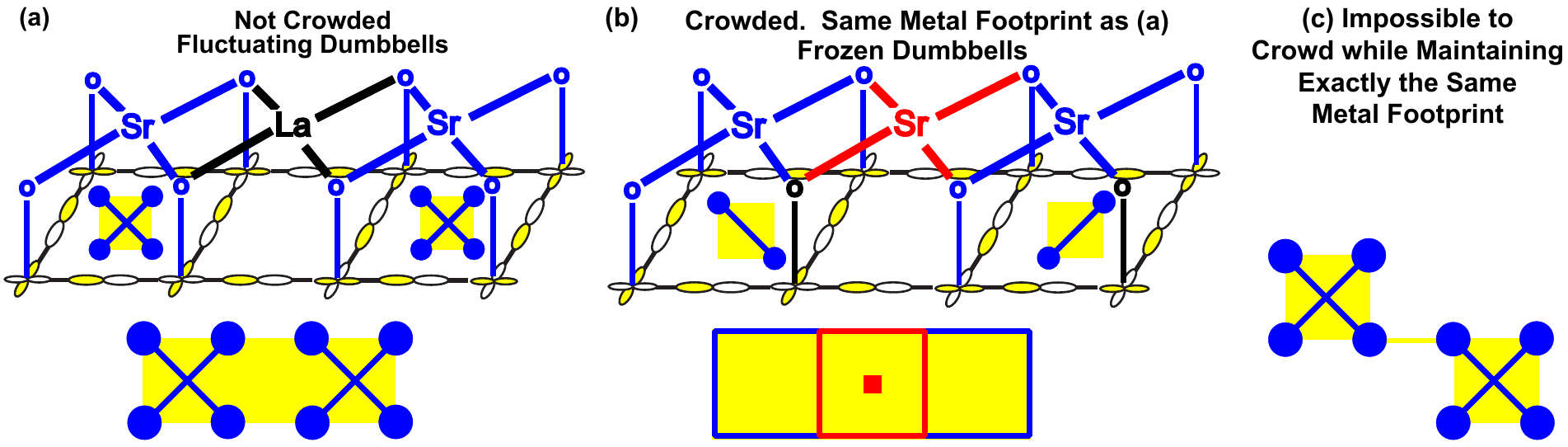}
\caption{(a) Two non-overlapped plaquettes with their fluctuating dumbbells and their metallic region (yellow) for the case of Sr dopants in \lsco. The bottom schematic shows the fluctuating dumbbells as crosses and the metallic region in yellow. (b) Adding a Sr dopant between the two plaquettes freezes the two fluctuating dumbbells while maintaining exactly the same metallic footprint. The S-wave \tc\ increases due to the decrease of the Coulomb pseudopotential while the D-wave \tc\ remains unchanged. The blue squares in the bottom schematic represent the frozen dumbbells. The added Sr dopant is shown by the red square. However, any atom that breaks the degeneracy of the dumbbell states will work instead. The yellow metallic overlay is unchanged from (a). The orientation of the frozen dumbbells is arbitrary in the figure because it depends on the environment (not shown). (c) Two adjacent plaquettes that are shifted by one lattice spacing. No plaquette can be added to freeze the dumbbells while maintaining the same metallic footprint.
}
\label{crowding}
\end{figure}

\begin{figure}[h!]
\centering \includegraphics[width=8.6cm]
{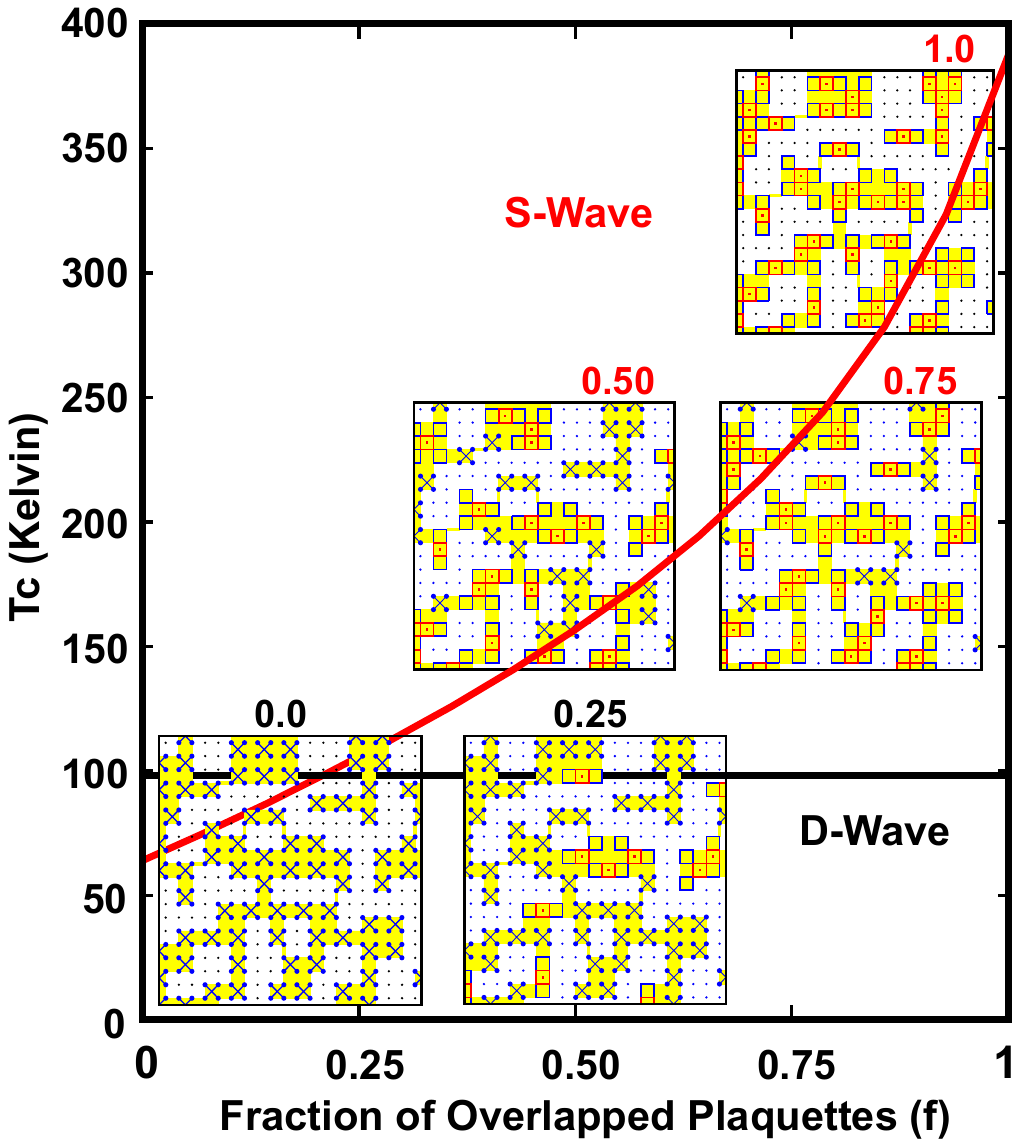}
\caption{The S-wave and D-wave \tc's as a function of overlapped plaquette fraction, $f$. The insets show the evolution of the frozen dumbbells (blue and red squares) for optimal doping of $x=0.16$. The metallic footprint is unchanged. The D-wave \tc\ is constant (the Coulomb pseudopotential, $\mu^*$, has no effect). The maximum S-wave \tc\ is 387.2 K and the D-wave \tc\ is 98.0 K. We assume $\mu^*$ varies linearly with $f$, $\mu^*(f,i\omega_n) = \mu^*_{Fluc}F(i\omega_n)(1-f)+\mu^*_{BCS}$, where $\mu^*_{Fluc}=7$ (the fluctuating dumbbell $\mu^*$), $\mu^*_{BCS}=0.1$ (a typical BCS value), and the cutoff $F(i\omega_n)=\omega_{Fluc}^2/(\omega_n^2+\omega_{Fluc}^2)$. We set $\omega_{Fluc}=60$ meV. The Eliashberg imaginary frequency, $i\omega_n$, is defined in Appendix \ref{eliashberg1}. In order to achieve 100\% frozen dumbbells, ``domino" doping as shown in Figures \ref{crowding}a and b was used. We set $\delta\epsilon=\delta t=0.30,\ 0.30,\ 0.15$ eV for Edge, Convex, and Concave orbital and hopping energies, respectively. See the definitions in Figure \ref{phonons}. We used a larger $\delta\epsilon$ than Figure \ref{tcdome} and Appendix \ref{parameters} because $\delta\epsilon$ should be greater than $\delta t$ due to its proximity to the planar O atom. Since the D-wave \tc\ is weakly dependent on $\delta\epsilon$, our choices in Figure \ref{tcdome} were very conservative. The S-wave \tc\ values for 100\% frozen dumbbells for the black, magenta, and red curves in Figure \ref{tcdome} at optimal doping are 280.1 K, 164.4 K, and 99.0 K, respectively. The corresponding D-wave \tc\ values are 92.2 K, 38.5 K, and 27.7 K. Hence, the S-wave \tc\ is in the range of $\approx280-390$ K.
}
\label{hightc}
\end{figure}

\noindent\textsc{\underline{\tc\ Concept 10}:} \textbf{Overlapping plaquettes (``crowding" the dopants) freeze the dumbbells, decrease the Coulomb pseudopotential, $\mathbf{\mu^*}$, and thereby raise the S-wave T$\mathbf{_c}$.} If the same metallic ``footprint" can be maintained, then there is no change in the phonon pairing. Only $\mu^*$ is reduced (see Figure \ref{S-D-wave}). If all the dumbbells can be frozen, then from Figures \ref{pairing} and \ref{S-D-wave}, the S-wave \tc\ will be larger than the D-wave \tc. Figure \ref{crowding} shows how two plaquettes with fluctuating dumbbells can be crowded by adding an additional dopant (Sr in the figure) while retaining exactly the same metallic footprint. For random doping, there will always exist adjacent plaquette pairs as shown in Figure \ref{crowding}c that cannot be overlapped by another plaquette within the existing metallic footprint. There are two ways to obtain an optimally doped metallic footprint and freeze 100\% of the dumbbells. First, dope ``dominoes" (adjacent pairs of plaquettes as in Figure \ref{crowding}a and b). Second, dope to less than optimum doping. Next, crowd all of the plaquettes in such a way as to end up with an optimally doped metallic footprint and 100\% frozen dumbbells.

\clearpage

\noindent\textsc{\underline{\tc\ Concept 11}:} \textbf{Crowding dopants while maintaining the optimal doping metallic footprint leads to room temperature S-wave $\mathbf{T_c}$.} See Figure \ref{hightc}.

\clearpage

\noindent\textsc{\underline{Cuprate Physics and Its Analogy to Chemical Dissociation}}

\begin{figure}[h]
\centering \includegraphics[width=11cm]{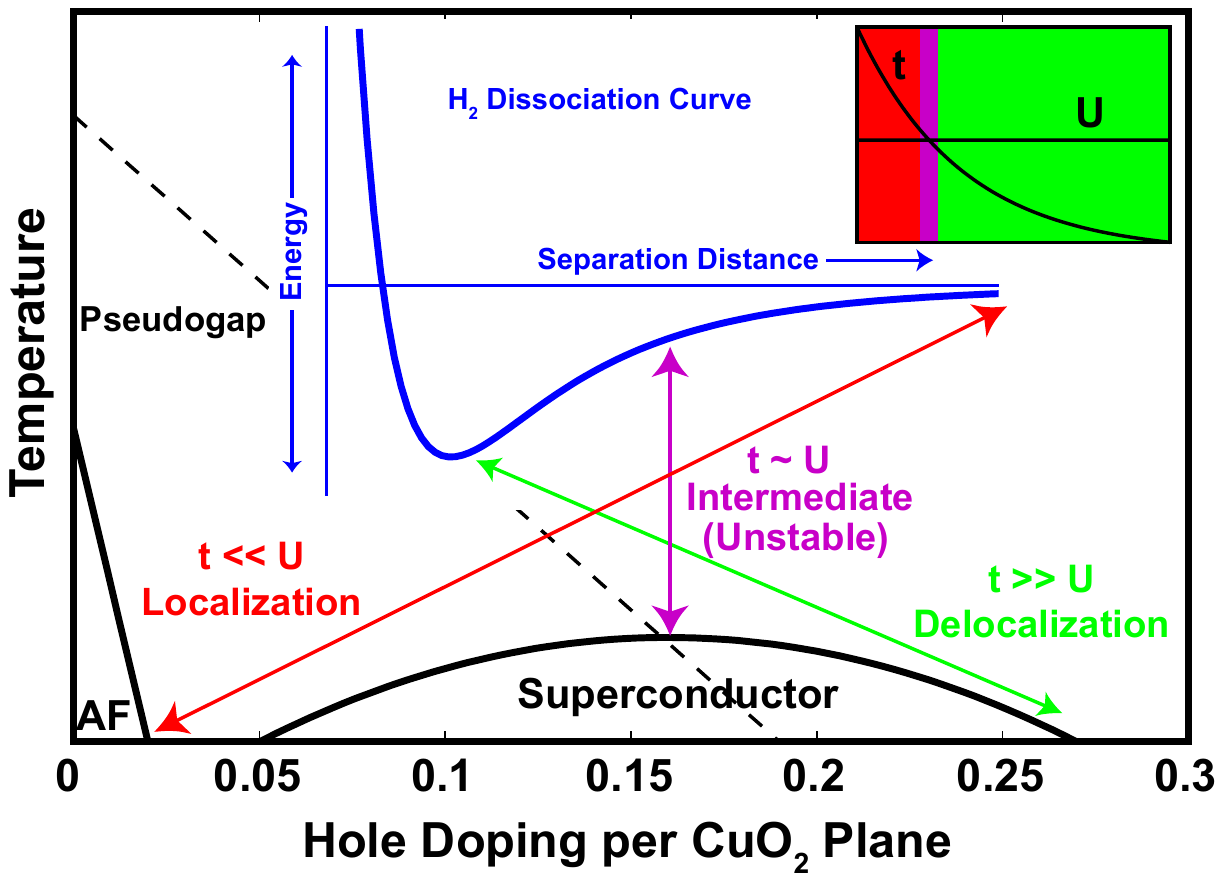}
\caption{The iconic H$\mathrm{_2}$ dissociation curve in chemistry (blue inset) and the iconic cuprate phase diagram (larger black figure). Both figures show the regimes of delocalized, intermediate, and localized electronic states. For H$\mathrm{_2}$, electron delocalization occurs near the equilibrium bond distance of 0.74 \AA, where the effective hopping, $t$, is larger than the effective Coulomb repulsion, $U$, of the electrons (see inset in upper right corner). Molecules do \emph{not} exist at the intermediate (magenta) region since the potential is not a local minimum. Similary, semiconductors can be classified into delocalized (covalent) and localized (ionic) with a sharp transition in-between based upon their Fermi level stabilization at a semiconductor-metal interface.\cite{Kurtin1969} In cuprates, the intermediate regime is where the optimal superconducting \tc\ occurs. In analogy to molecules and semiconductors, we believe cuprates are unstable here, and thereby ``phase separate" into delocalized metallic and insulating magnetic AF regions. Finally, we intentionally avoided using the terms ``weak correlation" and ``strong correlation" in the figure. The word correlation means different things to different scientists. For chemists, the delocalized region ($t\gg U$) is ``weak electron correlation," and the localized region ($t\ll U$) is ``strong electron correlation." For physicists, cuprates are considered a ``strong correlation" problem.
\label{H2}
}
\end{figure}

In Figure \ref{H2}, the ground state electronic wavefunction of H$\mathrm{_2}$ at the equilibrium bond separation of 0.74 \AA\ is well approximated by a restricted Hartree-Fock form (a spin up and spin down electron pair occupying the same bonding orbital). In the language of an effective electron hopping, $t$, and an onsite Coulomb repulsion, $U$, this region is represented by $t\gg U$. At dissociation ($t\ll U$), the ground state electronic wavefunction is highly correlated. The wavefunction is large only when there is one electron on each H atom.

From Figure \ref{H2}, the optimal superconducting \tc\ of cuprates is at ``intermediate" correlation. Molecules do not generally ``settle" at intermediate correlation. Since the dopants in cuprates are frozen in at high temperatures, the material avoids intermediate correlation by phase separating on an atomic-scale into a metallic (weak correlation) and an insulating AF (strong correlation) regions.

Atomic-scale inhomogeneity explains three important materials issues about cuprates. First, cuprates ``self-dope" to approximately optimal \tc. Since plaquette overlap occurs at $x=0.187$ doping, we believe it is energetically favorable for dopants to enter the crystal until their plaquettes begin to overlap. Adding further dopants is energetically unfavorable. The change in \tc\ between optimal doping ($x\approx0.16$) and plaquette overlap ($x=0.187$) is $\approx5\%$. Hence, cuprates ``self-dope" to approximately optimal \tc\ as a consequence of the energetics of overlapping plaquettes.

Second, YBa$\mathrm{_2}$Cu$\mathrm{_3}$O$\mathrm{_{7-\delta}}$ cannot be doped past $x\approx0.23$, as shown in Figure \ref{tcdome}. The phenomenon can be understood if it is energetically unfavorable to overlap plaquettes that share an edge (occuring at doping $x=0.226$). In the earliest days of cuprate superconductivity, materials scientists had difficulty observing superconductivity in \lsco\ above $\approx0.24$ doping.\cite{Takagi1989} We believe the difficulty was also due to the energetics of overlapping plaquettes with shared edges. Annealing in an O$\mathrm{_2}$ atmosphere solved the \lsco\ overdoping problem. However, the problem still remains for YBa$\mathrm{_2}$Cu$\mathrm{_3}$O$\mathrm{_{7-\delta}}$.

Third, it is known that a room-temperature thermopower measurement is one of the fastest ways to determine if a cuprate sample is near optimal doping for \tc\ because the room-temperature thermopower is very close to zero near optimal doping. This peculiar, but useful, observation can be understood because 2D percolation of the metallic region occurs at $x\approx0.15$ doping. Since the AF region thermopower is large ($\sim +100\ \mu V/K)$ and the metallic thermopower is $\sim -10\ \mu V/K$ at high overdoping, 2D metallic percolation ``shorts out" the AF thermopower and drives the thermopower close to zero near optimal \tc.

Finally, the potential energy curve in the intermediate correlation regime is hard to study for molecules. For H$\mathrm{_2}$, the equilibrium bond distance is 0.74 \AA.  The intermediate correlation regime is at $\approx2.0$ \AA\ bond separation. At this distance, the blue potential energy curve in Figure \ref{H2} can only be observed indirectly\cite{Herzberg-vol1,Herzberg-vol2,Wilson-book} because it is not at a local minimum. For H$\mathrm{_2}$, the ultraviolet spectrum of the vibrational modes (there are 14 discrete level below the continuum) can be fitted to a simple Morse potential to estimate the potential energy as a function of the $H-H$ separation distance. The $10^{\mathrm{th}}$ bond-stretching phonon mode probes the potential energy of the two of H atoms up to $\approx2.0$ \AA.

\vspace{.1in}

\noindent\textsc{\underline{Materials Approaches to Room-Temperature T$\mathrm{_c}$ and Large $\mathrm{J_c}$}}

There is enormous ``latent" \tc\ residing in the cuprate class of superconductors from converting the D-wave superconducting pairing wavefunction to an S-wave pairing wavefunction. The result is surprising because it has been assumed by most of the high-\tc\ cuprate community, including the author, that there was something special about the D-wave pairing symmetry that led to $T_c\sim100$ K.

The first thought that comes to mind given our finding is, ``With over 200,000 refereed papers\cite{Mann2011} and 30 years of intensive research in both academia and industry, surely someone overlapped plaquettes, and thereby created a room-temperature S-wave superconductor?"

Our answer is that plaquettes have been overlapped with regularity for 30 years. These materials are all overdoped with doping $x>0.187$, as shown in Figure \ref{dop19-32}. Hence, dumbbells have been frozen and the S-wave \tc\ has increased. However, our calculations find the S-wave \tc\ remains below the D-wave \tc\ for reasonable parameter choices. Unfortunately, the optimally doped metallic footprint is not obtained by naive dopant crowding. Instead, the size of the metallic footprint increases and its pairing interface decreases. The right side of the \tc-dome shown in Figure \ref{tcdome} is the result. Even the layer-by-layer Molecular Beam Epitaxy (MBE) of Bozovic et al.\cite{Pereiro2012} does not control the placement of the dopants in each layer, leading to the same result as above.

While almost everything that can be possibly be suggested for the mechanism for cuprate superconductivity has been suggested in over 200,000 papers (percolation, inhomogeneity, dynamic Jahn-Teller distortions, competing orders, quantum critical points at optimal doping or elsewhere, spin fluctuations, resonating valence bonds, gauge theories, blocked single electron interlayer hopping, stripes, mid-infrared scenarios, polarons, bipolarons, spin polarons, spin bipolarons, preformed Bose-Einstein pairs, spin bags, one-band Hubbard models, three-band Hubbard models, t-J models, t+U models, phonons, magnons, plasmons, anyons, Hidden Fermi liquids, Marginal Fermi liquids, Nearly Antiferromagnetic Fermi liquids, Gossamer Superconductivity, the Quantum Protectorate, etc.), we believe these ideas have lacked the microscopic detail necessary to guide experimental materials design, and in some instances, may have even led materials scientists down the wrong path.

We have shown above that freezing dumbbells in cuprates leads to room-temperature \tc\ (see Figure \ref{hightc}). However, the critical current density, $J_c$, is approximately two orders of magnitude smaller than the theoretical maximum, $J_c\sim10^{-2}J_{c,\mathrm{max}}$, where $J_{c,\mathrm{max}}$ is the depairing limit for Cooper pairs. $J_c$ is small because the conducting pathway in the CuO$\mathrm{_2}$ planes is \emph{extremely} tenuous (see the discussions in the captions of Figures \ref{pathway} and \ref{dop12-18}). For practical engineering, $J_c$ should be at least $\sim10^{-1}J_{c,\mathrm{max}}$.

In cuprates, \tc\ can be raised to room-temperature by freezing dumbbells while maintaining the random metallic footprint found at optimal doping. By fabricating wires (a wire is defined as a continuous 1D metallic pathway through the crystal), \tc\ remains large while $J_c$ increases to at least $\sim10^{-1}J_{c,\mathrm{max}}$.

Our results lead to the following approaches for achieving higher \tc\ and $J_c$. Unless explicitly stated, the bullet points below apply to any type of material (cuprate or non-cuprate).

\begin{itemize}
\item\textit{The material should be inhomogeneous with a metallic region and an insulating region.}

The insulating region does not have to be magnetic. However, we believe the antiferromagnetic insulating region helps maintain the sharp metal-insulator boundary seen in cuprates. An ordinary insulator or a semiconductor with a small number of mobile carriers is sufficient to obtain a longer ranged electron-phonon coupling at the interface because there is less electron screening in the semiconducting (or insulating) region compared to the metallic region.

\item\textit{The ratio of the number of metallic unit cells on the interface (adjacent to at least one insulating unit cell) to the total number of metallic unit cells must be larger than 20\%.}

We use the terms interface and surface interchangeably below.

The number of metallic unit cells on the interface (or surface) must be a large fraction of the total number of metallic unit cells in order for the enhanced electron-phonon pairing at the interface to have an appreciable affect on \tc. From our calculations in Figure \ref{tcdome}, 50\% of optimal \tc\ is obtained when the ratio is $\approx50\%$, and 25\% of optimal \tc\ occurs when the ratio is $\approx35\%$. Below a surface metal unit cells to total metal unit cells ratio of 20\%, \tc\ falls off exponentially, and therefore \tc\ is too low to be useful.

Metallic clusters that are smaller than approximately the coherence length do not contribute to \tc\ due to thermal fluctuations. The surface metal unit cells to total metal unit cells ratio above should only include surface metal unit cells in extended metallic clusters.

Inhomogeneous materials formed at eutectic points have a surface metal unit cells to total metal unit cells ratio of $\sim10^{-3}$ or less if the sizes of the metallic and insulating regions are on the order of microns. Standard materials fabrication methods do not lead to sufficient surface atomic sites for high \tc. Inhomogeneity on the atomic-scale is necessary.

It would appear that parallel 1D metallic wires that are one lattice constant wide (equal to one plaquette width in cuprates) would lead to the maximum surface unit cells to total metal unit cells ratio of 100\%, and thereby a large \tc\ increase. We were surprised to discover that at optimal doping of $x=0.16$, the surface metal unit cells to total metal unit cells ratio is 91\% in cuprates.

Increasing the ratio to 100\% increases \tc\ by only $\approx5\%$ because at higher \tc\ magnitudes, \tc\ no longer increases exponentially with the magnitude of the electron-phonon coupling, $\lambda$ (defined in Appendix \ref{eliashberg1}). Instead, \tc\ scales\cite{Allen1975} as $T_c\sim\sqrt{\lambda}$. A 10\% increase in the surface to total metal unit cells ratio increases $\lambda$ by 10\%, leading to a 5\% increase in \tc. Hence, there is negligible \tc\ to be gained by fabricating wires. However, wires lead to large $J_c$, as discussed below.

In fact, parallel wires that are a few lattice constants in width are bad superconductors because 1D superconductor-normal state thermal fluctuations lead to large resistances below the nominal \tc. By fabricating two (or more) sets of parallel wires that cross each other, the effect of resistive thermal fluctuations in a single wire are suppressed. In the figures below, we show perpendicularly crossed wires in 2D. The same pattern or a different pattern can be used in adjacent layers normal to the 2D wires. Crossing wires in 3D (two or more sets of parallel wires spanning the whole crystal) also leads to high \tc\ and $J_c$.

\item\textit{Add dopants to an insulating parent compound that leads to metallic regions. Doping a metallic parent compound to create insulating regions will work also. In cuprates, the parent compound is insulating and doping creates metallic regions.}

\item\textit{Avoid small disconnected metallic clusters. If they are smaller than the coherence length, they do not contribute to \tc\ due to thermal fluctuations.}

In cuprates, high \tc\ can be obtained at very low doping if all the dopants leading to isolated plaquettes and small plaquette clusters are arranged such that a single contiguous metallic cluster is formed. While the \tc\ may be high, $J_c$ will be low if the size of the metallic region is a small fraction of the total volume of the crystal.

\item\textit{Superconducting wires lead to a small increase of \tc\ and a large increase of $J_c$.}

Metallic wires lead to a tiny increase in \tc, as discussed in the second bullet above. However, metallic wires increase $J_c$ dramatically (up to a factor of $\sim100$) by eliminating the \emph{tortured} conduction pathways shown in Figure \ref{pathway}. For cuprates, optimal \tc\ doping at $x=0.16$ is barely above the 2D percolation threshold of $x\approx0.15$ doping. Hence, the conducting pathways in a single CuO$\mathrm{_2}$ plane are tenuous at optimal doping.

Current materials fabrication methods for cuprates have optimized the \tc\ at the expense of $J_c$. We find this point to be evidence that despite all the proposals in over 200,000 refereed publications,\cite{Mann2011} there has been little guidance to the materials synthesis community on what is relevant at the atomic level for optimizing \tc\ and $J_c$. See Figure \ref{Wires-Crowding}.

\begin{figure}[t]
\centering \includegraphics[width=16cm]{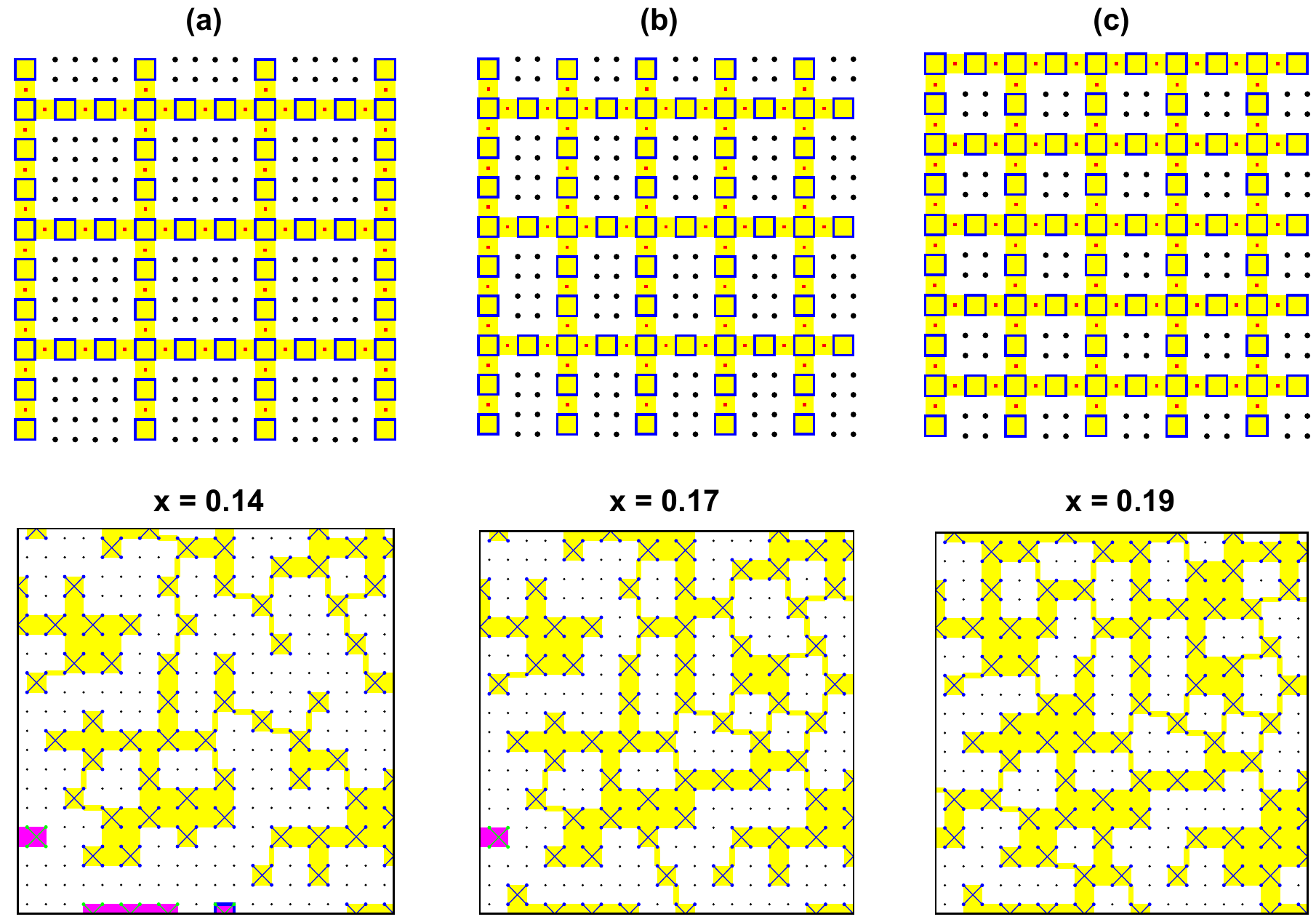}
\caption{``Crowded" crossed wires that have high \tc\ and $J_c$. The crossed metallic wires are one plaquette in width. The symbols in the metallic wires are defined in Figures \ref{crowding} and \ref{hightc}, with one change solely for clarity. Here, the added dopant (solid red square) that overlaps two plaquettes (blue squares) does not have a red square boundary. The black circles are undoped AF Cu spins. The first row of (a), (b), and (c) show crossed wires with $4\times4$, $2\times4$, and $2\times2$ AF spins, respectively, inside each interior region formed by the perpendicular wires on $20\times20$ CuO$\mathrm{_2}$ lattices. The second row shows the corresponding lattices that are randomly doped to approximately the same fraction of metallic unit cells as the crossed wires in the first row ($x=0.14,0.17$, and $0.19$). Larger $40\times40$ lattice figures for the second row can be found in Figures S14, S17, and S19 at the end of the paper. The \tc\ for each wire configuration is approximately equal to the S-wave \tc\ for an optimally doped system since the ratio of the surface metal cells to the total metal cells is 68\%, 75\%, and 80\% for (a), (b), and (c), respectively. The ratio is $\approx91\%$ at random optimal doping. The \tc's of (a), (b), and (c) are approximately 12\%, 8\%, and 6\% less than the S-wave \tc\ at optimal doping. The $J_c$ for (a) is $(1/2)J_{c,max}$, where $J_{c,max}$ is the theoretical maximum of $\sim 2\times10^{7}\ \mathrm{A/cm^2}$ for YBa$\mathrm{_2}$Cu$\mathrm{_3}$O$\mathrm{_{7-\delta}}$. In (b), $J_c=(1/3)J_{c,max}$ and $(1/2)J_{c,max}$ for currents along the x and y-axes, respectively. For (c), $J_c=(1/3)J_{c,max}$. Random optimal doping leads to $J_c\sim10^{-2}J_{c,max}$ (see Figure \ref{pathway}).
}
\label{Wires-Crowding}
\end{figure}

Crossed metallic wires with varying aspect ratios and widths provide many opportunities for optimizing \tc\ and $J_c$ for specific applications. For example, wires that are four metallic atoms wide (equal to two adjacent plaquettes in cuprates), would have $\approx1/2$ the surface to total metal ratio of metallic wires two atoms wide (or one plaquette in cuprates), leading to an $\approx50\%$ reduction in \tc\ compared to wires that are two metallic atoms wide. However, $J_c$ increases by a factor of two.

Generally, it is most favorable to fabricate the narrowest wires that are spaced closely together because both \tc\ and $J_c$ will be large. In addition, interfacial phonon modes will couple to both the closest wire and the next-nearest neighboring wire, leading to further increase in \tc.

For cuprates, the narrowest wire is one plaquette width (see Figure \ref{motif}). Other materials will have a different minimum width scale for wires.

\item\textit{Atomic-Scale metal-insulator inhomogeneity in a 3D material leads to a high-\tc\ 3D S-wave pairing wavefunction. A 3D material is more stable against defects and grain boundaries.}

\item\textit{Metallic and insulating regions provide new opportunities for pinning magnetic flux.}

Strong pinning of magnetic flux lines in superconductors is necessary to obtain large critical current densities, $J_c$. Insulating ``pockets" surrounded by metallic region are energetically favorable for magnetic flux to penetrate. The flux can be strongly bound inside these insulating regions by adding further pinning centers to the insulating region. Examples of insulating pockets are shown in Figure \ref{Wires-Crowding}.

\item\textit{In cuprates, freeze the fluctuating dumbbells in non-overlapping plaquettes while maintaining a metallic footprint with a large surface metallic unit cells to total metallic unit cells ratio.}

The ratio of the isotropic S-wave pairing wavefunction \tc\ to the corresponding D-wave \tc\ is $\approx2.8-4$ (see Figure \ref{hightc}).

\item\textit{In cuprates, fluctuating dumbbells in non-overlapping plaquettes can be frozen by breaking the symmetry inside each plaquette by an atomic substitution into the CuO$\mathrm{_2}$ plane, atomic substitution out of the CuO$\mathrm{_2}$ plane (such as the apical O atom sites), or interstitial atoms, as shown in Figure \ref{Wires-Asymmetry}.}

\begin{figure}[t]
\centering \includegraphics[width=16cm]{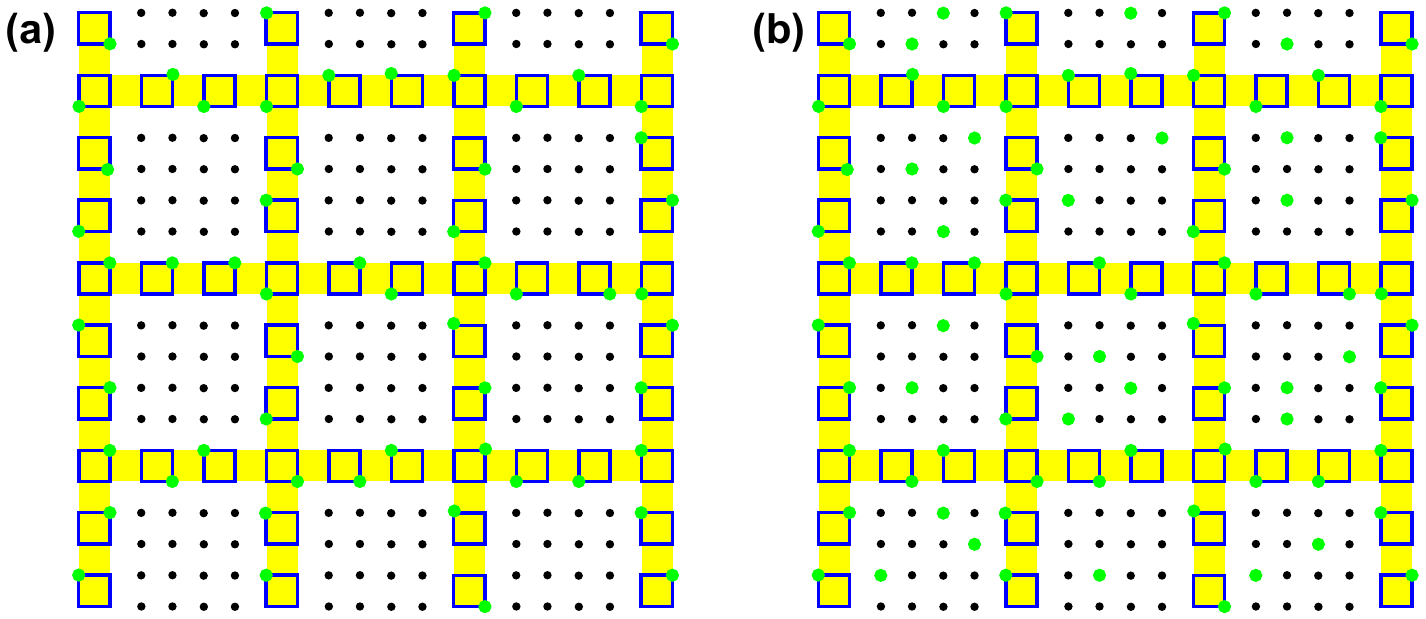}
\caption{Atomic substitution inside non-overlapping plaquettes can freeze dumbbells by breaking the dumbbell degeneracy inside the plaquette. (a) and (b) show $20\times20$ CuO$\mathrm{_2}$ lattices with $4\times4$ AF interiors and crossing metallic wires one plaquette in width. In (a), the green dots show random atomic substitutions or interstitial atoms inside the plaquettes. The atomic substitutions may occur at the Cu atom in the plaquette (shown here), the O atoms inside the CuO$\mathrm{_2}$ plaquette, or the apical O atom sites. The O atoms are not shown in the figure. (b), green dots representing atomic substitutions or interstitial atoms in the insulating AF region are also shown. So long as these green dots do not disrupt the insulating behavior of these regions, the superconductivity will not be disturbed. (b) may be easier to engineer than (a) because the green dots are dispersed more randomly.
}
\label{Wires-Asymmetry}
\end{figure}
\end{itemize}

\clearpage

\noindent\textsc{\underline{Conclusions}}

We have constructed a microscopic theory of cuprate superconductivity from the results of the chemist's ab initio \emph{hybrid} density functional methods (DFT). Hybrid DFT finds a \emph{localized} out-of-the-CuO$\mathrm{_2}$ hole is formed around a negatively charged dopant. The doped hole resides in a four-Cu-site plaquette. The out-of-plane hole destroys the antiferromagnetism inside the plaquette and creates a tiny piece of metal there. Hence, the crystal is inhomogeneous on the atomic-scale with metallic and insulating regions.

In contrast, the physicist's DFT methods (LDA and PBE) find a \emph{delocalized} hole residing in the CuO$\mathrm{_2}$ planes as a consequence of doping. We argue that the chemist's result is to be trusted over the physicist's result because it corrects the spurious self-Coulomb repulsion of the electrons found in the physicist's density functionals.

Due to dopant-dopant Coulomb repulsion, doped plaquettes do not overlap unless the doping is sufficiently high that overlap cannot be avoided.  Non-overlapping plaquettes have a dynamic Jahn-Teller distortion of the out-of-the-plane hole that we call a ``fluctuating dumbbell". The dumbbells inside an overlapped plaquette become static Jahn-Teller distortions, or ``frozen".

The above model explains a vast swath of normal state phenomenolgy using simple counting of the sizes of the metallic region, the insulating AF region, and the number of flutuating and frozen dumbbells. We show that superconducting pairing arises from planar Oxygen atoms near the interface between the metallic and insulating regions. These planar O atom phonon modes explain the large \tc\ $\sim100$ K, the \tc-dome as a function of doping, the changes in \tc\ as a function of the number of CuO$\mathrm{_2}$ layers per unit cell, the lack of a \tc\ isotope effect at optimal doping, and the D-wave superconducting pairing wavefunction (or superconducting gap symmetry).

Generally, with phonon superconducting pairing, an isotropic S-wave pairing wavefunction is favored over a D-wave pairing wavefunction. However, we show that the fluctuating dumbbells drastically raise the Cooper pair Coulomb repulsion, leading to the observed D-wave pairing wavefunction. By overlapping the plaquettes and freezing the dumbbells, the S-wave pairing wavefunction becomes favored over the D-wave pairing wavefunction. We show that the S-wave \tc\ is in the range of $\approx280-390$ K when the D-wave \tc\ $\approx100$ K.

Finally, we summarize the materials charateristics that are relevant for fabricating room-temperature superconductors and high current densities.

\vspace{.1in}

\noindent \textbf{Acknowledgments} 

\hspace{1.7in}\textit{``In Ogg's theory it was his intent}

\hspace{1.7in}\textit{That the current keep flowing, once sent;}

\hspace{1.7in}\textit{So to save himself trouble,}

\hspace{1.7in}\textit{He put them in double,}

\hspace{1.7in}\textit{And instead of stopping, it went."}

\hspace{1.7in}\textrm{George Gamov}\cite{Blatt-book}

\vspace{.1in}

The author thanks Professor Philip B. Allen for a discussion on estimating the electron-phonon coupling magnitude. The author is grateful to Professors William A. Goddard III and Carver A. Mead for discussions and encouragement.


\clearpage

\appendix
\section{Analysis of Physicist and Chemist DFT}
\label{DFT}
Since our prior normal state and current \tc\ results depend on the assumption that the chemist's out-of-the-CuO$_2$ hole is correct, we discuss the physicist's and chemist's DFT approaches here. 

Immediately following the discovery of cuprates in 1986, density functional (DFT) band structure calculations were performed using the local density approximation (LDA).\cite{Yu1987,Mattheiss1987,Pickett1989} These calculations all found the ground state of the undoped cuprates to be metallic rather than an insulating spin-1/2 antiferromagnet (AF) with a nonzero band gap of $\approx2$ eV.\cite{Ginder1988} The orbital character of the electrons near the Fermi level was a mixture of planar Cu \dxxyy\ and planar O \psigma, where the x and y-axes and the \psigma\ orbitals point along the planar Cu$-$O bonds. Hence, LDA found the correct orbital character at the Fermi level. The problem was the electrons were not localized.

The reason for this error of LDA and the more modern PBE (Perdew-Burke-Ernzerhof) density functional\cite{PBE} arises because both LDA and PBE contain unphysical Coulomb repulsion of an electron with itself\cite{Perdew1981} (known as the self-Coulomb repulsion).  The self-repulsion spreads the electron density out, leading to greater electron hopping, and hence an increase of the valence and conduction band widths. The increased band widths reduce the gap. In the case of cuprates, the gap is reduced to zero.

Given the failure of state-of-the-art ab initio methods for the undoped AF insulator, two different approaches were taken. In the first approach, the hole created by a dopant was assumed to ``knock out" one of the localized magnetic electrons by forming a spin singlet state (no magnetism) with the localized planar Cu \dxxyy\ spin and the neighboring planar O atoms \psigma\ orbitals to form a Zhang-Rice singlet.\cite{Zhang1988} This unmagnetized ``hole" in the AF spin background could hop from planar Cu site to planar Cu site. The interaction of these holes with each other and with the AF background spins comprise the $t-J$ and $t+U$ classes of Hubbard models for cuprates.

The second approach argued that the metallic band structure found by ab initio computations was a reasonable starting point for understanding cuprates because, in the superconducting range of dopings, the high temperature phase is metallic.  Hence, only the Fermi level needed to be lowered to the match the doping.

Both attacks assumed the only relevant orbitals for understanding cuprates are planar Cu \dxxyy\ and planar O \psigma.  These two pictures dominate the 30 year literature on cuprates.

Chemically, neither of these two approaches makes sense.  An out-of-plane hole dopant has a net negative charge relative to the undoped background. In analogy to acceptor (p-type) dopants in semiconductors, a \emph{localized} level surrounding the dopant should be pulled out of the valence band (the acceptor level). For cuprates, the hole orbital should be \emph{localized} and comprised of out-of-plane character pointing toward the dopant.  Instead, ab-initio DFT and the above two approaches remove a \emph{delocalized} electron with orbital character pointing away from the dopant (inside the CuO$_2$ plane).  If the acceptor impurity level was shallow (energy close to the top of the valence band), then at moderate temperatures, the hole will be in the valence band. In this case, the physicist's hole state would be correct, except at the lowest temperatures.  However, using the chemist's DFT methods described below, the impurity state is found to be $\approx0.75$ eV above the top of the valence band for a hole doping of $12.5\%$ ($x=0.125$) in \lsco\ (see Figure 4b of Perry et al.\cite{Perry2002}).  The acceptor level is a deep trap.

\begin{figure}[tbp]
\centering \includegraphics[width=0.5\linewidth]
{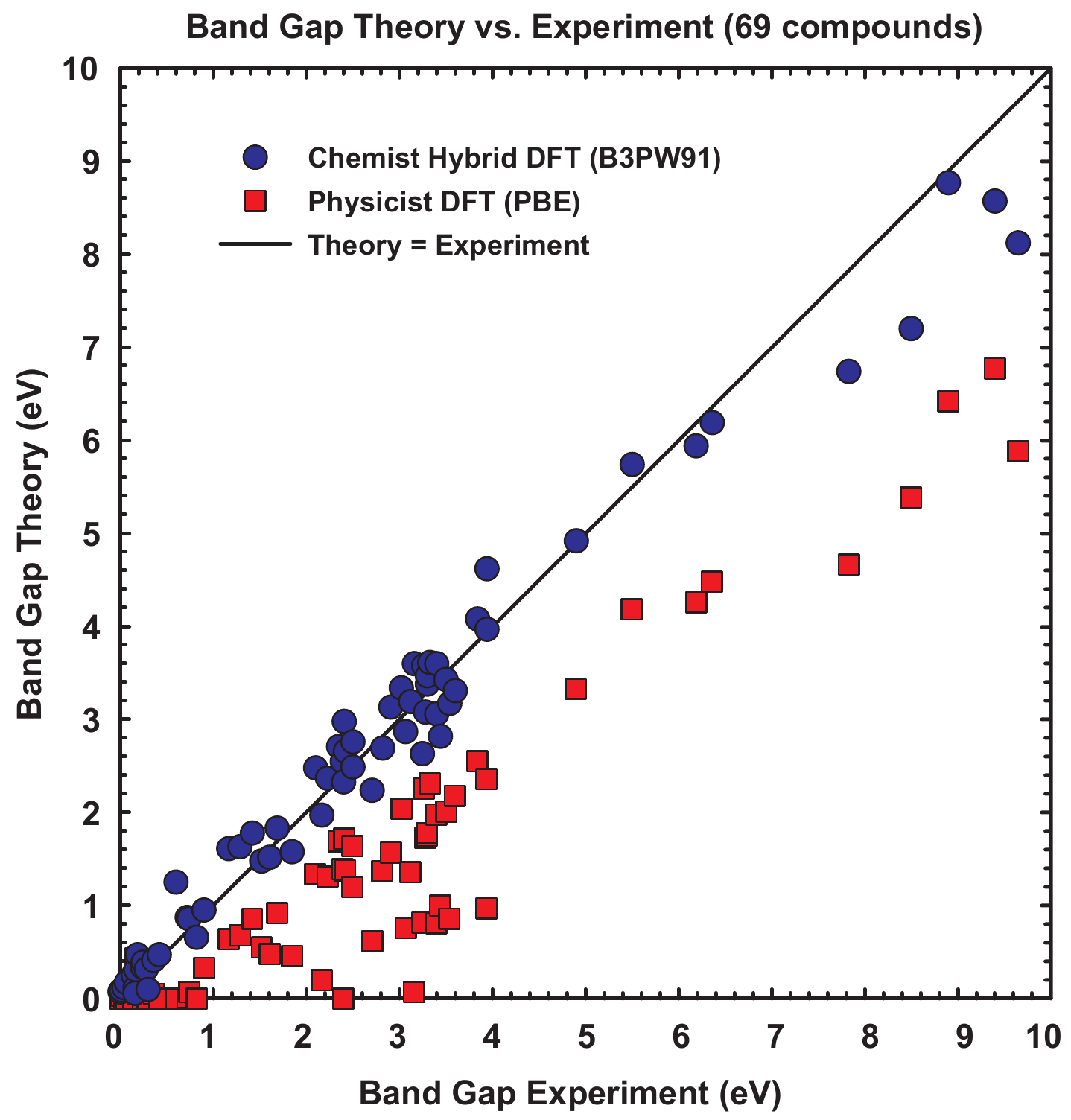}
\caption{
Superiority of the chemist's hybrid DFT functional, B3PW91,\cite{B3PW} (blue circles) to the physicist's PBE density functional (red squares) for band gaps of 69 compounds across the whole periodic table. The vertical distance from the data points to the  black diagonal line is the error between theory and experiment.  The data is taken from Crowley et al.\cite{Crowley2016} These compounds include thermoelectrics, topological insulators, transition-metal oxides, photovoltaics, transition-metal halides, and elemental and binary semiconductors. The mean absolute error for B3PW91 is 0.28 eV and 1.27 eV for PBE. The mean error is $-0.03$ eV and $-1.26$ eV for B3PW91 and PBE, respectively. PBE consistently returns a low band gap.  In 11 instances, PBE found zero gap (a metal) with the worst case being the Mott antiferromagnet FeO with an experimental gap of 2.4 eV. For the Mott antiferromagnet CoO, the experimental gap is 3.16 eV, while PBE is barely able to create a gap of 0.07 eV.  Results for the LDA functional are much worse than PBE, and are not shown here.
}
\label{bg}
\end{figure}

Obtaining the correct doped electronic structure for cuprates requires correcting the self-Coulomb repulsion error in the LDA and PBE functionals that led to an undoped metallic ground state.  A correction appeared in 1993 (seven years after the discovery of cuprates) by Becke\cite{B3PW} with the invention of hybrid density functionals. Hybrid density functionals reduce the self-Coulomb repulsion by including $20\%$ exact Hartree-Fock exchange. In molecular chemistry, hybrid density functionals are superior to LDA and PBE, and have been the workhorses for ab initio computations over the past two decades.  For crystalline materials, we have recently shown\cite{Crowley2016} that hybrid DFT functionals practically resolve the band gap prediction problem across the whole periodic table with a mean absolute deviation of $0.28$ eV, while PBE has an absolute error of $1.27$ eV. The LDA functional is worse than PBE and no numbers were reported.\cite{Crowley2016} These results are plotted in Figure \ref{bg}.

Unfortunately, the appearance of the superior hybrid DFT method had no influence on the cuprate field. One possible explanation is that the ab initio band structure codes familiar to most physicists construct the Hamiltonian in a plane-wave basis set and the Hartree-Fock exchange term in hybrid DFT is computationally impractical in this basis space.  Hybrid DFT becomes computationally practical using localized Gaussian basis sets. Localized basis sets are superior to plane-wave basis set for molecular problems and have been used extensively by ab initio quantum chemists.  The first software to successfully implement hybrid DFT for solids was CRYSTAL98,\cite{CRYSTAL98} which used localized Gaussian basis functions.  Regrettably, CRYSTAL98 appeared 12 years after the discovery of cuprate superconductivity.

Another possibility is historical.  The solid state community has had enormous success for over 80 years assuming any inhomogeneity (such as impurities) is a small perturbation to a zeroth order Hamiltonian with full translational symmetry. Atomic-scale inhomogeneity of the kind suggested by the chemist's DFT requires breaking translational symmetry at zeroth order.

Using the hybrid DFT functional, B3LYP (Becke-3-Lee-Yang-Parr),\cite{B3PW,LYP} we showed\cite{Perry2001} that a 2 eV band gap AF insulator is obtained
\footnote
{
CRYSTAL98 only had basic Fock Matrix mixing convergence (SCF) at the time of our calculation in 2001.\cite{Perry2001}  Using the most recent version of CRYSTAL (2015), we find the gap to be 3.1 eV using exactly the same basis set.  Improved SCF convergence algorithms, increased computing power, and memory indicates our result of 2001 had not fully converged. We know hybrid functionals generally overestimate the band gaps of Mott antiferromagnets by $\approx1$ eV,\cite{Crowley2016} perhaps because the unrestricted spin wavefunctions (UHF) do not represent the correct spin state. Regardless, the orbital character of the doped hole is unchanged. None of the conclusions of the current paper are altered.
}
for undoped La$_2$CuO$_4$, in agreement with experiment.\cite{Ginder1988} Explicitly doped \lsco\ has hole states pointing out of the CuO$_2$ plane towards the dopants.\cite{Perry2002} Thus, in the case of cuprates, hybrid DFT has led to a completely different electronic structure than the LDA or PBE hole, as shown in Figure \ref{physchem}. The success of hybrid DFT in molecular chemistry and for the band gaps in Figure \ref{bg} strongly suggests hybrid DFT should be trusted over LDA and PBE DFT results. We believe this mistake with the doped electronic structure led the entire field astray.

\vspace{.3in}
\hspace{1.0in}\textit{``If a mason building a wall puts the first stone in crooked,}

\hspace{1.0in}\textit{no matter how tall the wall becomes, it will still be crooked."}

\hspace{1.7in}\textrm{Saadi Shirazi, (Born $1210$, Died $\approx1291$)}

\clearpage

\section{Degeneracy at the Fermi Level in an Isolated Plaquette that Splits in the Crystal Environment and Leads to the Pseudogap}
\label{sec:PG}

\begin{figure}[h]
\centering \includegraphics[width=11cm]{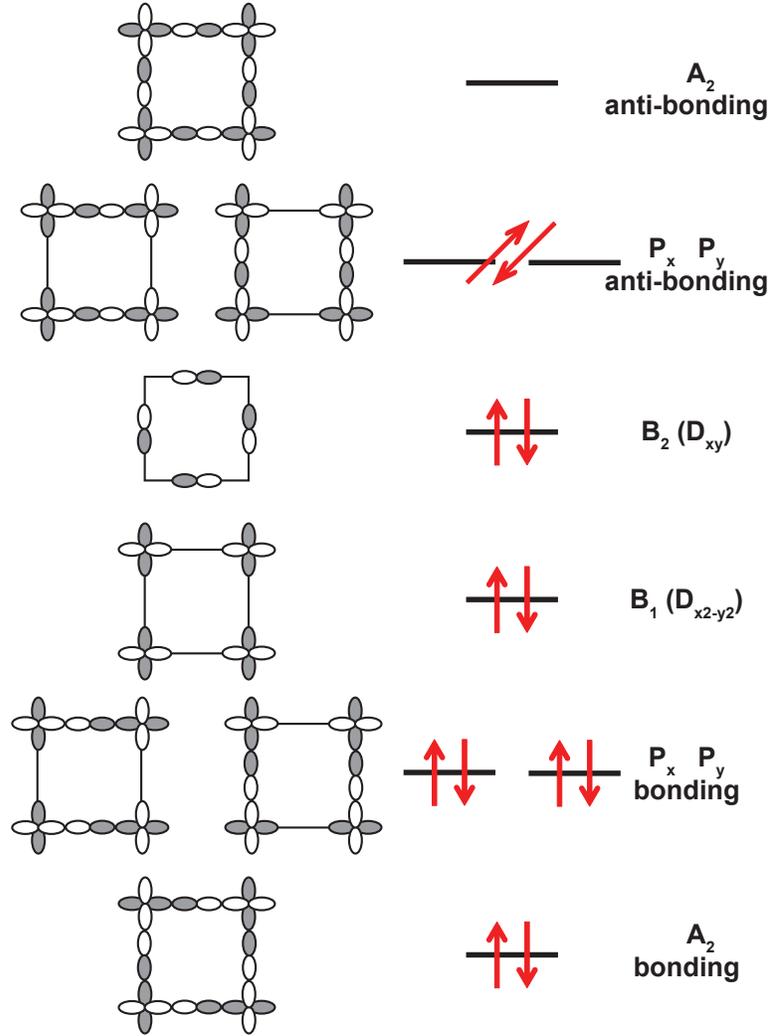}
\caption{The eight delocalized \emph{planar} metallic states inside an isolated plaquette. These states are comprised of the planar Cu \dxxyy\ and planar O \psigma\ orbitals. The states are arranged with the highest energy state at the top and the lowest energy state at the bottom. The right column shows the electron occupation in red. There are two electrons per O \psigma\ and one electron per Cu \dxxyy, leading to a total of twelve electrons. There are only two electrons to occupy the four degenerate $P_x$ and $P_y$ anti-bonding states. Theses states have dominant k-vector character at $(\pi,0)$ and $(0,\pi)$.  The splitting of the degnerate $P_x$ and $P_y$ states leads to the pseudogap.
}
\label{isolated-plaq}
\end{figure}
\clearpage

\section{Dumbbell Character of the Doped Plaquette Hole}
\label{sec:JT}

\begin{figure}[h]
\centering \includegraphics[width=12cm]{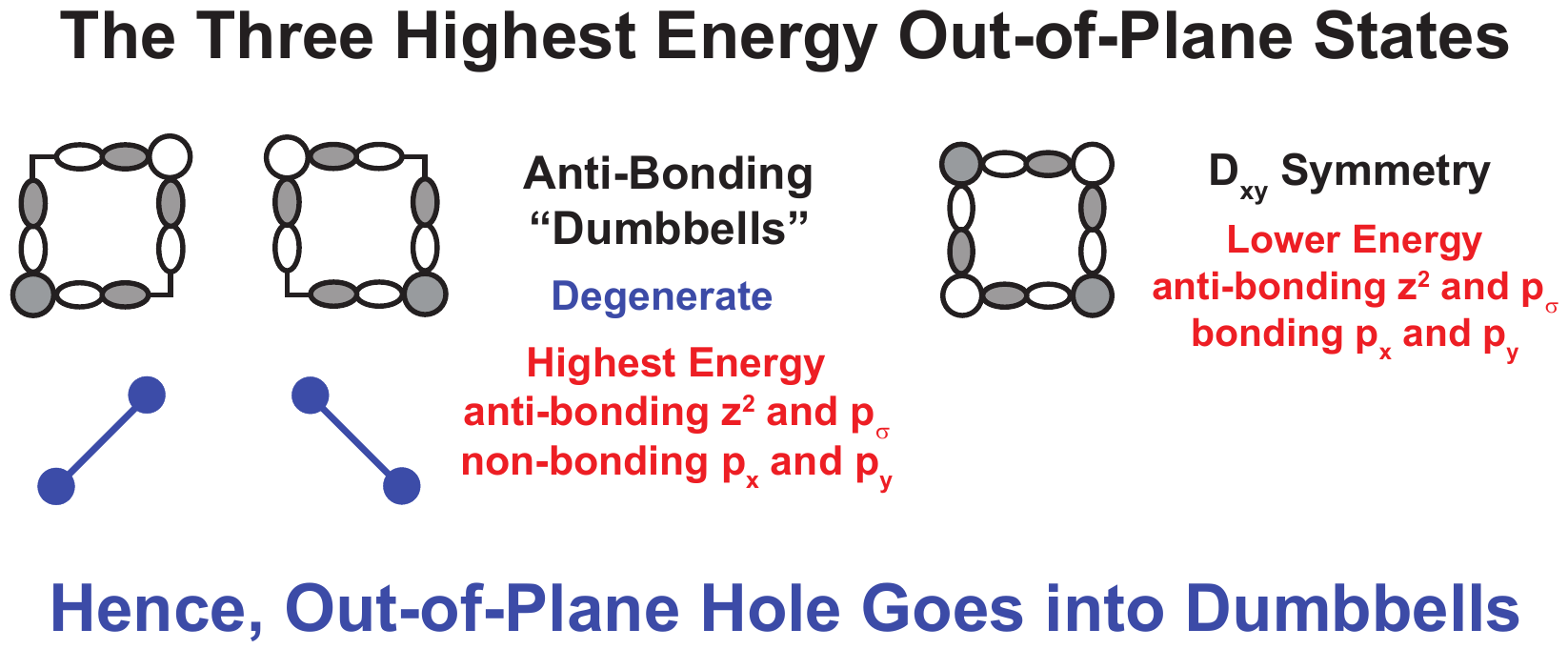}
\caption{The three highest energy out-of-the-CuO$\mathrm{_2}$ plane states. The apical O atom $p_z$ and the Cu \dzz\ orbitals are shown as a circle at the vertices of the plaquette. The two degenerate ``dumbbell" states are the highest in energy because they are non-bonding between the O $p_x$ and $p_y$ orbitals, while the D$\mathrm{_{xy}}$ state is bonding between these O orbitals. Since the hybridization of the $p_x$ and $p_y$ orbitals\cite{Hybertsen1992} is large ($\approx0.7$ eV), it pulls the D$\mathrm{_{xy}}$ energy below the P-type dumbbell energy. Hence, the out-of-plane hole surrounding a negatively charged dopant is a dumbbell state.
}
\label{hole-in-dumbbell}
\end{figure}

\section{Estimate of the Magnitude of the Electron-Phonon Coupling, g}
\label{sec:e-ph}

It is known to be qualitatively correct that $\hbar\omega_D/E_F\approx\sqrt{m/M}$ where $\hbar\omega_D$ is the Debye energy, $E_F$ is the Fermi energy, $m$ is the electron mass, and $M$ is the nuclear mass. One can quickly see that the form of the above expression is correct using $\omega_D\sim\sqrt{K/M}$ where $K$ is the spring constant and $K\sim E_F k_F^2\sim mE_F^2/\hbar^2$ due to metallic electron screening.

The electron-phonon coupling, $g$, is of the form $g\sim\sqrt{\hbar/2M\omega_D}\nabla V$, where $V$ is the nuclear potential energy. Substituting $\nabla V\sim k_F E_F$, leads to $g^2\sim(\hbar/2M\omega_D)k_F^2 E_F^2\sim (m/M)E_F^3/(\hbar\omega_D)\sim(\hbar\omega_D)E_F$. Hence, $g\approx\sqrt{\hbar\omega_D E_F}$.

Another derivation is dimensional. The coupling, $g$, has dimensions of energy and there are only two relevant energy scales, $\hbar\omega_D$ and $E_F$. Thus there are three possibilities for $g$: the mean, the geometric mean, and the harmonic mean of $\hbar\omega$ and $E_F$. Since $\hbar\omega_D\ll E_F$, the mean is $\approx E_F$, and the harmonic mean is $\approx\hbar\omega_D$. Neither of these two means makes intuitive sense because we know metallic electrons strongly screen the nuclear-nuclear potential. The only sensible choice is the geometric mean, $g\sim\sqrt{\hbar\omega_D E_F}$.
\clearpage

\section{Fluctuation T$\mathrm{_c}$: Plaquette Clusters Smaller than the Coherence Length}
\label{sec:fluc}

\begin{figure}[h]
\centering \includegraphics[width=11.5cm]
{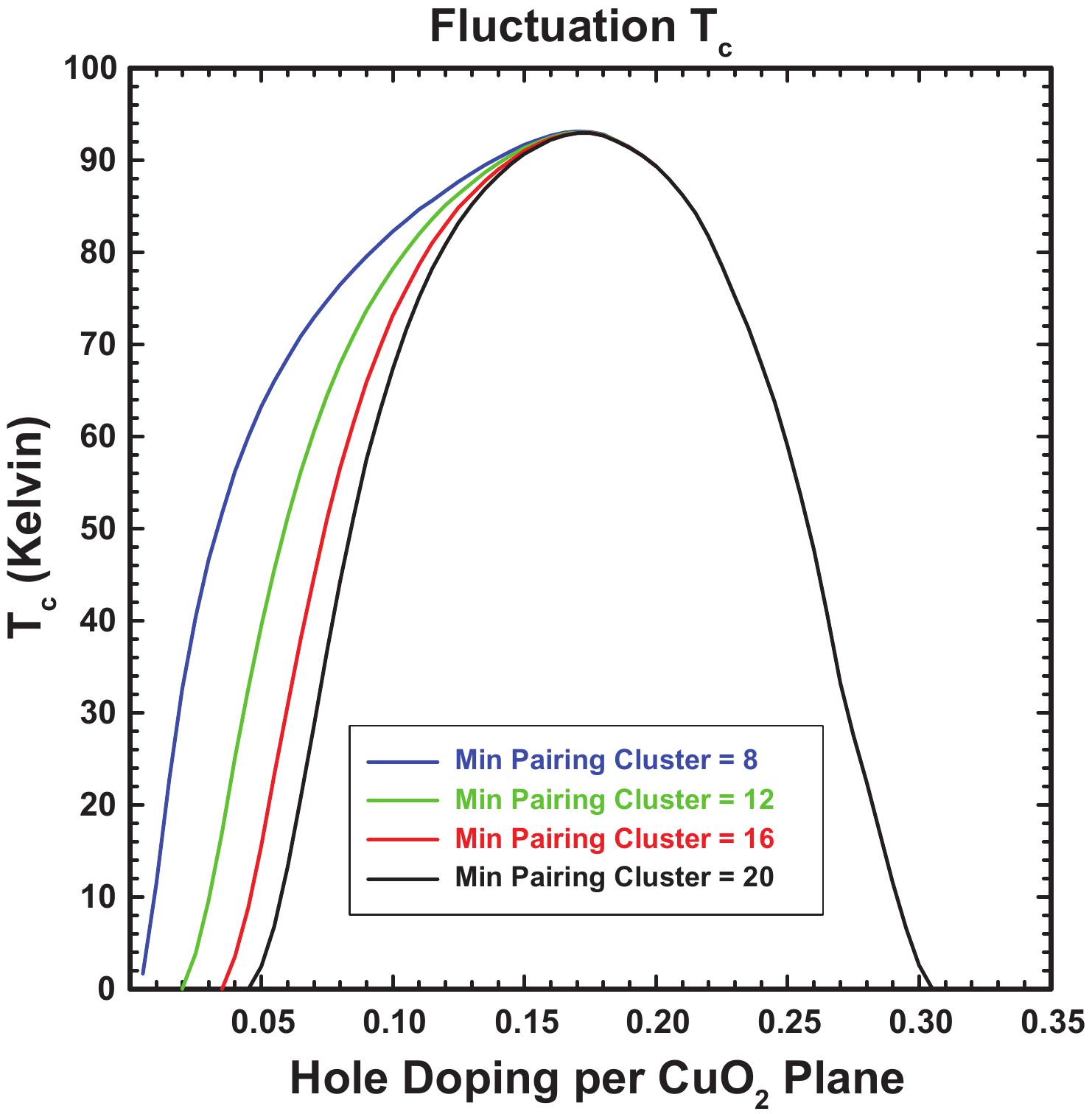}
\caption{The change in the \tc-dome as a function of the minimum plaquette cluster size for superconducting pairing. The black curve is identical to the black \tc\ curve in Figure \ref{tcdome}. All parameters are described in Appendix \ref{parameters}.
}
\label{fluc-tc}
\end{figure}

There are superconducting fluctuations above \tc\ at low dopings due to the fluctuating magenta plaquette clusters in Figures \ref{dop12}, \ref{dop00-10}, and \ref{dop12-18}. These plaquette clusters have superconducting pairing that does not contribute to the observed \tc\ because the clusters are smaller than the coherence length. Including these clusters into the \tc\ computation leads to an estimate of the temperature range where plaquette cluster superconducting fluctuations occur above \tc. The resulting ``fluctuation \tc\ domes" are plotted in Figure \ref{fluc-tc}. Of course, there exist superconducting fluctuations above \tc\ from the plaquettes clusters that are larger than the coherence length (yellow clusters in Figures \ref{dop12}, \ref{dop00-10}, and \ref{dop12-18}). The fluctuation \tc\ from the larger yellow clusters is not included in figure \ref{fluc-tc}.

\clearpage

\section{Parameters Used in the T$\mathbf{_c}$ Computations}
\label{parameters}

\begin{table}[h]
\caption{The planar Cu \dxxyy\ and O \psigma\ band structure used in all \tc\ computations. We use effective single band parameters derived from the angle-resolved photoemission (ARPES) Fermi surface for single layer Bi2201.\cite{Hashimoto2008} The 2D tight-binding band structure is $\epsilon_k=-2t[\cos(k_x a)+\cos(k_y a)]-4t'\cos(k_x a)\cos(k_y a)-2t''[\cos(2k_x a)+\cos(2k_y a)]$, where $a$ is the planar Cu$\mathrm{-}$Cu lattice size and the 2D momentum is $\mathbf{k}=(k_x,k_y)$. The ratio $t''/t'=-1/2$ is assumed.\cite{Hashimoto2008} The variable $n$ in the table is the number of metallic electrons per metallic Cu. At optimal doping, the number of holes per metallic Cu in the CuO$\mathrm{_2}$ plane is $x=0.16$, leading to $n=1.0-x=0.84$. The optimal doping Fermi level is used for \emph{all} dopings in order to keep the number of parameters to a minimum. For the multi-layer \tc\ calculations in Figure \ref{tclayer}, we assume the 2D band structure $\epsilon_k$ above. The 2D $\mathbf{k}$ states in adjacent layers, $l$ and $l\pm1$, are coupled by a momentum dependent matrix element equal to $<k,l\pm1|H_{inter}|k,l>=-\alpha t_{z}(1/4)(\cos(k_x a)-\cos(k_y a))^2$, where $\alpha$ is the product of the fraction of metallic sites in layers $l$ and $l\pm1$. Since interlayer hopping of an electron between Cu sites on different layers can only occur if the two Cu sites are metallic, $\alpha$ is the probability of two adjacent Cu sites in different layers being metallic. The dopings of the individual layers in the multi-layer \tc\ calculations are all less than the threshold for plaquette overlap at $x=0.187$. Hence, each plaquette covers 4 Cu sites. For example, between layers doped at $x=0.16$ and $x=0.11$, $\alpha=4(0.16)\times4(0.11)=0.2816$ leading to $\alpha t_z\approx0.023$ eV.
}
\label{band}
\begin{tabular}{cccc}
 $n$ & $t$ (eV) & $t'$ (eV) & $t_{z}$ (eV) \\ 
\hline
\hline
0.84 & 0.25 & -0.05625 & 0.08 \\
\hline
\end{tabular}
\end{table}

\begin{table}[h]
\caption{Parameters that remain the same for every \tc\ calculation. They are the mass-enhancement parameter derived from the high-temperature linear slope of the resistivity, $\lambda_{tr}$, the Debye energy, $\hbar\omega_{D}$, the minimum energy used in the low-temperature linear resistivity, $\hbar\omega_{min}$, the energy cutoff for Eliashberg sums, $\hbar\omega_{c}$, and the energy of the O atom phonon modes, $\hbar\omega_{ph}$.}
\label{fixedpars}
\begin{tabular}{ccccc}
$\lambda_{tr}$ (dimensionless)\ \ \ \ & $\hbar\omega_{D}$ (Kelvin)\ \ \ \ & $\hbar\omega_{min}$ (Kelvin)\ \ \ \ & $\hbar\omega_{c}$ (eV)\ \ \ \ & $\hbar\omega_{ph}$ (eV) \\
\hline
\hline
0.5 & 300 & 1.0 & 0.3 & 0.06 \\
\hline
\end{tabular}
\end{table}

\begin{table}[h]
\caption{Parameters for the \tc\ curves in Figures \ref{tcdome}, \ref{tclayer}, and \ref{fluc-tc}. The variable, $N_{min}$, is the number of metallic Cu sites inside the smallest plaquette cluster that is larger than the coherence length, and thereby contributes to \tc. The edge, convex, and concave couplings are chosen to be equal for the next layer couplings. All units are eV.}
\label{tcpars}
\begin{tabular}{llc|ccc|ccc|cc}
\hline
 &  & & \multicolumn{3}{c|}{$\delta\epsilon$} & \multicolumn{3}{c|}{$\delta t$} & \multicolumn{2}{c}{Next Layer} \\
Figure\ \  & Curve Color\ \  & $N_{min}$ & \ Edge\ & Convex & Concave & \ Edge\ & Convex & Concave & $\delta\epsilon_{NL}$\  & $\delta t_{NL}$ \\
\hline
\hline
\ref{tcdome} & Black & 20 & 0.150 & 0.150 & 0.075 & 0.240 & 0.240 & 0.120 & & \\
\ref{tcdome} & Magenta & 20 & 0.150 & 0.150 & 0.075 & 0.130 & 0.130 & 0.065 & & \\
\ref{tcdome} & Red & 100 & 0.000 & 0.000 & 0.000 & 0.132 & 0.132 & 0.000 & & \\
\ref{fluc-tc} & Blue & 8 & 0.150 & 0.150 & 0.075 & 0.240 & 0.240 & 0.120 & & \\
\ref{fluc-tc} & Green & 12 & 0.150 & 0.150 & 0.075 & 0.240 & 0.240 & 0.120 & & \\
\ref{fluc-tc} & Red & 16 & 0.150 & 0.150 & 0.075 & 0.240 & 0.240 & 0.120 & & \\
\ref{fluc-tc} & Black & 20 & 0.150 & 0.150 & 0.075 & 0.240 & 0.240 & 0.120 & & \\
\ref{tclayer} & Black & 20 & 0.150 & 0.150 & 0.075 & 0.240 & 0.240 & 0.120 & 0.0 & 0.2 \\
\ref{tclayer} & Blue & 20 & 0.050 & 0.000 & 0.000 & 0.130 & 0.130 & 0.065 &  0.05 & 0.13\\
\hline
\end{tabular}
\end{table}

\begin{table}[h]
\caption{Doping of each CuO$_2$ layer in the multi-layer \tc\ calculations. The outermost layers are always at optimal doping $(x=0.16)$. The adjacent layers are at $x=0.11$ doping. The innermost layers are all at $x=0.09$ doping. These dopings are obtained from Cu Knight shift measurements.\cite{Mukuda2012}}
\label{layer-doping}
\begin{tabular}{ccccccccccc}
Layers & \multicolumn{10}{c}{Hole Doping per CuO$_2$ Layer} \\
\hline
 1 & \ \ 0.16\ \ \ &  &  &  &  &  & &  &  & \\
 2 & \ \ 0.16\ \ \ & 0.16\ \ \ &  &  &  &  & &  &  & \\
 3 & \ \ 0.16\ \ \ & 0.11\ \ \ & 0.16\ \ \ &  &  &  & &  &  & \\
 4 & \ \ 0.16\ \ \ & 0.11\ \ \ & 0.11\ \ \ & 0.16\ \ \ &  &  & &  &  & \\
 5 & \ \ 0.16\ \ \ & 0.11\ \ \ & 0.09\ \ \ & 0.11\ \ \ & 0.16\ \ \ &  & &  &  & \\
 6 & \ \ 0.16\ \ \ & 0.11\ \ \ & 0.09\ \ \ & 0.09\ \ \ & 0.11\ \ \ & 0.16\ \ \ & &  &  & \\
 7 & \ \ 0.16\ \ \ & 0.11\ \ \ & 0.09\ \ \ & 0.09\ \ \ & 0.09\ \ \ & 0.11\ \ \ & 0.16\ \ \ &  &  & \\
 8 & \ \ 0.16\ \ \ & 0.11\ \ \ & 0.09\ \ \ & 0.09\ \ \ & 0.09\ \ \ & 0.09\ \ \ & 0.11\ \ \ & 0.16\ \ \ &  & \\
 9 & \ \ 0.16\ \ \ & 0.11\ \ \ & 0.09\ \ \ & 0.09\ \ \ & 0.09\ \ \ & 0.09\ \ \ & 0.09\ \ \ & 0.11\ \ \ & 0.16\ \ \ & \\
 10 & \ \ 0.16\ \ \ & 0.11\ \ \ & 0.09\ \ \ & 0.09\ \ \ & 0.09\ \ \ & 0.09\ \ \ & 0.09\ \ \ & 0.09\ \ \ & 0.11\ \ \ & 0.16 \\
\hline
\end{tabular}
\end{table}
\clearpage

\section{Description of the Eliashberg \tc\ Calculations}
\label{eliashberg1}

\subsection{The Eliashberg Equations}

The attractive electron-electron pairing mediated by phonons is not instantaneous in time due to the non-zero frequency of the phonon modes (phonon retardation). In addition, electrons are scattered by phonons leading to electron wavefunction renormalization (``lifetime effects") that decrease \tc. Any credible \tc\ prediction must incorporate \emph{both} of these effects.  All \tc\ calculations in this paper solve the Eliashberg equations for the superconducting pairing wavefunction (also called the gap function). It includes both the pairing retardation and the electron lifetime.\cite{Schrieffer-book,Allen1982,Scalapino1969}

The Eliashberg equations are non-linear equations for the superconducting gap function, $\Delta(\mathbf{k},\omega,T)$, and the wave function renormalization, $Z(\mathbf{k},\omega,T)$, as a function of momentum $\mathbf{k}$, frequency $\omega$, and temperature $T$. Usually, the $T$ dependence of $\Delta$ and $Z$ is assumed, and they are written as $\Delta(\mathbf{k},\omega)$ and $Z(\mathbf{k},\omega)$, respectively. We follow this convention here. Both $\Delta(\mathbf{k},\omega)$ and $Z(\mathbf{k},\omega)$ are a complex numbers. In this Appendix only, we will absorb Boltzmann's constant, $k_B$, into $T$. Thus $T$ has units of energy.

Both $\Delta$ and $Z$ are frequency dependent because of the non-instantaneous nature of the superconducting electron-electron pairing. If the pairing via phonons was instantaneous in time, then there would be no frequency dependence to $\Delta$ and $Z$. The simpler BCS\cite{Schrieffer-book} gap equation assumes an instantaneous pairing interaction ($\Delta$ is independent of $\omega$) and no wavefunction renormalization ($Z=1$).

The Eliashberg equations may be solved in momentum and frequency space $(\mathbf{k},\omega)$, or in momentum and discrete imaginary frequency space, $(\mathbf{k},i\omega_n)$, where $n$ is an integer and $\omega_n=(2n+1)\pi T$. In the imaginary frequency space representation, the temperature dependence and the retardation of the phonon induced pairing are both absorbed into the imaginary frequency dependence, $i\omega_n$. In theory, both $\Delta(\mathbf{k},\omega)$ and $Z(\mathbf{k},\omega)$ can be obtained by analytic continuation of their $(\mathbf{k},i\omega_n)$ counterparts. In practice, the analytic continuation is fraught with numerical difficulties.\cite{Vidberg1977,Leavens1985,Beach2000,Ostlin2012} However, the symmetry of the gap can be extracted from either the real or imaginary frequency representations of $\Delta$.

In the pioneering work of Schrieffer, Scalapino, and Wilkins,\cite{Schrieffer1963,Scalapino1966,Schrieffer-book,Scalapino1969} the goal was to obtain the isotropic (in $k$-space) gap function at zero temperature, $\Delta(\omega)$, as a function of $\omega$ in order to compute the superconducting tunneling of lead ($T_c=7.2$ K). Hence, they solved the full non-linear Eliashberg equations in frequency space.

Above \tc, $\Delta(\mathbf{k},\omega)$ is zero. For $T\approx T_c$, $\Delta$ is small. Since our interest in this paper is on the magnitude of \tc\ and the symmetry of the superconducting gap, we can linearize the gap, $\Delta$, in the Eliashberg equations for temperatures, $T$, close to \tc. The result is a temperature dependent real symmetric matrix eigenvalue equation with $\Delta(\mathbf{k},\omega)$ as the eigenvector. The eigenvalues are dimensionless and the largest eigenvalue monotonically increases as $T$ decreases. For $T>T_c$, the largest eigenvalue of the real symmetric matrix is less than 1. At $T=T_c$, the largest eigenvalue equals 1, signifying the onset of superconductivity.

The non-linear Eliashberg equations (or the linearized version) are easier to solve in imaginary frequency space.\cite{Allen1982} Hence, we solve the linearized Eliashberg equations in imaginary frequency space to obtain \tc.

We use the linearized Eliashberg equations as derived in the excellent chapter by Allen and Mitrovic.\cite{Allen1982} Prior Eliashberg formulations assume translational symmetry (momentum $\mathbf{k}$ is a good quantum number for the metallic states). Our metallic wavefunctions are not $\mathbf{k}$ states because they are only non-zero in the percolating metallic region. We write the wavefunction and energy for the state with index $l$ as $\psi_l$ and $\epsilon_l$, respectively. Since $\psi_l$ is only delocalized over the metallic region and is normalized, $\psi_l\sim 1/\sqrt{N_M}$, where $N_M$ is the total number of metallic Cu sites. Rather than Cooper pairing occuring between $\mathbf{k}\uparrow$ and its time-reversed partner, $\mathbf{-k}\downarrow$, a Cooper pair here is comprised of $(\psi_l\uparrow,\overline{\psi_l}\downarrow)$, where $\overline{\psi_l}$ is the complex conjugate of $\psi_l$.

The linearized Eliashberg equations for $\Delta(l,i\omega_n)$ and $Z(l,i\omega_n)$ are obtained from the $\mathbf{k}$-vector equations\cite{Allen1982} simply by replacing $\mathbf{k}$ with the index $l$ everywhere

\begin{eqnarray}
\label{e1}
Z(l,i\omega_n) & = & 1 +
      \frac{\pi T}{|\omega_n|}\sum_{l'n'}^{|\omega_{n'}|<\omega_c}
      \left[\frac{\delta(\epsilon_{l'}-\epsilon_F)}{N(0)}\right]
      \lambda(l,l',\omega_n-\omega_{n'})s_n s_{n'}, \\
\label{e2}
Z(l,i\omega_n)\Delta(l,i\omega_n) & = &
       \pi T \nonumber \\
 & \times &
       \sum_{l'n'}^{|\omega_{n'}|<\omega_c}\frac{1}{|\omega_{n'}|}
       \left[\frac{\delta(\epsilon_{l'}-\epsilon_F)}{N(0)}\right]
       [\lambda(l,l',\omega_n-\omega_{n'})-\mu^*(\omega_c)]
       \Delta(l',i\omega_{n'}),
\end{eqnarray}
\noindent where $\epsilon_F$ is the Fermi energy, $N(0)$ is the total metallic density of states per spin per energy, $s_n=\omega_n/|\omega_n|=\mathrm{sgn}(\omega_n)$ is the sign of $\omega_n$, $\omega_c$ is the cutoff energy for the frequency sums, $\lambda(l,l',\omega_n)$ is the dimensionless phonon pairing strength (defined below), and $\mu^*(\omega_c)$ is the dimensionless Morel-Anderson Coulomb pseudopotential at cutoff energy $\omega_c$. It is a real number. The wavefunction renormalization, $Z(l,i\omega_n)$, is dimensionless. In the non-linear Eliashberg equations, $\Delta(l,i\omega_n)$ has units of energy. In the linearized equations above, $\Delta(l,i\omega_n)$ is an eigenvector and is arbitrary up to a constant factor.

The ``electron-phonon spectral function" $\alpha^2 F(l,l',\Omega)$ is defined

\begin{equation}
\label{aF}
\alpha^2 F(l,l',\Omega)=N(0)\sum_{\sigma}|<l|H_{ep}^{\sigma}|l'>|^2
        \delta(\omega_\sigma-\Omega),
\end{equation}

\noindent and the phonon pairing strength $\lambda(l,l',\omega_n)$ is defined

\begin{eqnarray}
\label{lam}
\lambda(l,l',\omega_n) &= & \int_0^{+\infty}d\Omega\ \alpha^2 F(l,l',\Omega)
             \left(\frac{2\Omega}{\omega_n^2+\Omega^2}\right) \\
                       &= & N(0)\sum_{\sigma}|<l|H_{ep}^{\sigma}|l'>|^2
 \left(\frac{2\omega_{\sigma}}{\omega_n^2+\omega_{\sigma}^2}\right),
\end{eqnarray}

\noindent where $<l|H_{ep}^{\sigma}|l'>$ is the matrix element (units of energy) between initial and final states $l'$ and $l$, respectively of the electron-phonon coupling, and $H_{ep}^{\sigma}$ is the electron-phonon coupling for the phonon mode $\sigma$ with energy $\omega_\sigma$. Both $\alpha^2 F(l,l',\Omega)$ and $\lambda(l,l',\omega_n)$ are real positive numbers. Hence, $Z(l,i\omega_n)$ is a real positive number. From \ref{e2}, the gap $\Delta(l,i\omega_n)$ can always be chosen to be real. Since $\lambda(l,l',\omega_n)=\lambda(l,l',-\omega_n)$ from equation \ref{lam},

\begin{eqnarray}
 Z(l,-i\omega_n)       & = & Z(l,i\omega_n) = \mathrm{Real\ Number}, \\
 \Delta(l,-i\omega_n)  & = & \Delta(l,i\omega_n) = \mathrm{Real\ Number}.
\end{eqnarray}

$\alpha^2 F(l,l',\Omega)$ and $\lambda(l,l',\omega_n)$ are dimensionless because $(\mathrm{eV})^{-1}(\mathrm{eV})^2(\mathrm{eV})^{-1}\sim 1$. Physically, they should be independent of the number of metallic Cu sites, $N_M$, as $N_M$ becomes infinite. The independence with respect to $N_M$ is shown below.

The electron-phonon Hamiltonian for phonon mode $\sigma$, $H_{ep}^{\sigma}$, is

\begin{equation}
H_{ep}^{\sigma}=\left(\frac{\hbar}{2M\omega_\sigma}\right)^{\frac{1}{2}}\nabla V(a_\sigma+a_\sigma^{\dagger}),
\end{equation}
\noindent where $M$ is the nuclear mass. $a_\sigma$ and $a_\sigma^\dagger$ destroy and create $\sigma$ phonon modes, respectively. $V$ is the potential energy of the electron. For localized phonon modes, $\nabla V$ is independent of the number of metallic sites, $N_M$. The $l$ and $l'$ metallic states each scale as $1/\sqrt{N_M}$, leading to $<l|H_{ep}^{\sigma}|l'>\sim 1/N_M$. Since the number of localized phonon modes scales as $N_M$, the $N_M$ scaling of the sum $\sum_\sigma|<l|H_{ep}^{\sigma}|l'>|^2$ is $\sim N_M(1/N_M)^2\sim 1/N_M$. Hence, we have shown that $\alpha^2 F(l,l',\Omega)$ and $\lambda(l,l',\omega_n)$ are dimensionless and independent of $N_M$ because the density of states per spin, $N(0)$, is proportional to $N_M$. In fact, $\alpha^2 F$ and $\lambda$ are independent of $N_M$ even when the phonon modes $\sigma$ are delocalized. In this case, $\nabla V\sim 1/\sqrt{N_M}$. The electron-phonon matrix element $<l|H_{ep}^{\sigma}|l'>$ is now summed over the crystal, and thereby picks up a factor of $N_M$. Hence, $<l|H_{ep}^{\sigma}|l'>\sim N_M\times \sqrt{1/N_M}\times \sqrt{1/N_M}\times \sqrt{1/N_M}\sim 1/\sqrt{N_M}$. For delocalized phonons, the sum over phonon modes $\sigma$ in $\sum_\sigma|<l|H_{ep}^{\sigma}|l'>|^2$ does not add another factor of $N_M$. The claim is obvious when $l$ and $l'$ are momentum states $\mathbf{k}$ and $\mathbf{k'}$ because the only phonon mode that connects these two states has momentum $\mathbf{q=k-k'}$. Therefore, $\alpha^2 F(l,l',\Omega)$ and $\lambda(l,l',\omega_n)$ are always dimensionless and independent of $N_M$.

The atomic-scale inhomogeneity of cuprates implies translation is not a perfect symmetry of the crystal. However, the dopants are distributed \emph{randomly}, and therefore \emph{on average} $\mathbf{k}$ becomes a good quantum number. Hence, we may work with Green's functions in $\mathbf{k}$ space and approximate the Cooper pairing to occur between $(\mathbf{k}\uparrow,\mathbf{-k}\downarrow)$ states. The approximation is identical to the very successful Virtual Crystal Approximation (VCA) and the Coherent Potential Approximation (CPA) for random alloys.\cite{CPA}

In the VCA and CPA, the Green's function between two distinct $\mathbf{k}$ states, $\mathbf{k}$ and $\mathbf{k'}$ is zero

\begin{equation}
 G(\mathbf{k},\mathbf{k'},i\omega_n) \approx 0,\ \mathrm{if\ \mathbf{k}\ne\mathbf{k'}}.
\end{equation}

\noindent The fact that $\mathbf{k}$ is \emph{not} a good quantum number of the crystal is incorporated by including a self-energy correction, $\Sigma(\mathbf{k},i\omega_n)$ at zeroth order into the metallic Green's function

\begin{equation}
 G(\mathbf{k},i\omega_n)=
   \frac{1}{i\omega_n-\epsilon_{bare}(\mathbf{k})-\Sigma(\mathbf{k},i\omega_n)}.
\end{equation}

\noindent Here, $\epsilon_{bare}(\mathbf{k})$ is the bare electron energy. $\Sigma(\mathbf{k},i\omega_n)$ can be written as the sum of two terms, $\Sigma(\mathbf{k},i\omega_n)=\Sigma_0(\mathbf{k},i\omega_n)+i\omega_n\Sigma_1(\mathbf{k},i\omega_n)$. Both $\Sigma_0$ and $\Sigma_1$ are even powers of $\omega_n$, $\Sigma_i(\mathbf{k},-i\omega_n)=\Sigma_i(\mathbf{k},i\omega_n)$, for $i=1,2$. $\Sigma_0$ adds a shift to the bare electron energy, $\epsilon_{bare}(\mathbf{k})$, and a lifetime broadening to the electronic state. $\Sigma_1$ leads to wavefunction renormalization of the bare electron state.

The shift of $\epsilon_{bare}(\mathbf{k})$ due to $\Sigma_0(\mathbf{k},i\omega_n)$ leads to the observed angle-resolved photoemission (ARPES) band structure in cuprates,\cite{Hashimoto2008} $\epsilon_{\mathbf{k}}$, and its lifetime broadening. The lifetime broadening integrates out of the Eliashberg equations because the integral of a Lorentzian across the Fermi energy is independent of the width of the Lorentzian.\cite{Allen1982} Hence, we may use the ARPES band structure, $\epsilon_{\mathbf{k}}$, in the Eliashberg equations and absorb $\Sigma_1(\mathbf{k},i\omega_n)$ into $Z(\mathbf{k},i\omega_n)$ in the Eliashberg equations.

Hence, we are right back to the standard Eliashberg equations\cite{Allen1982,Schrieffer-book,Schrieffer1963,Scalapino1966,Scalapino1969}

\begin{eqnarray}
\label{e1a}
Z(\mathbf{k},i\omega_n) & = & 1 +
    \frac{\pi T}{|\omega_n|}\sum_{\mathbf{k'}n'}^{|\omega_{n'}|<\omega_c}
    \left[\frac{\delta(\epsilon_{\mathbf{k'}}-\epsilon_F)}{N(0)}\right]
    \lambda(\mathbf{k},\mathbf{k'},\omega_n-\omega_{n'})s_n s_{n'}, \nonumber \\
\label{e2a}
Z(\mathbf{k},i\omega_n)\Delta(\mathbf{k},i\omega_n) & = &
       \pi T \\
\label{e3a}
 & \times &
       \sum_{\mathbf{k'}n'}^{|\omega_{n'}|<\omega_c}\frac{1}{|\omega_{n'}|}
 \left[\frac{\delta(\epsilon_{\mathbf{k'}}-\epsilon_F)}{N(0)}\right]
  [\lambda(\mathbf{k},\mathbf{k'},\omega_n-\omega_{n'})-\mu^*(\omega_c)]
       \Delta(\mathbf{k'},i\omega_{n'}), \nonumber
\end{eqnarray}

\begin{eqnarray}
\label{eZ}
 Z(\mathbf{k},-i\omega_n) & = & Z(\mathbf{k},i\omega_n) =
                                        \mathrm{Real\ Number}, \\
\label{eD}
 \Delta(\mathbf{k},-i\omega_n) & = &
            \Delta(\mathbf{k},i\omega_n) = \mathrm{Real\ Number}.
\end{eqnarray}

The Eliashberg equations above are completely general for a single band crossing the Fermi level. The only inputs into the equations are the Fermi surface, Fermi velocity (in order to obtain the local density of states), the dimensionless electron-phonon pairing, $\lambda(\mathbf{k},\mathbf{k'},\omega_n)$, and the dimensionless Morel-Anderson Coulomb pseudopotential at the cutoff energy (typically, chosen to be five times larger than the highest phonon mode, $\omega_c=5\omega_{ph}$), $\mu^*(\omega_c)$. We apply the standard methods\cite{Allen1982} to map the above equations into a matrix equation for the highest eigenvalue as a function of $T$. The highest eigenvalue monotonically increases at $T$ decreases. When the highest eigenvalue crosses 1, \tc\ is found.

Equations \ref{e1a}, \ref{eZ}, \ref{eD} need to be modified when more than one band crosses the Fermi level. Phonons can scatter electron pairs from one band to another in addition to scattering within a single band. The modification to the single Fermi surface Eliashberg equations above require changing the $\mathbf{k}$ and $\mathbf{k'}$ labels to $\mathbf{bk}$ and $\mathbf{b'k'}$ where $\mathbf{b}$ and $\mathbf{b'}$ refer to the band index. $\mathbf{k}$ and $\mathbf{k'}$ remain vectors in 2D so long as we assume the coupling of CuO$\mathrm{_2}$ layers in different unit cells is weak. The number of bands is equal to the number of CuO$\mathrm{_2}$ layers per unit cell, $L$. We derive the electron-phonon pairing $\lambda$ for a single layer cuprate in sections \ref{lamO} and \ref{lamR}. In section \ref{eliashberg-multi}, we derive the multi-layer $\lambda$.

The total electron-phonon spectral function is the sum of four terms
\begin{equation}
\alpha^2 F=\alpha^2 F_1 + \alpha^2 F_2 +
                 \alpha^2 F_{surf} + \alpha^2 F_{\perp},
\end{equation}

\noindent where $\alpha^2 F_1$ and $\alpha^2 F_2$ are the spectral functions from phonons that contribute to the resistivity. $\alpha^2 F_1$ is due to the phonons that lead to the low-temperature linear term in the resistivity, and $\alpha^2 F_2$ is due to the phonons that lead to the the low-temperature $T^2$ resistivity term.\cite{Hussey2011} $\alpha^2 F_{surf}$ is the component due to the planar O atom at the surface between the metal and insulating regions. It is the O atom phonon shown in Figure \ref{phonons}a. $\alpha^2 F_{\perp}$ is the contribution from the planar O atom adjacent to the metal-insulator surface on the insulating AF side. It is shown in Figure \ref{phonons}b. Since the energy of these two O phonons modes is $\approx 60$ meV,\cite{Pintschovius2005} their contribution to the resistivity is very small.

Sections \ref{lamO} and \ref{lamR} in this Appendix derive the four $\alpha^2 F$ terms above in order to obtain the total phonon pairing, $\lambda=\lambda_1+\lambda_2+\lambda_{surf}+\lambda_{\perp}$, that is used in the Eliashberg equations \ref{e1a}, \ref{eZ}, and \ref{eD} for \tc.

\subsection{Contribution to $\mathbf{\lambda}$ from the Interface O Atom Phonons in Figure \textbf{\ref{phonons}}}
\label{lamO}

\subsubsection{Surface O Atom Mode in Figure \ref{phonons}a}
\label{lam0s}

\begin{figure}[h]
\centering \includegraphics[width=6cm]
{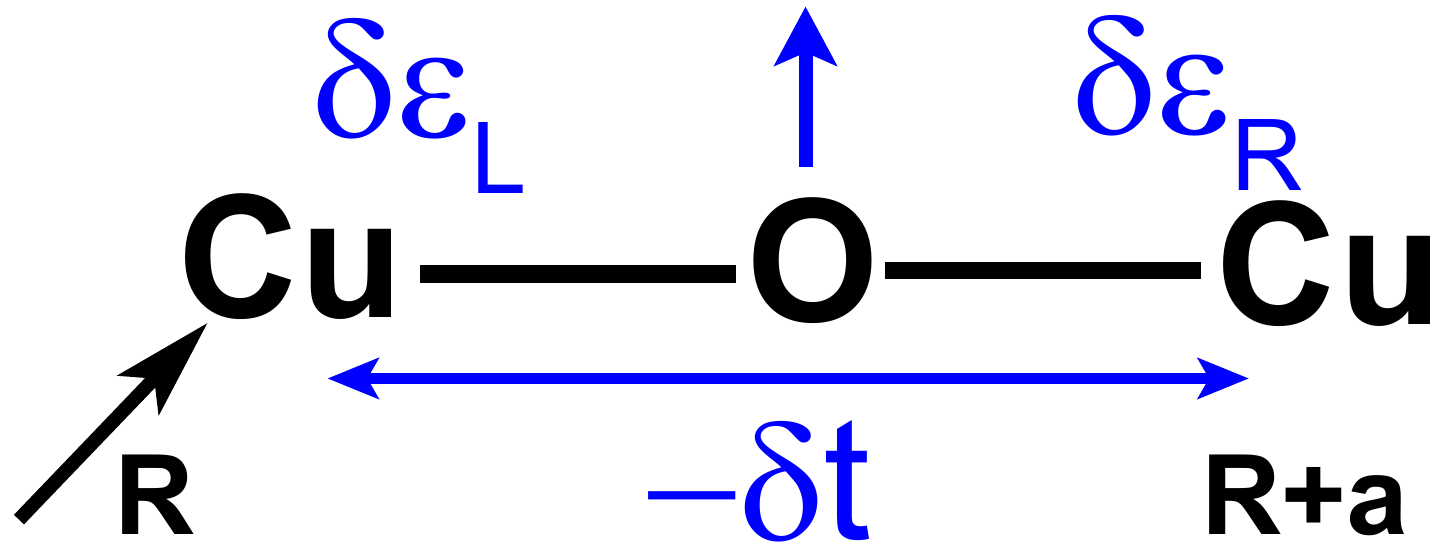}
\caption{A surface O atom phonon mode. The change in Cu orbital energy for the Cu atom, $\delta\epsilon_L$ for the left Cu atom, and $\delta\epsilon_R$, for the right Cu atom due to the displacement of the surface O atom on the metal-insulator interface. $-\delta t$ is the change in the hopping energy. The position of the left Cu atom is at \textbf{R} and the right Cu atom is at \textbf{R+a}. $\delta\epsilon_L$, $\delta\epsilon_R$, and $\delta t$ are functions of \textbf{R} as seen in Figure \ref{phonons}a. We choose the energy of all the surface phonon modes to be the same and equal to $\omega_{ph}$.
}
\label{surfO}
\end{figure}

The Hamiltonian for Figure \ref{surfO} is

\begin{equation}
\label{ham-surf}
H_{surf}(\mathbf{R})=\delta\epsilon_L c_L^\dagger c_L
 +\delta\epsilon_R c_R^\dagger c_R
 -\delta t\left(c_L^\dagger c_R + c_R^\dagger c_L\right),
\end{equation}

\noindent where $c_L^\dagger$ and $c_L$ create and destroy an electron at the $L$ Cu site. $c_R^\dagger$ and $c_R$ are defined similarly. Since there is no electron spin coupling to the O atom phonon mode, the electron spin index is dropped in equation \ref{ham-surf}.

The $\mathbf{k}$ state $\phi(\mathbf{k})$ is defined as

\begin{equation}
\phi(\mathbf{k})=N_M^{-\frac{1}{2}}\sum_R
   \mathrm{e}^{i\mathbf{k}\cdot\mathbf{R}}\phi(\mathbf{R}),
\end{equation}

\noindent where $\phi(\mathbf{R})$ is the localized effective Cu \dxxyy\ orbital at position $\mathbf{R}$, and $N_M$ is the number of metallic Cu sites. The matrix element between $\mathbf{k'}$ and $\mathbf{k}$ is

\begin{equation}
\left<\phi(\mathbf{k'})|H_{surf}(\mathbf{R})|\phi(\mathbf{k})\right>=
N_M^{-1}\mathrm{e}^{-i(k'-k)(R+\frac{1}{2}a)}\left[
\delta\epsilon_L\mathrm{e}^{\frac{i}{2}(k'-k)a}+
\delta\epsilon_R\mathrm{e}^{-\frac{i}{2}(k'-k)a}-
2\delta t\cos\frac{1}{2}(k'+k)a
\right]
\end{equation}

\noindent The modulus squared is

\begin{eqnarray}
\left|\left<\phi(\mathbf{k'})|H_{surf}(\mathbf{R})|\phi(\mathbf{k})
\right>\right|^2 
& = & N_M^{-2}
\left\{\left[(\delta\epsilon_L+\delta\epsilon_R)\cos\frac{1}{2}(k'-k)a-
2\delta t\cos\frac{1}{2}(k'+k)a\right]^2\right. \nonumber \\
 & & +\left.(\delta\epsilon_L-\delta\epsilon_R)^2\sin^2\frac{1}{2}(k'-k)a
\right\}
\end{eqnarray}

Define the two functions of $\mathbf{k}$ and $\mathbf{k'}$, $J_{surf}^{(x)}(\mathbf{k},\mathbf{k'})$ and $J_{surf}^{(y)}(\mathbf{k},\mathbf{k'})$ as

\begin{eqnarray}
\label{Jsx}
J_{surf}^{(x)}(\mathbf{k},\mathbf{k'}) & = &
\left<(\delta\epsilon_L+\delta\epsilon_R)^2\right>_x
\cos^2\frac{1}{2}(k'_x-k_x)a \nonumber \\
 & & -2\left<\delta t(\delta\epsilon_L+\delta\epsilon_R)\right>_x
\cos\frac{1}{2}(k'_x-k_x)a\cos\frac{1}{2}(k'_x+k_x)a  \nonumber \\
 & & +\left<\delta t^2\right>_x\cdot4\cos^2\frac{1}{2}(k'_x+k_x)a+
\left<(\delta\epsilon_L-\delta\epsilon_R)^2\right>_x
\sin^2\frac{1}{2}(k'_x-k_x)a,
\end{eqnarray}

\begin{eqnarray}
\label{Jsy}
J_{surf}^{(y)}(\mathbf{k},\mathbf{k'}) & = &
\left<(\delta\epsilon_L+\delta\epsilon_R)^2\right>_y
\cos^2\frac{1}{2}(k'_y-k_y)a \nonumber \\
 & & -2\left<\delta t(\delta\epsilon_L+\delta\epsilon_R)\right>_y
\cos\frac{1}{2}(k'_y-k_y)a\cos\frac{1}{2}(k'_y+k_y)a  \nonumber \\
 & & +\left<\delta t^2\right>_y\cdot4\cos^2\frac{1}{2}(k'_y+k_y)a+
\left<(\delta\epsilon_L-\delta\epsilon_R)^2\right>_y
\sin^2\frac{1}{2}(k'_y-k_y)a, 
\end{eqnarray}

\noindent where $<F(\mathbf{R_\sigma})>_x$ is the average of the function $F(\mathbf{R_\sigma})$ defined for each planar surface O on the x-axis with position $\mathbf{R_\sigma}$ as shown in Figure \ref{surfO},

\begin{equation}
\label{defF}
\left<F(\mathbf{R_\sigma})\right>=\frac{1}{N_M}
\sum_{\mathbf{R_\sigma}}F(\mathbf{R_\sigma}).
\end{equation}

\noindent Similary, $<F(\mathbf{R_\sigma})>_y$ is the average of $F(\mathbf{R_\sigma})$ over the y-axis surface O atoms. The expression in equation \ref{Jsy} for $J_{surf}^{(y)}$ is identical to the expression for $J_{surf}^{(x)}$ in equation \ref{Jsx} with $x$ replaced by $y$.

From the $\mathbf{k}$-space versions of equations \ref{aF} and \ref{lam}

\begin{equation}
\lambda_{surf}(\mathbf{k},\mathbf{k'},\omega_n)=\left(\frac{N(0)}{N_M}\right)
\left[J_{surf}^{(x)}(\mathbf{k},\mathbf{k'})+
      J_{surf}^{(y)}(\mathbf{k},\mathbf{k'})\right]
\left(\frac{2\omega_{ph}}{\omega_n^2+\omega_{ph}^2}\right).
\end{equation}

\subsubsection{O Atom Mode Perpendicular to Surface in Figure \ref{phonons}b}
\label{lam0perp}

\begin{figure}[h]
\centering \includegraphics[width=6cm]
{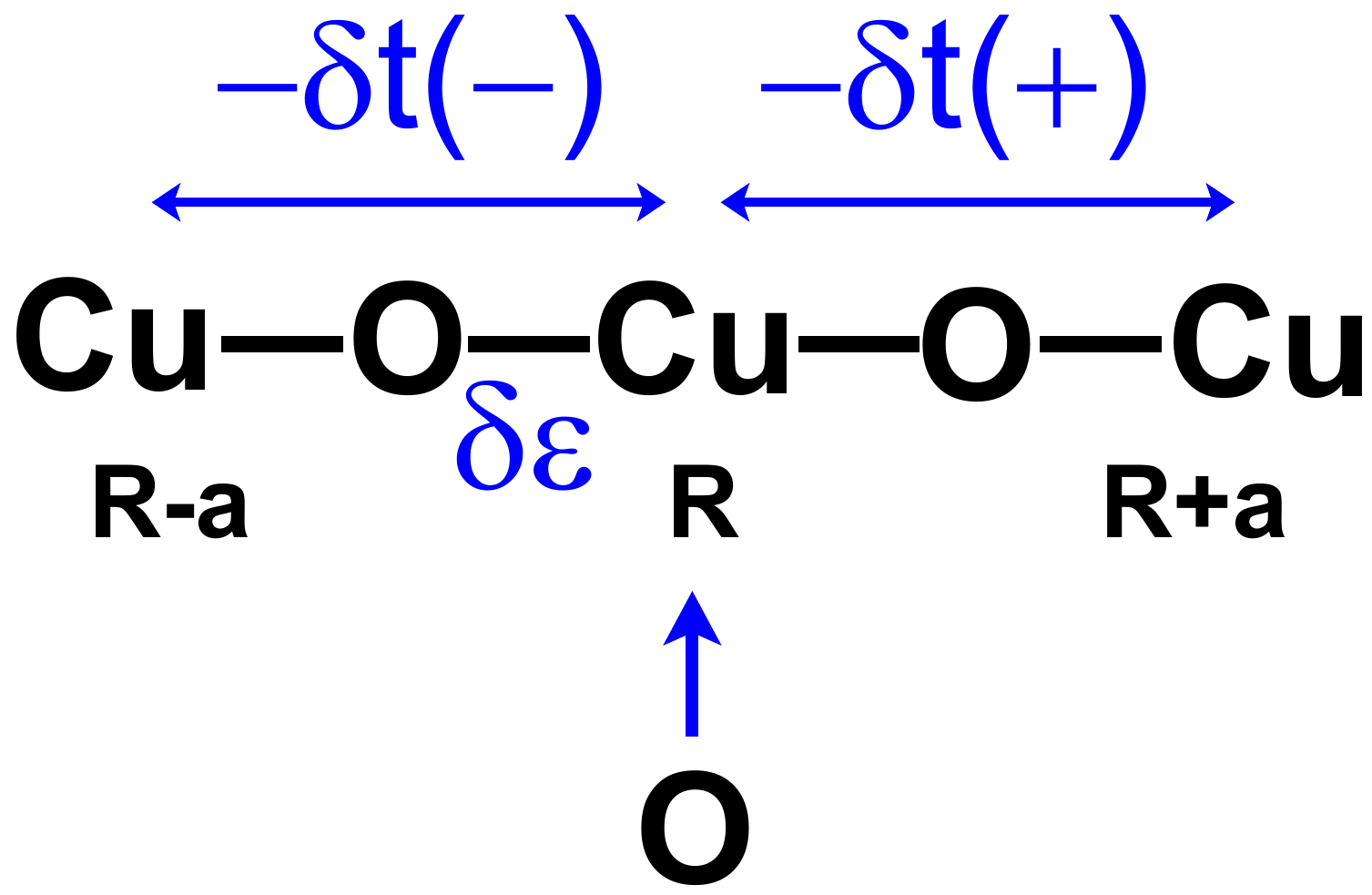}
\caption{The Perpendicular O atom phonon mode. The O atom is in the insulating AF region. The chain Cu$-$O$-$Cu$-$O$-$Cu is on the metal-insulator surface. The change in Cu orbital energy for the central Cu atom located at position $\mathbf{R}$ is $\delta\epsilon$. $-\delta t(-)$ is the change in the hopping between the Cu site at $\mathbf{R-a}$ and $\mathbf{R}$. $-\delta t(+)$ is the change in the hopping between the Cu site at $\mathbf{R+a}$ and $\mathbf{R}$. $\delta\epsilon$, $\delta t(-)$, and $\delta t(+)$ are functions of position \textbf{R} as seen in Figure \ref{phonons}b. We choose the energy of all the perpendicular surface phonon modes to be the same and equal to $\omega_{ph}$. In this figure, the displaced O atom is on the y-axis while the Cu$-$O$-$Cu$-$O$-$Cu chain is along the x-axis. We choose the convention of labeling the phonon mode by the axis of the AF O atom. Hence, this figure is a y-axis O phonon mode.
}
\label{perpO}
\end{figure}

The Hamiltonian for Figure \ref{perpO} is

\begin{equation}
\label{ham-perp}
H_{\perp}(\mathbf{R})=\delta\epsilon c_R^\dagger c_R
 -\delta t(-)\left(c_{R-a}^\dagger c_R + c_R^\dagger c_{R-a}\right)
 -\delta t(+)\left(c_{R+a}^\dagger c_R + c_R^\dagger c_{R+a}\right),
\end{equation}

\noindent where $c_R^\dagger$ and $c_R$ create and destroy an electron at the $R$ Cu site. $c_{R\pm a}^\dagger$ and $c_{R\pm a}$ are defined similarly. Since there is no electron spin coupling to the O atom phonon mode, the electron spin index is dropped in equation \ref{ham-perp}.

\noindent The matrix element between $\mathbf{k'}$ and $\mathbf{k}$ is

\begin{equation}
\left<\phi(\mathbf{k'})|H_{\perp}(\mathbf{R})|\phi(\mathbf{k})\right>=
N_M^{-1}\mathrm{e}^{-i(k'-k)R}\left[
\delta\epsilon
-\delta t(+)\left(\mathrm{e}^{-ik'a}+\mathrm{e}^{ika}\right)
-\delta t(-)\left(\mathrm{e}^{ik'a}+\mathrm{e}^{-ika}\right)
\right]
\end{equation}

\noindent The modulus squared is

\begin{eqnarray}
\left|\left<\phi(\mathbf{k'})|H_{\perp}(\mathbf{R})|\phi(\mathbf{k})
\right>\right|^2 
& = & N_M^{-2}\left\{
\left\{\delta\epsilon-\left[\delta t(+)+\delta t(-)\right]
\left[\cos(k'a)+\cos(ka)\right]\right\}^2\right. \nonumber \\
 & & +\left.\left[\delta t(+)-\delta t(-)\right]^2
\left[\sin(k'a)-\sin(ka)\right]^2
\right\}
\end{eqnarray}

Define the two functions of $\mathbf{k}$ and $\mathbf{k'}$, $J_{\perp}^{(x)}(\mathbf{k},\mathbf{k'})$ and $J_{\perp}^{(y)}(\mathbf{k},\mathbf{k'})$ as

\begin{eqnarray}
\label{Jpx}
J_{\perp}^{(x)}(\mathbf{k},\mathbf{k'}) & = &
\left<\delta\epsilon^2\right>_x
-2\left<\delta\epsilon\left[\delta t(+)+\delta t(-)\right]\right>_x
\left(\cos k'_xa+\cos k_xa\right) \nonumber \\
 & & +\left<\left[\delta t(+)+\delta t(-)\right]\right>_x
      \left(\cos k'_xa+\cos k_xa\right)^2 \nonumber \\
 & & +\left<\left[\delta t(+)-\delta t(-)\right]\right>_x
      \left(\sin k'_xa-\sin k_xa\right)^2,
\end{eqnarray}

\begin{eqnarray}
\label{Jpy}
J_{\perp}^{(y)}(\mathbf{k},\mathbf{k'}) & = &
\left<\delta\epsilon^2\right>_y
-2\left<\delta\epsilon\left[\delta t(+)+\delta t(-)\right]\right>_y
\left(\cos k'_ya+\cos k_ya\right) \nonumber \\
 & & +\left<\left[\delta t(+)+\delta t(-)\right]\right>_y
      \left(\cos k'_ya+\cos k_ya\right)^2 \nonumber \\
 & & +\left<\left[\delta t(+)-\delta t(-)\right]\right>_y
      \left(\sin k'_ya-\sin k_ya\right)^2,
\end{eqnarray}

\noindent where $<F(\mathbf{R_\sigma})>_x$ is the average, defined in equation \ref{defF}, of the function $F(\mathbf{R_\sigma})$ for each x-axis O phonon mode with position $\mathbf{R_\sigma}$ as shown in Figure \ref{perpO}. Similary, $<F(\mathbf{R_\sigma})>_y$ is the average of $F(\mathbf{R_\sigma})$ over the y-axis O atoms. The expression in equation \ref{Jpy} for $J_{\perp}^{(y)}$ is identical to the expression for $J_{\perp}^{(x)}$ in equation \ref{Jpx} with $x$ replaced by $y$.

From the $\mathbf{k}$-space versions of equations \ref{aF} and \ref{lam}

\begin{equation}
\lambda_{\perp}(\mathbf{k},\mathbf{k'},\omega_n)=\left(\frac{N(0)}{N_M}\right)
\left[J_{\perp}^{(x)}(\mathbf{k},\mathbf{k'})+
      J_{\perp}^{(y)}(\mathbf{k},\mathbf{k'})\right]
\left(\frac{2\omega_{ph}}{\omega_n^2+\omega_{ph}^2}\right).
\end{equation}

\subsection{Contribution to $\mathbf{\lambda}$ from the Phonons Responsible for the the Resistivity}
\label{lamR}

The low-temperature resistivity of \lsco\ is the sum of two terms.\cite{Hussey2011} One term is linear in $T$ and the other is proportional to $T^2$. At high temperatures, both terms become linear in $T$. Previously, we showed\cite{Tahir-Kheli2013} that the doping evolution of these two terms can be explained by phonon scattering and simple counting of the number of metallic sites and the number of overlapped plaquettes, as a function of doping. The contribution of these phonons on \tc\ must be included in our Eliashberg calculation.

The power law dependence of the two terms in the resistivity restricts the form of their electron-phonon spectral functions, $\alpha^2 F_1$ and $\alpha^2 F_2$ for the linear and $T^2$ contributions, respectively. From Fermi's Golden Rule, the electron scattering rate is

\begin{equation}
\frac{1}{\tau(\mathbf{k})}=\frac{2\pi}{\hbar}\sum_{\mathbf{k}}
  2\int_0^{+\infty}d\Omega\ \alpha^2 F(\mathbf{k},\mathbf{k'},\Omega)
                                     n_B(\Omega),
\end{equation}

\noindent where $n_B(\Omega)$ is the Bose-Einstein distribution $n_B(\omega)=1/[\exp(\omega/T)-1]$. The factor of two in front of the integral comes from the absorption and emission of phonons. $\alpha^2 F$ is zero for $\Omega$ greater than the highest phonon energy.

At high temperatures, $n_B(\Omega)\approx T/\Omega$ leading to $\hbar/\tau(\mathbf{k})\approx 2\pi\lambda_{\mathbf{k}}T$, where

\begin{equation}
\lambda_{\mathbf{k}}=
  2\int_0^{+\infty}d\Omega\ \frac{\alpha^2 F(\mathbf{k},\Omega)}{\Omega},
\end{equation}

\noindent and $\alpha^2 F(\mathbf{k},\Omega)=\sum_{\mathbf{k'}}\alpha^2 F(\mathbf{k},\mathbf{k'},\Omega)$. $\lambda_{\mathbf{k}}$ is called the mass-enhancement factor.\cite{Allen1982} The slope of the high-temperature scattering rate can be obtained from the resistivity. Hence, the mass-enhancement can be computed from experiment.

At low-temperatures, the Bose-Einstein distribution cuts the integral in the scattering rate off at $\Omega\sim T$. If $\alpha^2 F\sim\Omega^n$, then

\begin{equation}
\label{a2Fscaling}
\frac{1}{\tau(\mathbf{k})}\sim\int_0^T d\Omega
       \ \frac{\alpha^2 F(\mathbf{k},\Omega)}{\Omega}(T)\sim T^{n+1}.
\end{equation}

The low-temperature $T^2$ scattering rate is known to be isotropic in $\mathbf{k}$-space,\cite{Abdel-Jawad2006} and thereby it must scale as $\sim\Omega$ from equation \ref{a2Fscaling}. From the low-temperature resistivity experiments\cite{Hussey2011}, we showed the $T^2$ resistivity term was proportional to $(1-N_{4M}/N_{Tot})$, where $N_{Tot}$ is the total number of Cu sites (metallic plus insulating AF sites) and $N_{4M}$ is the number of metallic Cu sites that are in non-overlapping plaquettes. Therefore, $\alpha^2 F_2(\mathbf{k},\mathbf{k'},\Omega)$ is of the form

\begin{equation}
\alpha^2 F_2(\mathbf{k},\mathbf{k'},\Omega)=
  C_2\left(\frac{\Omega}{\omega_D}\right)\left(1-\frac{N_{4M}}{N_{Tot}}\right),
\end{equation}

\noindent where $C_2$ is a constant to be determined. $\omega_D$ is the Debye energy. $\alpha^2 F_2=0$, for $\Omega>\omega_D$.

The low-temperature $T$ scattering rate is zero along the diagonals, $k_x=\pm k_y$, and large at $\mathbf{k}=(0,\pm\pi),(\pm\pi,0)$.\cite{Abdel-Jawad2006} $\alpha^2 F_1$ is independent of $\Omega$ from equation \ref{a2Fscaling}. The scattering rate in equation \ref{a2Fscaling} logarithmically diverges for small $\Omega$. Hence, it must be cutoff at some minimum, $\omega_{min}$. For temperatures below $\omega_{min}$, the scattering rate cannot be linear in $T$. Previously, we showed that $\omega_{min}\approx 1$ K.\cite{Tahir-Kheli2013} In this paper, we fix $\omega_{min}=1$ K. See Appendix \ref{parameters}.

The spectral function, $\alpha^2 F_1(\mathbf{k},\mathbf{k'},\Omega)$, is of the form

\begin{equation}
\alpha^2 F_1(\mathbf{k},\mathbf{k'},\Omega)=
  C_1\left(\frac{N_{4M}}{N_{Tot}}\right)D(\mathbf{k})D(\mathbf{k'}),
\ \ \mathrm{for}\ \omega_{min}<\Omega<\omega_D,
\end{equation}

\noindent where $C_1$ is a constant, and $\alpha^2 F_1=0$ outside of the range $\omega_{min}<\Omega<\omega_D$.

The anisotropy factor, $D(\mathbf{k})$, is

\begin{equation}
\label{Dfac}
D(\mathbf{k})=\frac{(\cos(k_x a)-\cos(k_y a))^2}
    {\left<(\cos(k_x a)-\cos(k_y a))^2\right>}
\end{equation}

\noindent where the denominator is the average over the Fermi surface of the numerator.

The average of a function, $f(\mathbf{k})$, over the Fermi surface is defined as

\begin{equation}
\left<f(\mathbf{k})\right>=
\frac{
\sum_{k'}\left[\frac{\delta\left(\epsilon_{\mathbf{k'}}-\epsilon_F\right)}{N(0)}\right]f(\mathbf{k'})
}
{
\sum_{k'}\left[\frac{\delta\left(\epsilon_{\mathbf{k'}}-\epsilon_F\right)}{N(0)}\right]
}.
\end{equation}

\noindent Thus $<D(\mathbf{k})>=1$.

The constants $C_1$ and $C_2$ can be determined as follows. The average around the Fermi surface of the scattering rate at high-temperatures is $1/\tau=2\pi\lambda_{tr}T$. From resistivity measurements,\cite{Cooper2009} $\lambda_{tr}\approx 0.5$. A fraction $(N_{4M}/N_{Tot})\lambda_{tr}$ of $\lambda_{tr}$ comes from $<\alpha^2 F_1>$ and the fraction $(1-N_{4M}/N_{Tot})\lambda_{tr}$ comes from $<\alpha^2 F_2>$ leading to

\begin{eqnarray}
\lambda_{tr}\left(1-\frac{N_{4M}}{N_{Tot}}\right) & = & 2\int_0^{+\infty}
 d\Omega\ \frac{\left<\alpha^2 F_2(\mathbf{k},\mathbf{k'},\Omega)\right>}
                                                     {\Omega}, \nonumber \\
\lambda_{tr} & = & 2C_2\int_0^{\omega_D}
               \frac{d\Omega}{\omega_D}, \nonumber \\
\lambda_{tr} & = & 2C_2,
\end{eqnarray}

\begin{eqnarray}
\lambda_{tr}\left(\frac{N_{4M}}{N_{Tot}}\right)  & = & 
   2\int_{\omega_{min}}^{\omega_D}d\Omega
   \ \frac{\left<\alpha^2 F_1(\mathbf{k},\mathbf{k'},\Omega)\right>}
        {\Omega}, \nonumber \\
\lambda_{tr} & = &
 2C_1\int_{\omega_{min}}^{\omega_D}\frac{d\Omega}{\Omega}, \nonumber \\
\lambda_{tr} & = &
     C_1\left[2\ln\left(\frac{\omega_D}{\omega_{min}}\right)\right].
\end{eqnarray}

\noindent Substituting $C_1$ and $C_2$ in terms of $\lambda_{tr}$ back into $\alpha^2 F_1$ and $\alpha^2 F_2$ yields

\begin{equation}
\alpha^2 F_1(\mathbf{k},\mathbf{k'},\Omega) = \lambda_{tr}
\left[\ln\left(\frac{\omega_D}{\omega_{min}}\right)^2\right]^{-1}
D(\mathbf{k})D(\mathbf{k'})\left(\frac{N_{4M}}{N_{Tot}}\right),
\ \ \omega_{min}<\Omega<\omega_D, 
\end{equation}

\noindent and

\begin{equation}
\alpha^2 F_2(\mathbf{k},\mathbf{k'},\Omega) =
  \frac{1}{2}\lambda_{tr}
  \left(\frac{\Omega}{\omega_D}\right)\left(1-\frac{N_{4M}}{N_{Tot}}\right),
 \ \ 0<\Omega<\omega_D.
\end{equation}

\noindent $\alpha^2 F_1=0$ outside of the range $\omega_{min}<\Omega<\omega_D$ and $\alpha^2 F_2=0$ for $\Omega>\omega_D$.

We solve for $\lambda_i(\mathbf{k},\mathbf{k'},\omega_n)$, for $i=1,2$ using the $\mathbf{k}$-space version of equation \ref{lam}

\begin{equation}
\label{lamk}
\lambda_i(\mathbf{k},\mathbf{k'},\omega_n) =
\int_0^{+\infty}d\Omega\ \alpha^2 F_i(\mathbf{k},\mathbf{k'},\Omega)
             \left(\frac{2\Omega}{\omega_n^2+\Omega^2}\right),
\end{equation}

\noindent leading to

\begin{equation}
\lambda_1(\mathbf{k},\mathbf{k'},\omega_n) = 
\left(\frac{N_{4M}}{N_{Tot}}\right)\lambda_{tr}D(\mathbf{k})D(\mathbf{k'})
\left[
\frac{
  \ln\left(\frac{\omega_n^2+\omega_D^2}{\omega_n^2+\omega_{min}^2}\right)
}
{
  \ln\left(\frac{\omega_D^2}{\omega_{min}^2}\right)
}
\right]
\end{equation}

\noindent and

\begin{equation}
\lambda_2(\mathbf{k},\mathbf{k'},\omega_n) = 
\left(1-\frac{N_{4M}}{N_{Tot}}\right)\lambda_{tr}
  \left[1-\frac{|\omega_n|}{\omega_D}
         \tan^{-1}\left(\frac{\omega_D}{|\omega_n|}\right)\right]
\end{equation}

\subsection{Generalization of the Eliashberg Equations for Multi-Layer Cuprates}
\label{eliashberg-multi}

The Eliashberg equations \ref{e1a}, \ref{eZ}, and \ref{eD} for a single CuO$\mathrm{_2}$ layer per unit cell are generalized to multi-layer cuprates by changing $\mathbf{k}$ and $\mathbf{k'}$ to $\mathbf{bk}$ and $\mathbf{b'k'}$, respectively, in the single layer Eliashberg equations.

\begin{eqnarray}
\label{e1ab}
Z(\mathbf{bk},i\omega_n) & = & 1 +
    \frac{\pi T}{|\omega_n|}\sum_{\mathbf{b'k'}n'}^{|\omega_{n'}|<\omega_c}
    \left[\frac{\delta(\epsilon_{\mathbf{b'k'}}-\epsilon_F)}{N(0)}\right]
    \lambda(\mathbf{bk},\mathbf{b'k'},\omega_n-\omega_{n'})s_n s_{n'},
\nonumber \\
\label{e2ab}
Z(\mathbf{bk},i\omega_n)\Delta(\mathbf{bk},i\omega_n) & = &
       \pi T \\
\label{e3ab}
 & \times &
       \sum_{\mathbf{b'k'}n'}^{|\omega_{n'}|<\omega_c}\frac{1}{|\omega_{n'}|}
 \left[\frac{\delta(\epsilon_{\mathbf{b'k'}}-\epsilon_F)}{N(0)}\right]
  [\lambda(\mathbf{bk},\mathbf{b'k'},\omega_n-\omega_{n'})-\mu^*(\omega_c)]
       \Delta(\mathbf{b'k'},i\omega_{n'}), \nonumber
\end{eqnarray}

\begin{eqnarray}
\label{eZb}
 Z(\mathbf{bk},-i\omega_n) & = & Z(\mathbf{bk},i\omega_n) =
                                        \mathrm{Real\ Number}, \\
\label{eDb}
 \Delta(\mathbf{bk},-i\omega_n) & = &
            \Delta(\mathbf{bk},i\omega_n) = \mathrm{Real\ Number},
\end{eqnarray}

\noindent where $\mathbf{b}$ and $\mathbf{b'}$ are band indicies. They vary from $1$ to $L$, where $L$ is the number of CuO$\mathrm{_2}$ layers per unit cell. A unit cell contains $L$ Cu atoms, one in each layer. The $\mathbf{k}$ vector is a 2D vector. $N(0)$ is the total density of states per spin

\begin{equation}
N(0)= \sum_{\mathbf{b'k'}}\delta(\epsilon_{\mathbf{b'k'}}-\epsilon_F).
\end{equation}

There is a Bloch $\mathbf{k}$ state for each layer, $l$, given by $\phi(l\mathbf{k})$. The band eigenfunctions are

\begin{equation}
\label{psi-evec}
\psi(\mathbf{bk}) = \sum_l A_{bl}(\mathbf{k})\phi(l\mathbf{k}).
\end{equation}

\noindent The coefficients, $A_{bl}(\mathbf{k})$, are real since the inter-layer hopping matrix elements are real. The matrix element for hopping between adjacent layers is

\begin{equation}
\label{phi-inter}
<\phi(l\pm 1\mathbf{k'})|H_{inter}|\phi(l\mathbf{k})>=
-t(l\pm1,l,\mathbf{k})\delta_{\mathbf{k'k}}
\end{equation}

\noindent where

\begin{equation}
t(l\pm1,l,\mathbf{k})=\alpha t_z(1/4)[\cos(k_xa)-\cos(k_ya)]^2,
\end{equation}

\noindent and $\alpha$ is the product of the fraction of metallic sites in layers $l$ and $l\pm1$. See Appendix \ref{parameters} Table \ref{band}.

The eigenvectors $\psi(\mathbf{bk})$ of equations \ref{psi-evec} and \ref{phi-inter} are independent of the magnitude of $t(l\pm1,l,\mathbf{k})$. Thus $A_{bl}(\mathbf{k})$ is independent of $\mathbf{k}$,

\begin{equation}
A_{bl}(\mathbf{k})=A_{bl}. 
\end{equation}

The eigenstates, $\psi(\mathbf{bk})$, are normalized leading to

\begin{equation}
\sum_{\mathbf{b}} A_{bl'}A_{bl} = \delta_{ll'},
\end{equation}

\begin{equation}
\sum_{l} A_{b'l}A_{bl} = \delta_{b'b}.
\end{equation}

The electron-phonon spectral function $\alpha^2 F_2(\mathbf{bk},\mathbf{b'k'},\Omega)$ is

\begin{equation} 
\alpha^2 F_2(\mathbf{bk},\mathbf{b'k'},\Omega)=
\frac{1}{2}\lambda_{tr}\left(\frac{\Omega}{\omega_D}\right)
\left(1-n_{4M}\right),\ \ 0<\Omega<\omega_D,
\end{equation}

\noindent where

\begin{equation}
n_{4M}(l)=\frac{N_{4M}(l)}{N_{xy}},
\end{equation}

\begin{equation}
n_{4M}=\frac{1}{L}\sum_l n_{4M}(l).
\end{equation}

\noindent $N_{4M}(l)$ is the number of metallic Cu sites in layer $l$ that are in non-overlapping plaquettes, $L$ is the total number of CuO$\mathrm{_2}$ layers per unit cell, and $N_{xy}$ is the total number of Cu sites (metallic plus insulating AF) in a single CuO$\mathrm{_2}$ layer. Hence, $LN_{xy}$ is the total number of Cu sites in the crystal and $n_{4M}$ is the total fraction of metallic Cu sites over all the CuO$\mathrm{_2}$ layers. $\alpha^2 F_2=0$ for $\Omega>\omega_D$.

For the electron-phonon spectral function, $\alpha^2 F_1$, that leads to the low-temperature linear resistivity, define the anisotropy factor, $D(\mathbf{bk})$ as

\begin{equation}
\label{Dfacb}
D(\mathbf{bk})=\frac{(\cos(k_x a)-\cos(k_y a))^2}
    {\left<(\cos(k_x a)-\cos(k_y a))^2\right>}
\end{equation}

\noindent where the denominator is the average over all the $L$ Fermi surfaces of the numerator.

The average of a function, $f(\mathbf{bk})$, over all the Fermi surfaces is defined as

\begin{equation}
\left<f(\mathbf{bk})\right>=
\frac{
\sum_{b'k'}\left[\frac{\delta\left(\epsilon_{\mathbf{b'k'}}-\epsilon_F\right)}{N(0)}\right]f(\mathbf{b'k'})
}
{
\sum_{b'k'}\left[\frac{\delta\left(\epsilon_{\mathbf{b'k'}}-\epsilon_F\right)}{N(0)}\right]
}.
\end{equation}

The phonon modes in $\alpha^2 F_1$ are 2D. Hence, the form of the spectral function between layers $l$ and $l'$ is of the form,

\begin{equation}
\alpha^2 F_1(lk,l'k',\Omega) = \delta_{ll'}
 \left[\ln\left(\frac{\omega_D^2}{\omega_{min}^2}\right)\right]^{-1}
n_{4M}(l)D(\mathbf{k})D(\mathbf{k'}).
\end{equation}

Expanding the eigenstates $\psi(\mathbf{bk})$ in terms of $\phi(l\mathbf{k})$ from equation \ref{psi-evec} leads to

\begin{equation}
\alpha^2 F_1(\mathbf{bk},\mathbf{b'k'},\Omega)=
\left\{
\sum_l|A_{bl}|^2|A_{b'l'}|^2 n_{4M}(l)
\right\}
\lambda_{tr}\left[\ln\left(\frac{\omega_D^2}{\omega_{min}^2}\right)\right]^{-1}
D(\mathbf{bk})D(\mathbf{b'k'}),
\end{equation}

\noindent where $\omega_{min}<\Omega<\omega_D$. $\alpha^2 F_1=0$, for $\Omega<\omega_{min}$ or $\Omega>\omega_D$.

Hence,

\begin{equation}
\lambda_1(\mathbf{bk},\mathbf{b'k'},\omega_n) = 
\left\{
\sum_l|A_{bl}|^2|A_{b'l'}|^2 n_{4M}(l)
\right\}
\lambda_{tr}D(\mathbf{bk})D(\mathbf{bk'})
\left[
\frac{
  \ln\left(\frac{\omega_n^2+\omega_D^2}{\omega_n^2+\omega_{min}^2}\right)
}
{
  \ln\left(\frac{\omega_D^2}{\omega_{min}^2}\right)
}
\right]
\end{equation}

\noindent and

\begin{equation}
\lambda_2(\mathbf{bk},\mathbf{b'k'},\omega_n) = 
\left(1-n_{4M}\right)\lambda_{tr}
  \left[1-\frac{|\omega_n|}{\omega_D}
         \tan^{-1}\left(\frac{\omega_D}{|\omega_n|}\right)\right]
\end{equation}

The multi-layer expressions for $\lambda_{surf}(\mathbf{bk},\mathbf{b'k'},\omega_n)$ and $\lambda_{\perp}(\mathbf{bk},\mathbf{b'k'},\omega_n)$ are similar to their single-layer counterparts with a modified definition for the averaging in their respective $J^{(x)}$ and $J^{(y)}$ functions.

\begin{eqnarray}
\label{Jsxb}
J_{surf}^{(x)}(\mathbf{bk},\mathbf{b'k'}) & = &
\left<(\delta\epsilon_L+\delta\epsilon_R)^2\right>_x
\cos^2\frac{1}{2}(k'_x-k_x)a \nonumber \\
 & & -2\left<\delta t(\delta\epsilon_L+\delta\epsilon_R)\right>_x
\cos\frac{1}{2}(k'_x-k_x)a\cos\frac{1}{2}(k'_x+k_x)a  \nonumber \\
 & & +\left<\delta t^2\right>_x\cdot4\cos^2\frac{1}{2}(k'_x+k_x)a+
\left<(\delta\epsilon_L-\delta\epsilon_R)^2\right>_x
\sin^2\frac{1}{2}(k'_x-k_x)a,
\end{eqnarray}

\begin{eqnarray}
\label{Jsyb}
J_{surf}^{(y)}(\mathbf{bk},\mathbf{b'k'}) & = &
\left<(\delta\epsilon_L+\delta\epsilon_R)^2\right>_y
\cos^2\frac{1}{2}(k'_y-k_y)a \nonumber \\
 & & -2\left<\delta t(\delta\epsilon_L+\delta\epsilon_R)\right>_y
\cos\frac{1}{2}(k'_y-k_y)a\cos\frac{1}{2}(k'_y+k_y)a  \nonumber \\
 & & +\left<\delta t^2\right>_y\cdot4\cos^2\frac{1}{2}(k'_y+k_y)a+
\left<(\delta\epsilon_L-\delta\epsilon_R)^2\right>_y
\sin^2\frac{1}{2}(k'_y-k_y)a, 
\end{eqnarray}

\noindent where $<F(\mathbf{R_\sigma})>_x$ is the multi-layer average of the function $F(\mathbf{R_\sigma})$ defined for each planar surface O on the x-axis with position $\mathbf{R_\sigma}$ as shown in Figure \ref{surfO},

\begin{equation}
\label{defFb}
\left<F(\mathbf{R_\sigma})\right>=
\sum_l|A_{bl}|^2|A_{b'l}|^2
\left<F_l(\mathbf{R_\sigma})\right>,
\end{equation}

\noindent and $<F_l(\mathbf{R_\sigma})>_x$ is the average over layer $l$, as defined in equation \ref{defF}. Similarly for $<F(\mathbf{R_\sigma})>_y$. The expression in equation \ref{Jsyb} for $J_{surf}^{(y)}$ is identical to the expression for $J_{surf}^{(x)}$ in equation \ref{Jsxb} with $x$ replaced by $y$.

Hence, $\lambda_{surf}(\mathbf{bk},\mathbf{b'k'},\omega_n)$ is

\begin{equation}
\lambda_{surf}(\mathbf{k},\mathbf{k'},\omega_n)=\left(\frac{N(0)}{N_M}\right)
\left[J_{surf}^{(x)}(\mathbf{k},\mathbf{k'})+
      J_{surf}^{(y)}(\mathbf{k},\mathbf{k'})\right]
\left(\frac{2\omega_{ph}}{\omega_n^2+\omega_{ph}^2}\right),
\end{equation}

\noindent where $N_M$ is the total number of metallic Cu sites, $N_M=\sum_l N_{lM}$, and $N_{lM}$ is the total number of metallic Cu sites in layer $l$.

For $\lambda_\perp(\mathbf{bk},\mathbf{b'k'},\omega_n)$, the corresponding $J_\perp^{(x)}$ and $J_\perp^{(y)}$ functions are

\begin{eqnarray}
\label{Jpxb}
J_{\perp}^{(x)}(\mathbf{bk},\mathbf{b'k'}) & = &
\left<\delta\epsilon^2\right>_x
-2\left<\delta\epsilon\left[\delta t(+)+\delta t(-)\right]\right>_x
\left(\cos k'_xa+\cos k_xa\right) \nonumber \\
 & & +\left<\left[\delta t(+)+\delta t(-)\right]\right>_x
      \left(\cos k'_xa+\cos k_xa\right)^2 \nonumber \\
 & & +\left<\left[\delta t(+)-\delta t(-)\right]\right>_x
      \left(\sin k'_xa-\sin k_xa\right)^2,
\end{eqnarray}

\begin{eqnarray}
\label{Jpyb}
J_{\perp}^{(y)}(\mathbf{bk},\mathbf{b'k'}) & = &
\left<\delta\epsilon^2\right>_y
-2\left<\delta\epsilon\left[\delta t(+)+\delta t(-)\right]\right>_y
\left(\cos k'_ya+\cos k_ya\right) \nonumber \\
 & & +\left<\left[\delta t(+)+\delta t(-)\right]\right>_y
      \left(\cos k'_ya+\cos k_ya\right)^2 \nonumber \\
 & & +\left<\left[\delta t(+)-\delta t(-)\right]\right>_y
      \left(\sin k'_ya-\sin k_ya\right)^2.
\end{eqnarray}

\noindent All averages in equations \ref{Jpxb} and \ref{Jpyb} are defined in equation \ref{defFb}.

Hence, $\lambda_{\perp}(\mathbf{bk},\mathbf{b'k'},\omega_n)$ is

\begin{equation}
\lambda_{\perp}(\mathbf{bk},\mathbf{b'k'},\omega_n)=\left(\frac{N(0)}{N_M}
\right)
\left[J_{\perp}^{(x)}(\mathbf{bk},\mathbf{b'k'})+
      J_{\perp}^{(y)}(\mathbf{bk},\mathbf{b'k'})\right]
\left(\frac{2\omega_{ph}}{\omega_n^2+\omega_{ph}^2}\right).
\end{equation}

\subsection{Computational Details}
\label{comp}

The band structure, $\epsilon_{\mathbf k}$, and all the parameters used the solve the Eliashberg equations for \tc\ are described in Appendix \ref{parameters}. Here, we discuss the computational issues necessary to obtain an accurate \tc.

The two planar interface O atom phonon modes in Figure \ref{phonons} require averaging products of pairs of parameters ($\delta\epsilon_L$, $\delta\epsilon_R$, and $\delta t$ for $\lambda_{surf}$, and $\delta\epsilon$ and $\delta t(\pm)$ for $\lambda_{\perp}$) over the lattice as seen in equations \ref{Jsx}, \ref{Jsy}, \ref{Jpx}, \ref{Jpy}, \ref{Jsxb}, \ref{Jsyb}, \ref{Jpxb}, and \ref{Jpyb}. These parameters vary depending on the environment of the Cu atoms, as shown in Figure \ref{phonons}.

For each doping value, we generate a $2000\times2000$ lattice of doped plaquettes. All O atoms that contribute to $\lambda_{surf}$ and $\lambda_{\perp}$ are identified along with the nature of the corresponding Cu sites (edge, convex, or concave, as shown in Figure \ref{phonons}). All the product averages are computed. Ensembles of $2000\times2000$ lattices can be generated to obtain more accurate product averages. We found that a single $2000\times2000$ doped lattice is large enough to obtain all the products to an accuracy of less than 1 \%.

All four electron-phonon pairing functions, $\lambda_1$, $\lambda_2$, $\lambda_{surf}$, and $\lambda_{\perp}$ can be written in the following product form
$\lambda(\mathbf{k},\mathbf{k'},\omega_n)= \lambda'(\mathbf{k},\mathbf{k'})F(\omega_n)$. The product separation, $\lambda=\lambda'F$, leads to a large reduction in the storage requirements because $\lambda'$ and $F$ can be computed once and saved, and the product computed on the fly.

We discretize the Fermi surface by choosing 10 uniformly spaced (in angle) \textbf{k}-points in the 45$^\circ$ wedge bounded by the vectors along the x-axis, $(\pi,0)$, and the diagonal, $(\pi,\pi)$, leading to a total of 80 \textbf{k}-points over the full Fermi surface. Increasing the number of \textbf{k}-points further led to $<0.1$ K change in the calculated \tc.

Fermi surface weights, $W_{\mathbf{bk}}$, are computed at each \textbf{bk}-point using the Fermi velocity evaluated from the band structure, $\epsilon_{\mathbf{k}}$. By rescaling the gap function, $\Delta(\mathbf{bk},\omega_n)$,

\begin{equation}
\Delta'(\mathbf{bk},\omega_n)=
\left[\frac{W_{\mathbf{bk}}}{|2n+1|}\right]^{\frac{1}{2}}
\Delta(\mathbf{bk},\omega_n)
\end{equation}

\noindent the Eliashberg equations can be turned into an eigenvalue equation with a real symmetric matrix.\cite{Allen1982} Since \tc\ occurs when the largest eigenvalue reaches one, we can perform a Lanczos projection. We compute \tc\ by bracketing. All the \tc\ values found in this paper are accurate to $\pm 0.3$K. For approximate timings, a full \tc-dome is computed on a small workstation in $\approx5-10$ minutes.

\clearpage

\begin{figure}[tbp]
\centering \includegraphics[width=15cm]{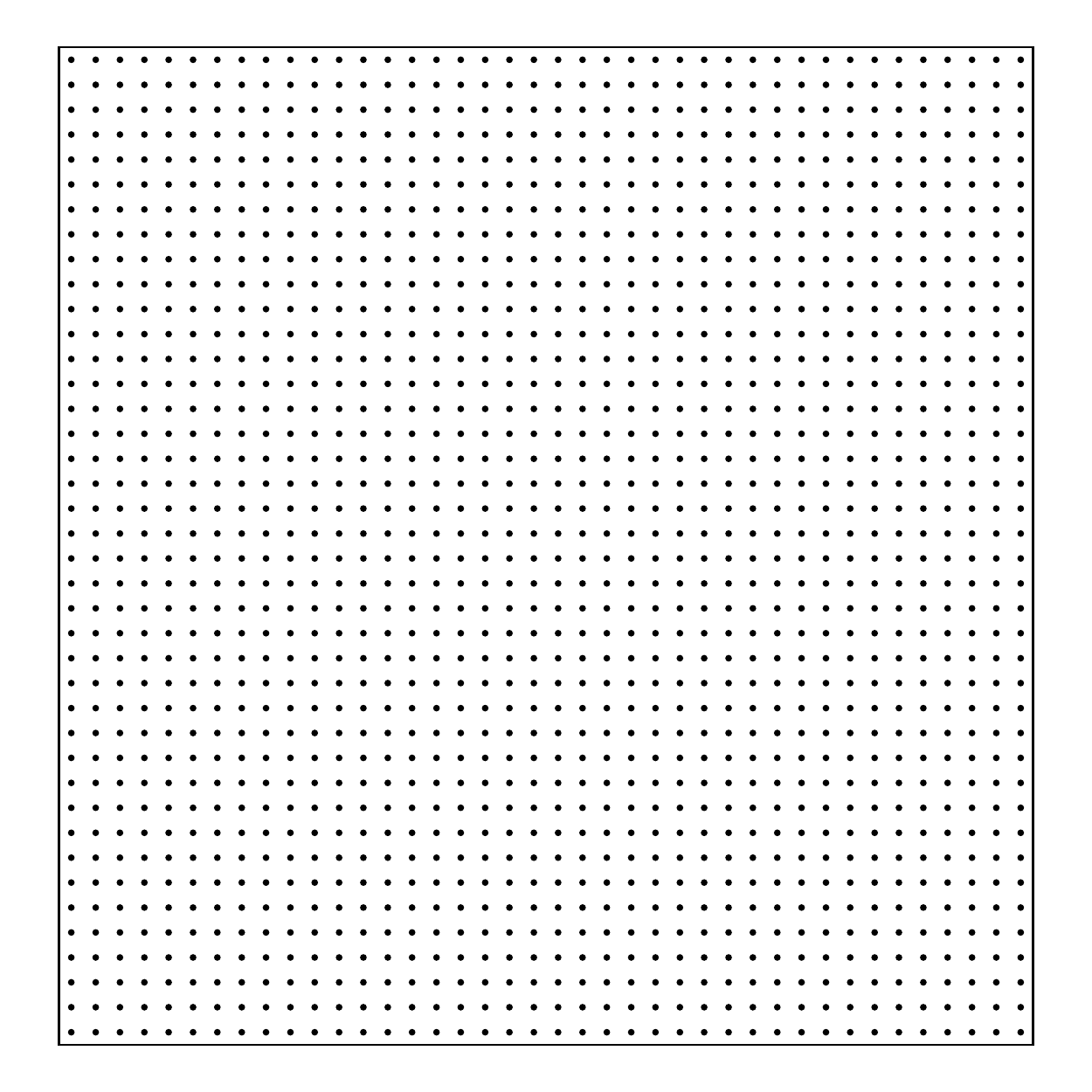}
\end{figure}
\noindent \textbf{Figure S0 (x = 0.00)}: Plaquette doping of a $40\times40$ square CuO$_2$ lattice. Only the Cu sites are shown. The black dots are undoped AF Cu sites. The material is a spin-1/2 antiferromagnet (AF).
\clearpage

\begin{figure}[tbp]
\centering \includegraphics[width=15cm]{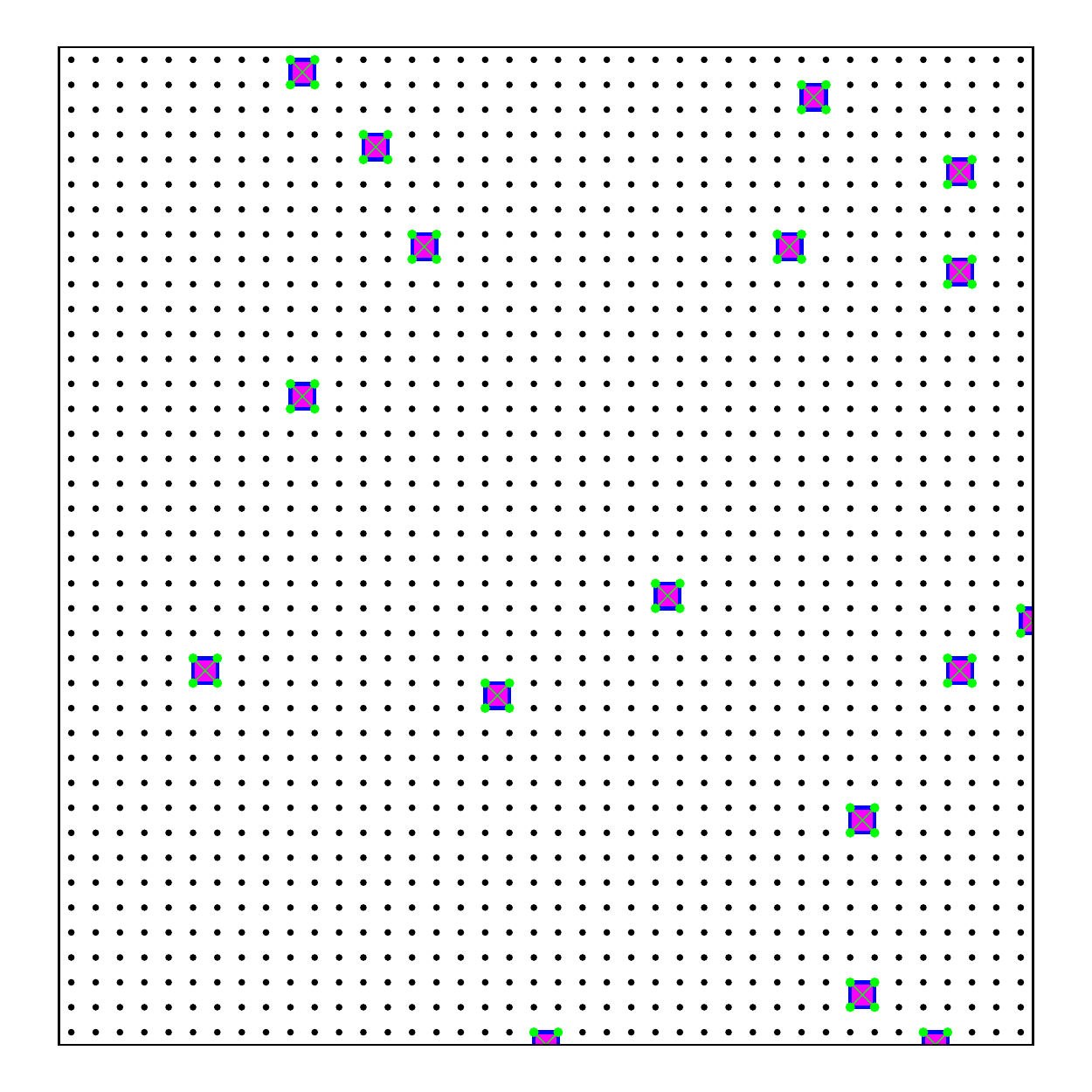}
\end{figure}
\noindent \textbf{Figure S1 (x = 0.01)}: Plaquette doping of a $40\times40$ square CuO$_2$ lattice. Only the Cu sites are shown. The black dots are undoped AF Cu sites. The blue squares are isolated plaquettes (no neighboring plaquette). There are only isolated plaquettes. The magenta overlay represents metallic delocalization of the planar Cu \dxxyy\ and O \psigma\ orbitals inside each plaquette. Fluctuating dumbbells are shown in every plaquette. There is no plaquette overlap at this doping.
\clearpage

\begin{figure}[tbp]
\centering \includegraphics[width=15cm]{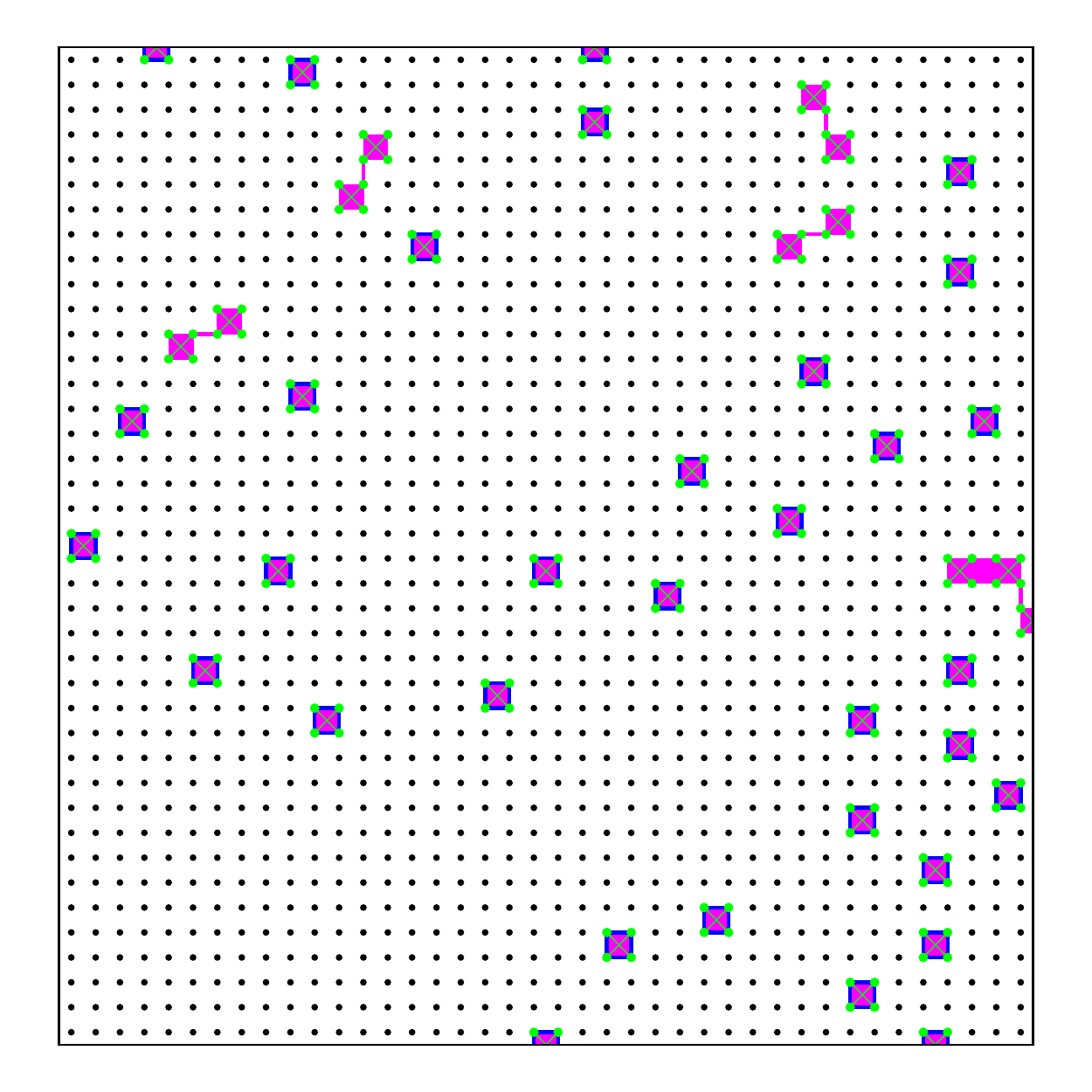}
\end{figure}
\noindent \textbf{Figure S2 (x = 0.02)}: Plaquette doping of a $40\times40$ square CuO$_2$ lattice. Only the Cu sites are shown. The black dots are undoped AF Cu sites. The blue squares are isolated plaquettes (no neighboring plaquette). A few magenta plaquette clusters are formed. These magenta clusters are smaller than the Cooper pair coherence length and do not contribute to superconducting pairing. The magenta overlay represents metallic delocalization of the planar Cu \dxxyy\ and O \psigma\ orbitals. Fluctuating dumbbells are shown in every plaquette. There is no plaquette overlap at this doping.
\clearpage

\begin{figure}[tbp]
\centering \includegraphics[width=15cm]{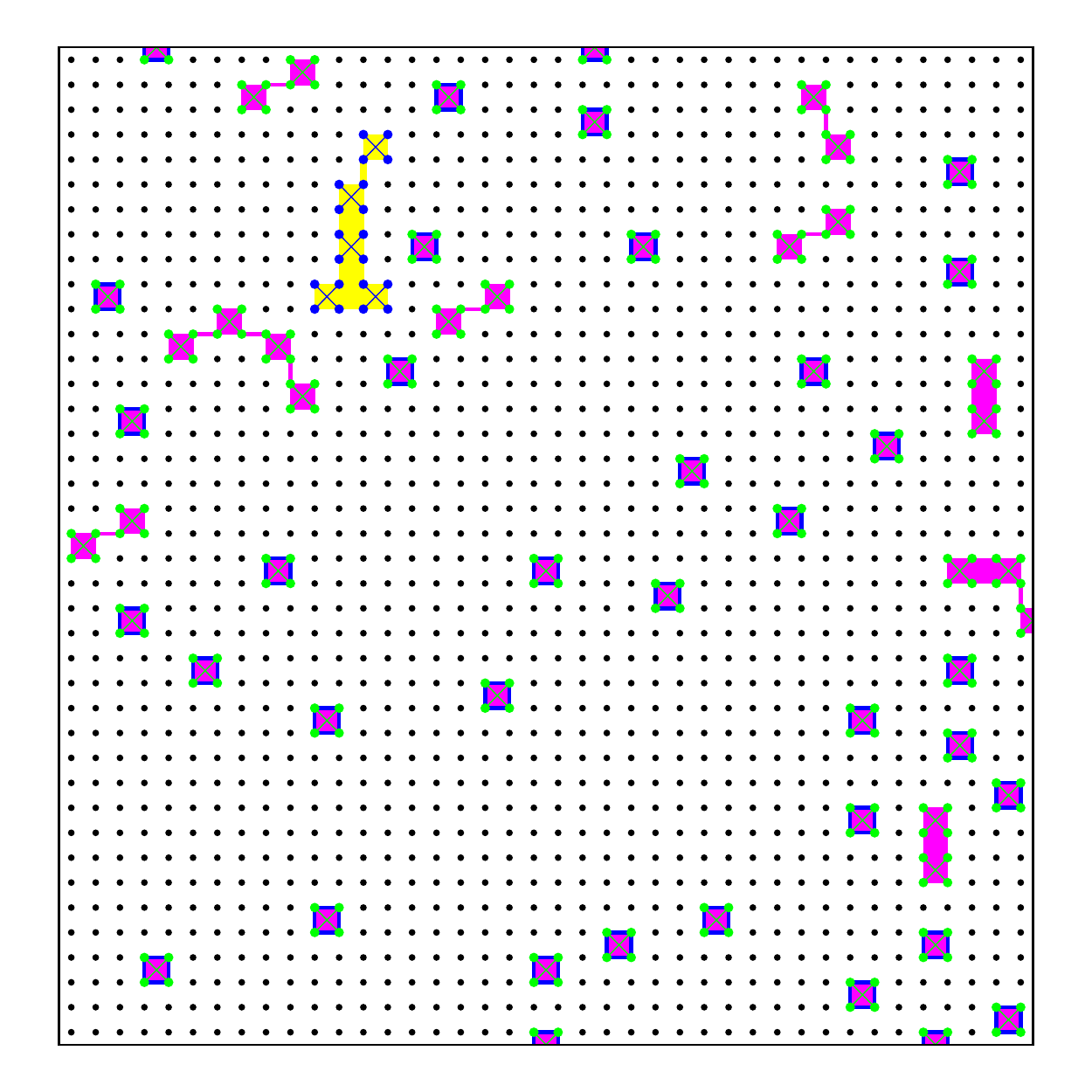}
\end{figure}
\noindent \textbf{Figure S3 (x = 0.03)}: Plaquette doping of a $40\times40$ square CuO$_2$ lattice. Only the Cu sites are shown. The black dots are undoped AF Cu sites. The blue squares are isolated plaquettes (no neighboring plaquette). The yellow plaquette clusters are comprised of more than 4 plaquettes and contribute to the superconducting pairing because they are larger than the coherence length. The yellow overlay represents the metallic region comprised of planar Cu \dxxyy\ and planar O \psigma\ character. The magenta clusters are smaller than the Cooper pair coherence length and do not contribute to superconducting pairing. The magenta overlay represents metallic delocalization of the planar Cu \dxxyy\ and O \psigma\ orbitals. Fluctuating dumbbells are shown in every plaquette. There is no plaquette overlap at this doping.
\clearpage

\begin{figure}[tbp]
\centering \includegraphics[width=15cm]{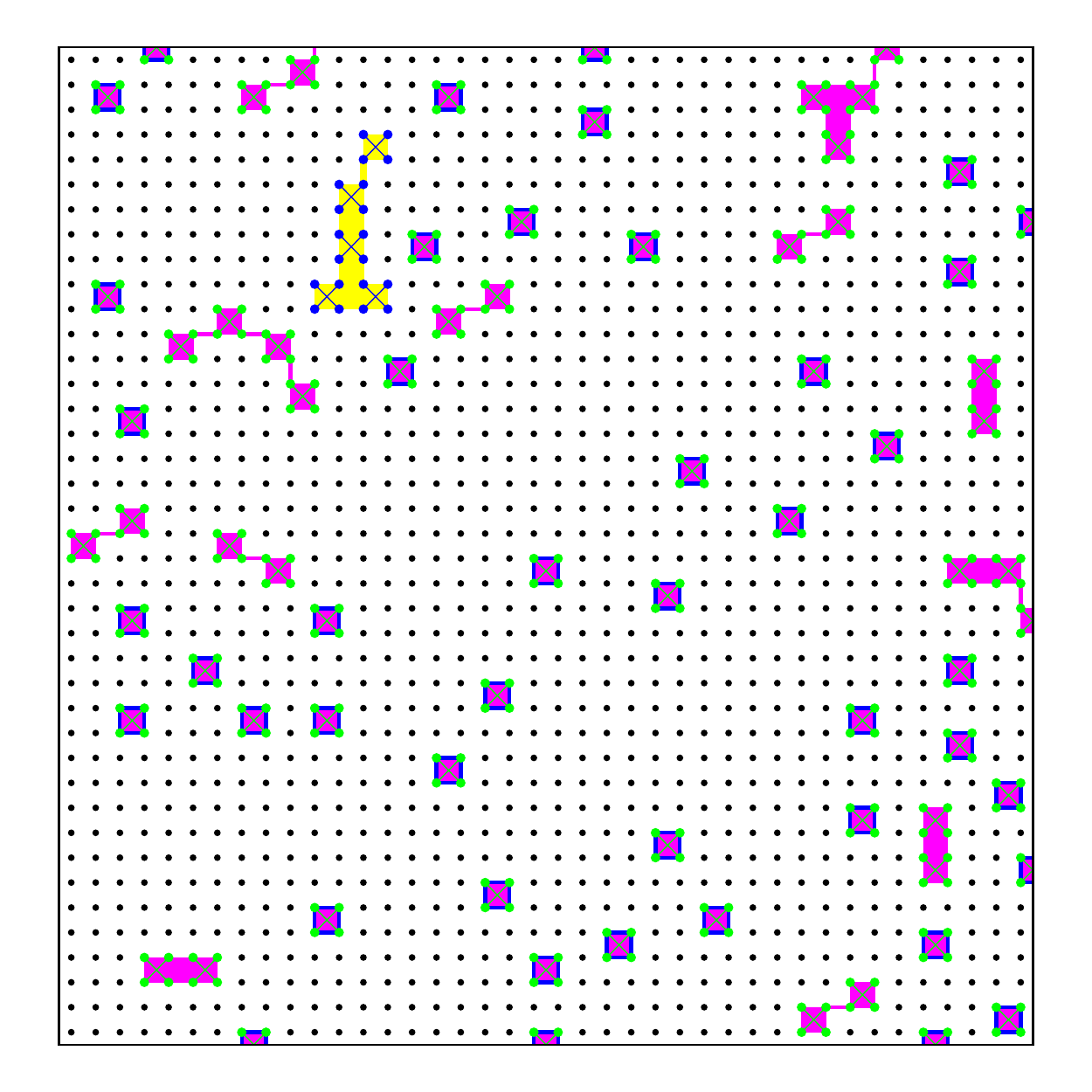}
\end{figure}
\noindent \textbf{Figure S4 (x = 0.04)}: Plaquette doping of a $40\times40$ square CuO$_2$ lattice. Only the Cu sites are shown. The black dots are undoped AF Cu sites. The blue squares are isolated plaquettes (no neighboring plaquette). The yellow plaquette clusters are comprised of more than 4 plaquettes and contribute to the superconducting pairing because they are larger than the coherence length. The yellow overlay represents the metallic region comprised of planar Cu \dxxyy\ and planar O \psigma\ character. The magenta clusters are smaller than the Cooper pair coherence length and do not contribute to superconducting pairing. The magenta overlay represents metallic delocalization of the planar Cu \dxxyy\ and O \psigma\ orbitals. Fluctuating dumbbells are shown in every plaquette. There is no plaquette overlap at this doping.
\clearpage

\begin{figure}[tbp]
\centering \includegraphics[width=15cm]{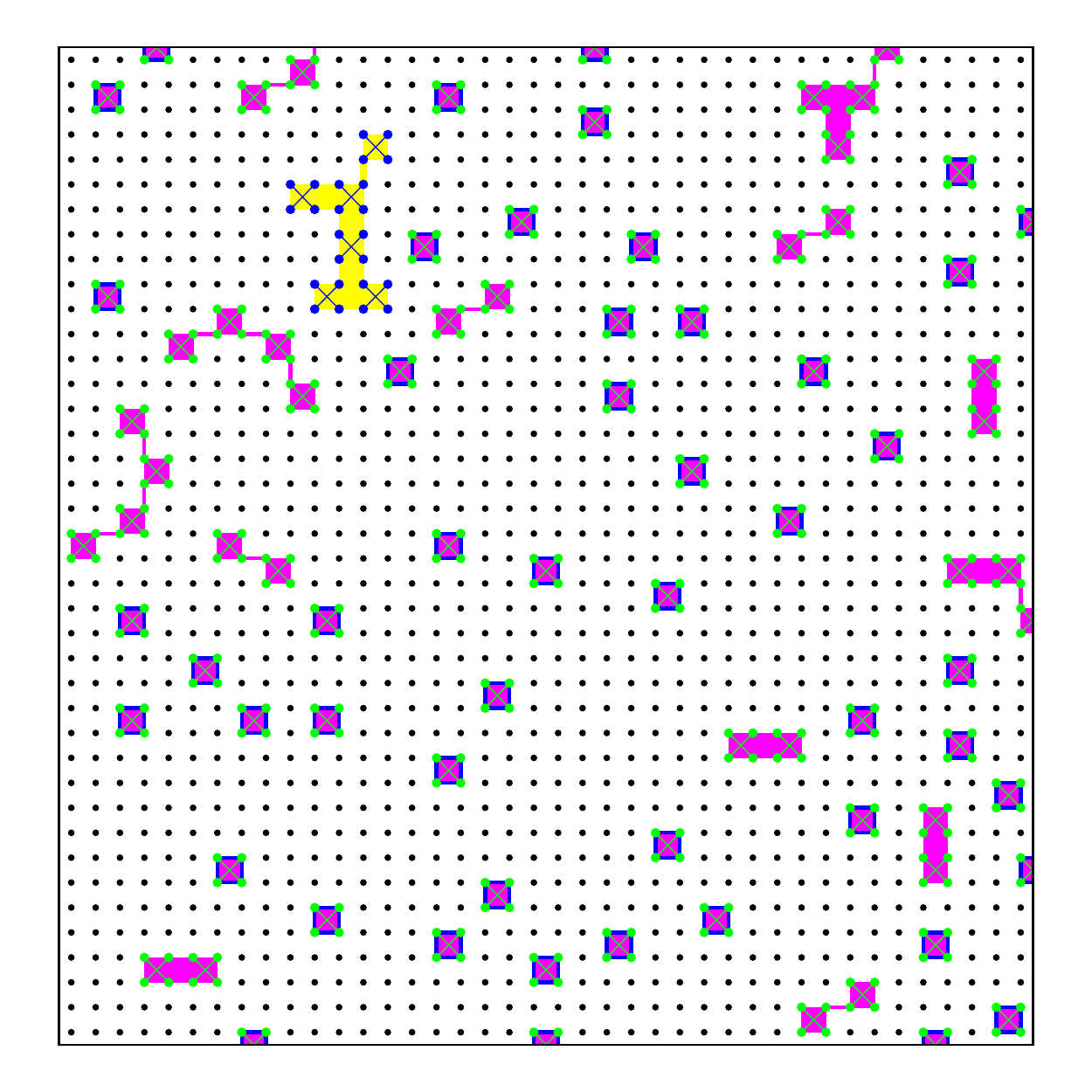}
\end{figure}
\noindent \textbf{Figure S5 (x = 0.05)}:
Plaquette doping of a $40\times40$ square CuO$_2$ lattice. Only the Cu sites are shown. The black dots are undoped AF Cu sites. The blue squares are isolated plaquettes (no neighboring plaquette). The yellow plaquette clusters are comprised of more than 4 plaquettes and contribute to the superconducting pairing because they are larger than the coherence length. The yellow overlay represents the metallic region comprised of planar Cu \dxxyy\ and planar O \psigma\ character. The magenta clusters are smaller than the Cooper pair coherence length and do not contribute to superconducting pairing. The magenta overlay represents metallic delocalization of the planar Cu \dxxyy\ and O \psigma\ orbitals. Fluctuating dumbbells are shown in every plaquette. There is no plaquette overlap at this doping. The number of isolated plaquettes peaks at this doping.
\clearpage

\begin{figure}[tbp]
\centering \includegraphics[width=15cm]{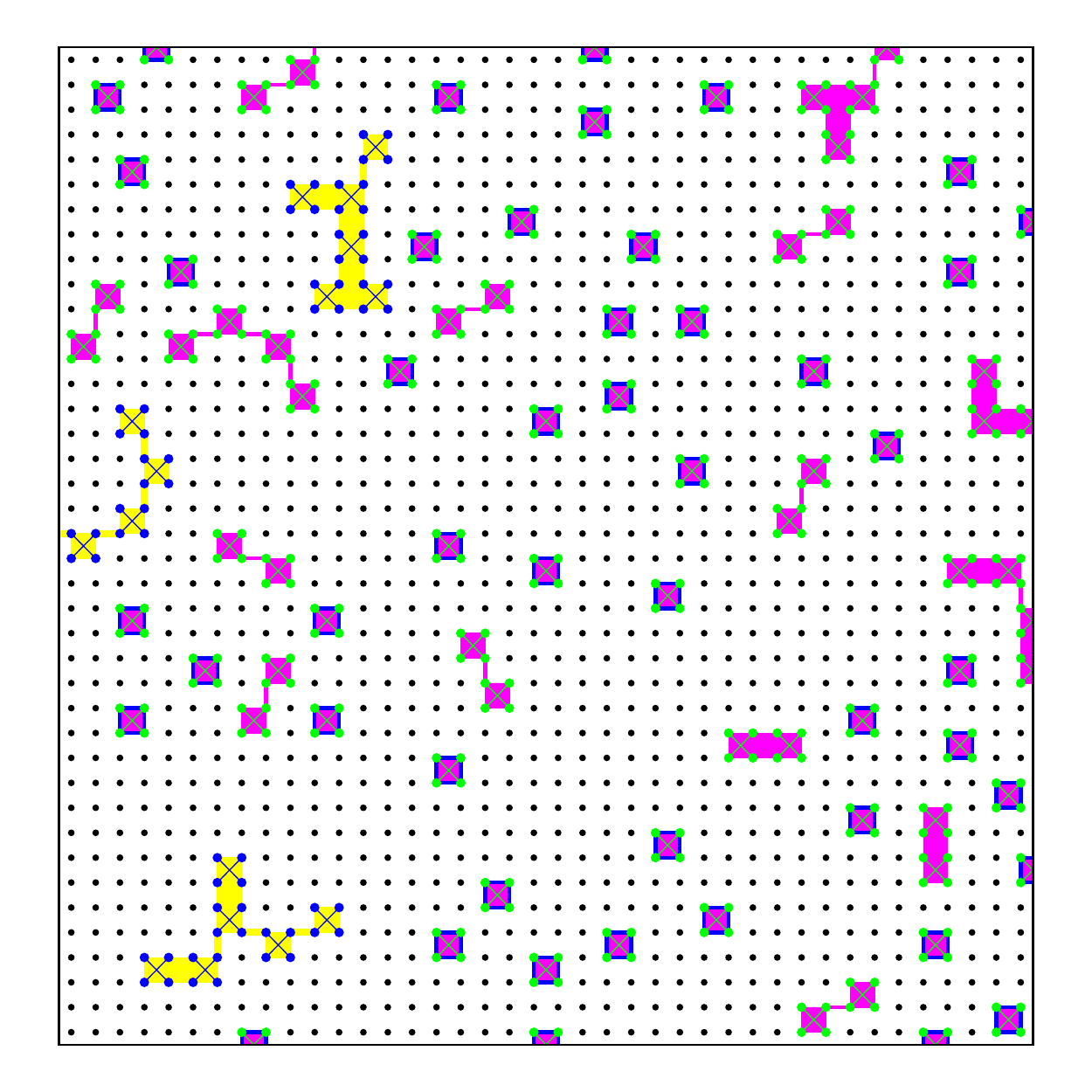}
\end{figure}
\noindent \textbf{Figure S6 (x = 0.06)}:
Plaquette doping of a $40\times40$ square CuO$_2$ lattice. Only the Cu sites are shown. The black dots are undoped AF Cu sites. The blue squares are isolated plaquettes (no neighboring plaquette). The yellow plaquette clusters are comprised of more than 4 plaquettes and contribute to the superconducting pairing because they are larger than the coherence length. The yellow overlay represents the metallic region comprised of planar Cu \dxxyy\ and planar O \psigma\ character. The magenta clusters are smaller than the Cooper pair coherence length and do not contribute to superconducting pairing. The magenta overlay represents metallic delocalization of the planar Cu \dxxyy\ and O \psigma\ orbitals. Fluctuating dumbbells are shown in every plaquette. There is no plaquette overlap at this doping.
\clearpage

\begin{figure}[tbp]
\centering \includegraphics[width=15cm]{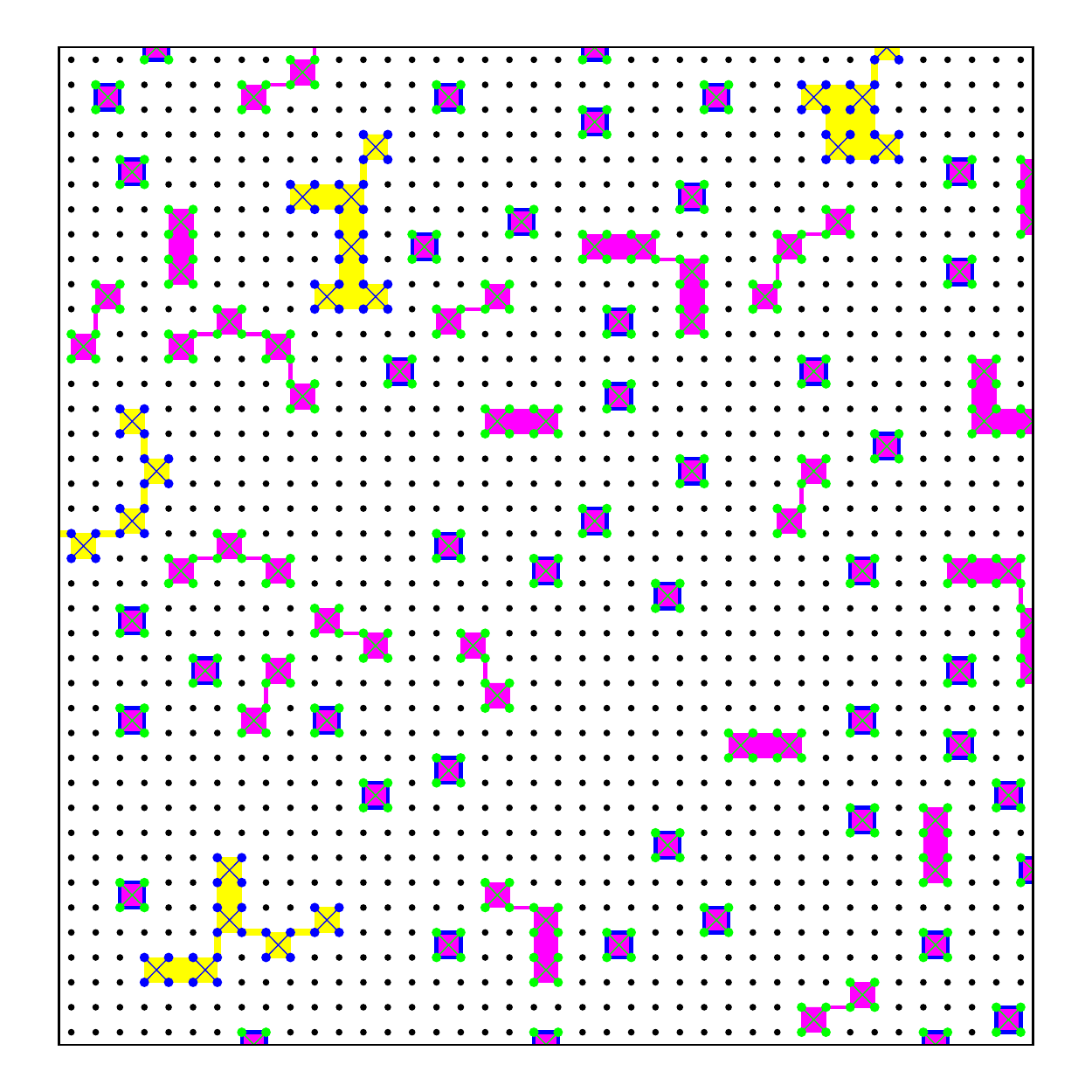}
\end{figure}
\noindent \textbf{Figure S7 (x = 0.07)}:
Plaquette doping of a $40\times40$ square CuO$_2$ lattice. Only the Cu sites are shown. The black dots are undoped AF Cu sites. The blue squares are isolated plaquettes (no neighboring plaquette). The yellow plaquette clusters are comprised of more than 4 plaquettes and contribute to the superconducting pairing because they are larger than the coherence length. The yellow overlay represents the metallic region comprised of planar Cu \dxxyy\ and planar O \psigma\ character. The magenta clusters are smaller than the Cooper pair coherence length and do not contribute to superconducting pairing. The magenta overlay represents metallic delocalization of the planar Cu \dxxyy\ and O \psigma\ orbitals. Fluctuating dumbbells are shown in every plaquette. There is no plaquette overlap at this doping.
\clearpage

\begin{figure}[tbp]
\centering \includegraphics[width=15cm]{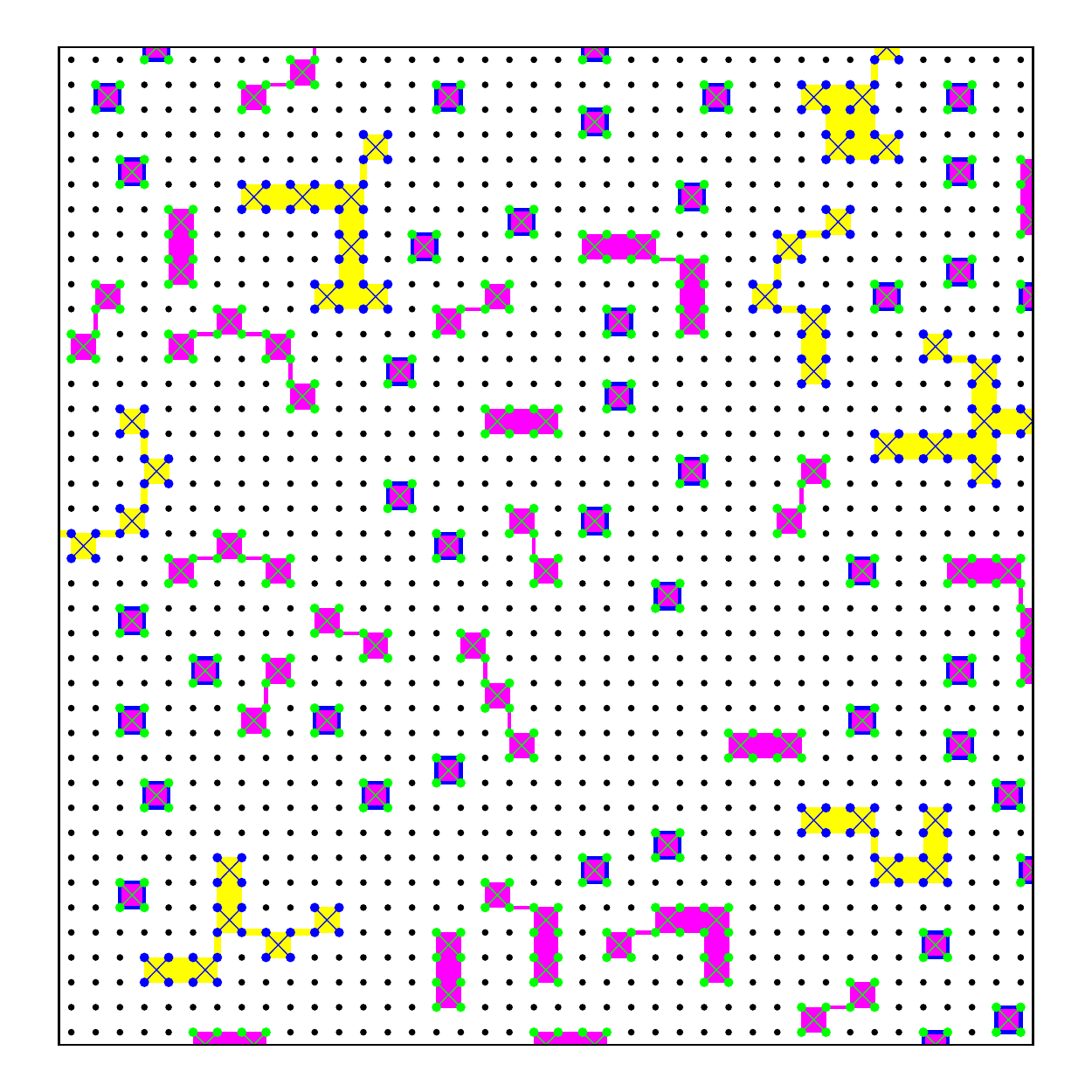}
\end{figure}
\noindent \textbf{Figure S8 (x = 0.08)}:
Plaquette doping of a $40\times40$ square CuO$_2$ lattice. Only the Cu sites are shown. The black dots are undoped AF Cu sites. The blue squares are isolated plaquettes (no neighboring plaquette). The yellow plaquette clusters are comprised of more than 4 plaquettes and contribute to the superconducting pairing because they are larger than the coherence length. The yellow overlay represents the metallic region comprised of planar Cu \dxxyy\ and planar O \psigma\ character. The magenta clusters are smaller than the Cooper pair coherence length and do not contribute to superconducting pairing. The magenta overlay represents metallic delocalization of the planar Cu \dxxyy\ and O \psigma\ orbitals. Fluctuating dumbbells are shown in every plaquette. There is no plaquette overlap at this doping.
\clearpage

\begin{figure}[tbp]
\centering \includegraphics[width=15cm]{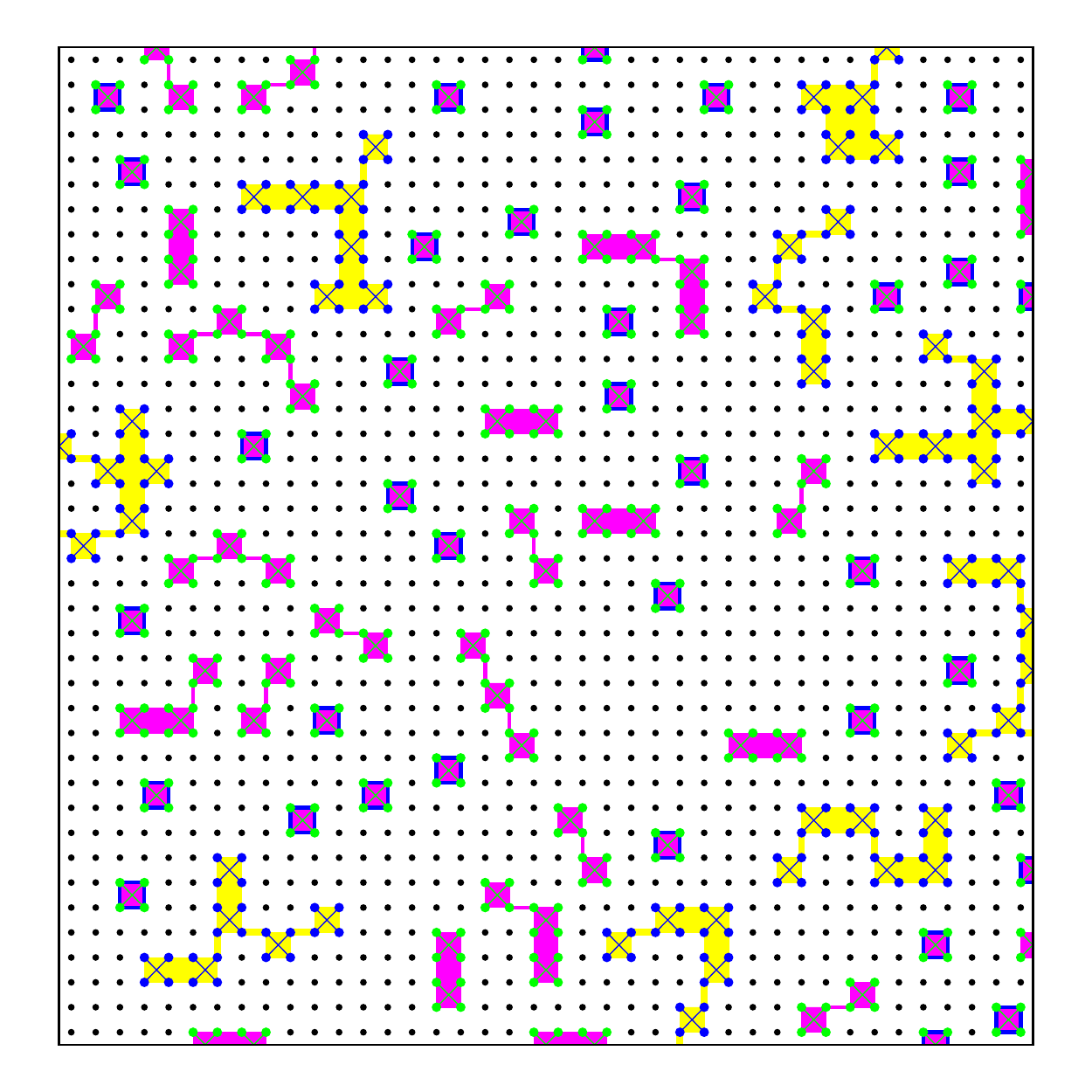}
\end{figure}
\noindent \textbf{Figure S9 (x = 0.09)}:
Plaquette doping of a $40\times40$ square CuO$_2$ lattice. Only the Cu sites are shown. The black dots are undoped AF Cu sites. The blue squares are isolated plaquettes (no neighboring plaquette). The yellow plaquette clusters are comprised of more than 4 plaquettes and contribute to the superconducting pairing because they are larger than the coherence length. The yellow overlay represents the metallic region comprised of planar Cu \dxxyy\ and planar O \psigma\ character. The magenta clusters are smaller than the Cooper pair coherence length and do not contribute to superconducting pairing. The magenta overlay represents metallic delocalization of the planar Cu \dxxyy\ and O \psigma\ orbitals. Fluctuating dumbbells are shown in every plaquette. There is no plaquette overlap at this doping.
\clearpage

\begin{figure}[tbp]
\centering \includegraphics[width=15cm]{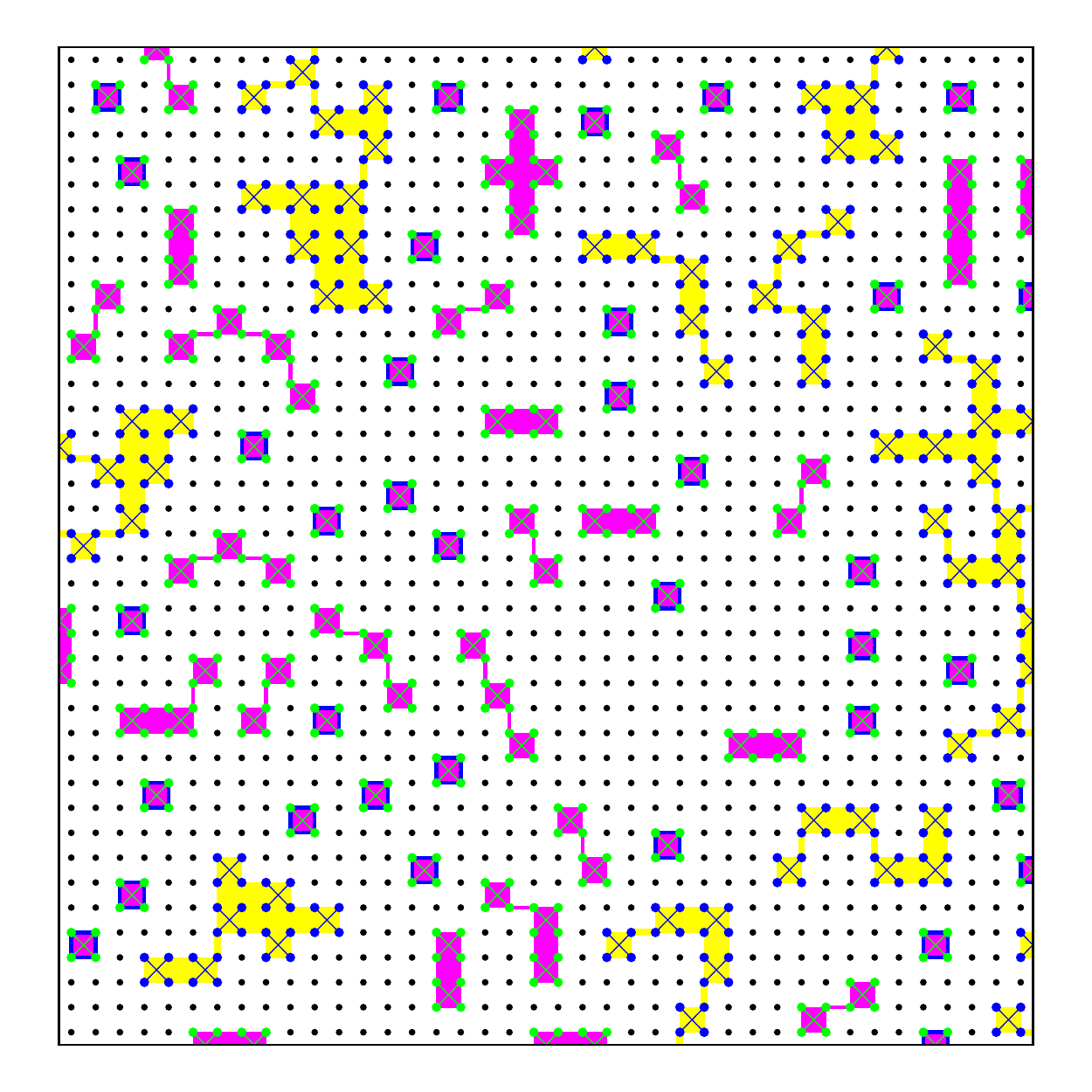}
\end{figure}
\noindent \textbf{Figure S10 (x = 0.10)}:
Plaquette doping of a $40\times40$ square CuO$_2$ lattice. Only the Cu sites are shown. The black dots are undoped AF Cu sites. The blue squares are isolated plaquettes (no neighboring plaquette). The yellow plaquette clusters are comprised of more than 4 plaquettes and contribute to the superconducting pairing because they are larger than the coherence length. The yellow overlay represents the metallic region comprised of planar Cu \dxxyy\ and planar O \psigma\ character. The magenta clusters are smaller than the Cooper pair coherence length and do not contribute to superconducting pairing. The magenta overlay represents metallic delocalization of the planar Cu \dxxyy\ and O \psigma\ orbitals. Fluctuating dumbbells are shown in every plaquette. There is no plaquette overlap at this doping.
\clearpage

\begin{figure}[tbp]
\centering \includegraphics[width=15cm]{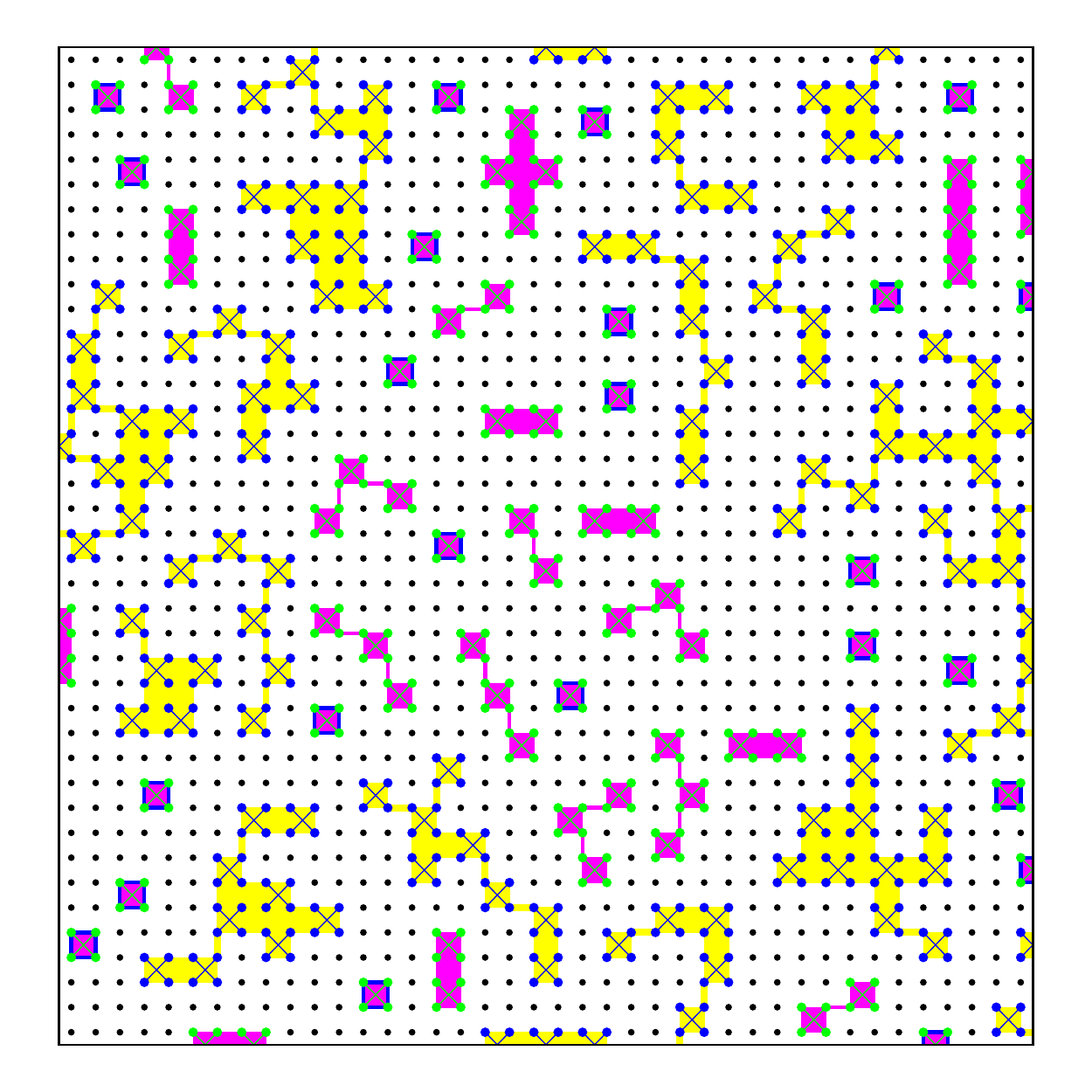}
\end{figure}
\noindent \textbf{Figure S11 (x = 0.11)}:
Plaquette doping of a $40\times40$ square CuO$_2$ lattice. Only the Cu sites are shown. The black dots are undoped AF Cu sites. The blue squares are isolated plaquettes (no neighboring plaquette). The yellow plaquette clusters are comprised of more than 4 plaquettes and contribute to the superconducting pairing because they are larger than the coherence length. The yellow overlay represents the metallic region comprised of planar Cu \dxxyy\ and planar O \psigma\ character. The magenta clusters are smaller than the Cooper pair coherence length and do not contribute to superconducting pairing. The magenta overlay represents metallic delocalization of the planar Cu \dxxyy\ and O \psigma\ orbitals. Fluctuating dumbbells are shown in every plaquette. There is no plaquette overlap at this doping.
\clearpage

\begin{figure}[tbp]
\centering \includegraphics[width=15cm]{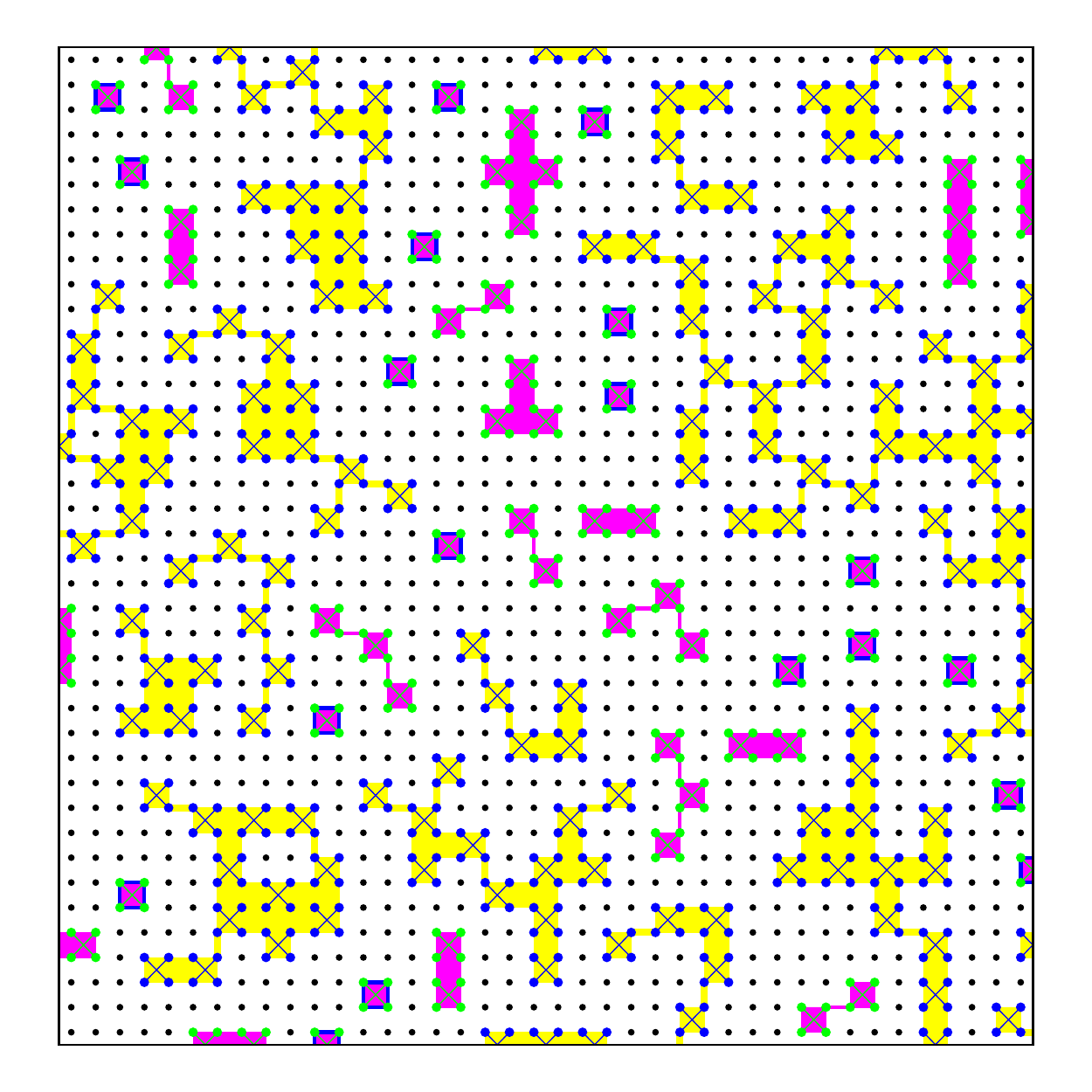}
\end{figure}
\noindent \textbf{Figure S12 (x = 0.12)}:
Plaquette doping of a $40\times40$ square CuO$_2$ lattice. Only the Cu sites are shown. The black dots are undoped AF Cu sites. The blue squares are isolated plaquettes (no neighboring plaquette). The yellow plaquette clusters are comprised of more than 4 plaquettes and contribute to the superconducting pairing because they are larger than the coherence length. The yellow overlay represents the metallic region comprised of planar Cu \dxxyy\ and planar O \psigma\ character. The magenta clusters are smaller than the Cooper pair coherence length and do not contribute to superconducting pairing. The magenta overlay represents metallic delocalization of the planar Cu \dxxyy\ and O \psigma\ orbitals. Fluctuating dumbbells are shown in every plaquette. There is no plaquette overlap at this doping.
\clearpage

\begin{figure}[tbp]
\centering \includegraphics[width=15cm]{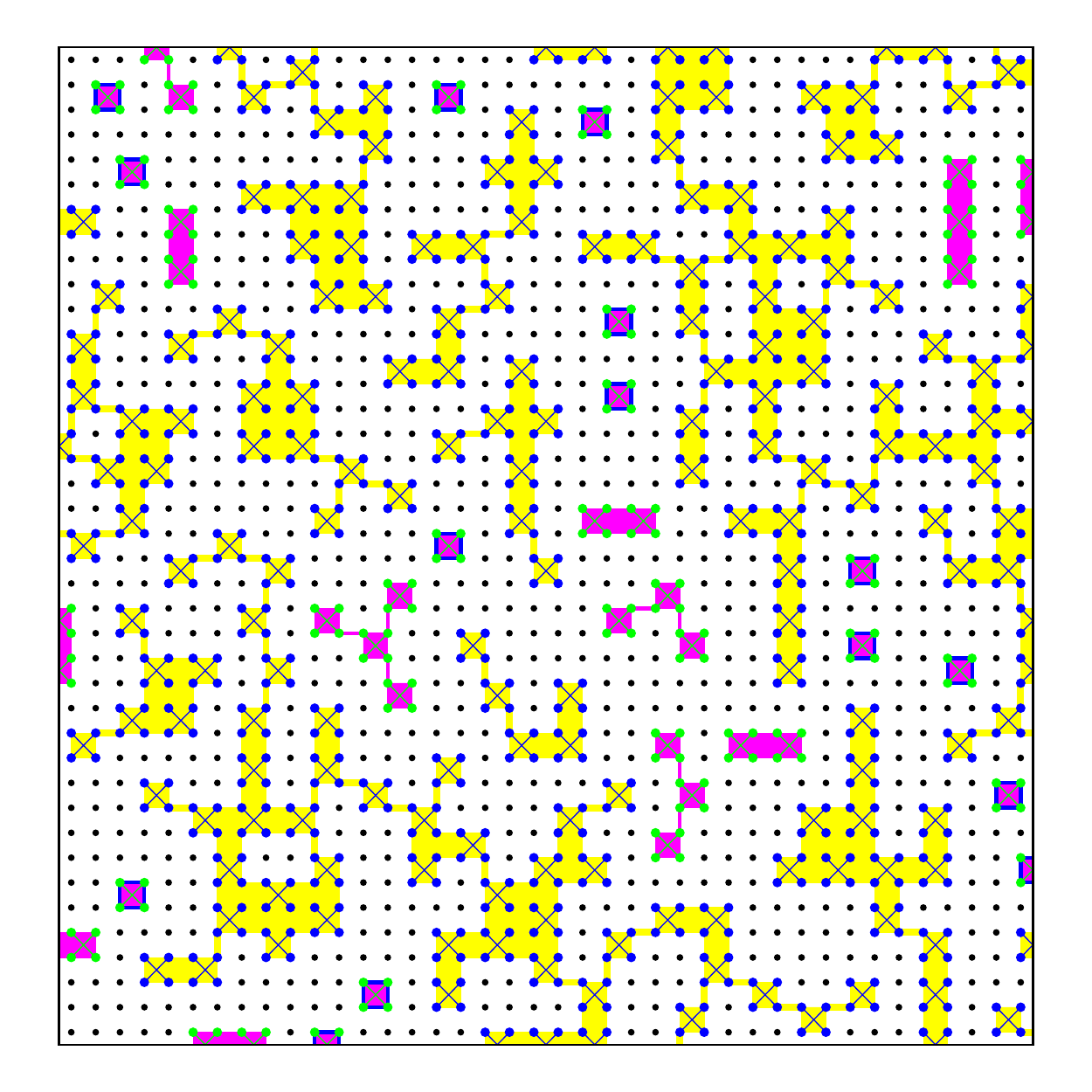}
\end{figure}
\noindent \textbf{Figure S13 (x = 0.13)}:
Plaquette doping of a $40\times40$ square CuO$_2$ lattice. Only the Cu sites are shown. The black dots are undoped AF Cu sites. The blue squares are isolated plaquettes (no neighboring plaquette). The yellow plaquette clusters are comprised of more than 4 plaquettes and contribute to the superconducting pairing because they are larger than the coherence length. The yellow overlay represents the metallic region comprised of planar Cu \dxxyy\ and planar O \psigma\ character. The magenta clusters are smaller than the Cooper pair coherence length and do not contribute to superconducting pairing. The magenta overlay represents metallic delocalization of the planar Cu \dxxyy\ and O \psigma\ orbitals. Fluctuating dumbbells are shown in every plaquette. There is no plaquette overlap at this doping.
\clearpage

\begin{figure}[tbp]
\centering \includegraphics[width=15cm]{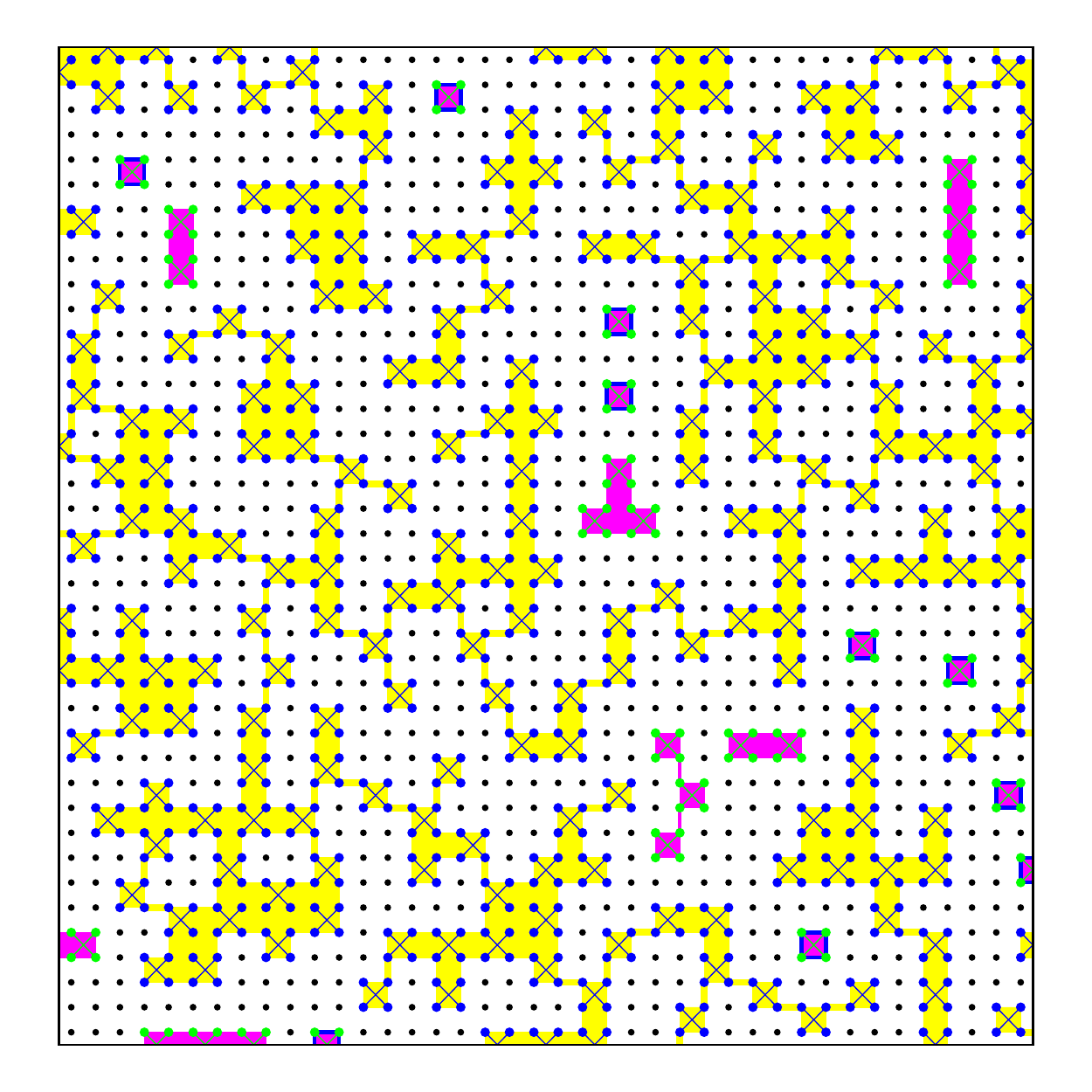}
\end{figure}
\noindent \textbf{Figure S14 (x = 0.14)}:
Plaquette doping of a $40\times40$ square CuO$_2$ lattice. Only the Cu sites are shown. The black dots are undoped AF Cu sites. The blue squares are isolated plaquettes (no neighboring plaquette). The yellow plaquette clusters are comprised of more than 4 plaquettes and contribute to the superconducting pairing because they are larger than the coherence length. The yellow overlay represents the metallic region comprised of planar Cu \dxxyy\ and planar O \psigma\ character. The magenta clusters are smaller than the Cooper pair coherence length and do not contribute to superconducting pairing. The magenta overlay represents metallic delocalization of the planar Cu \dxxyy\ and O \psigma\ orbitals. Fluctuating dumbbells are shown in every plaquette. There is no plaquette overlap at this doping.
\clearpage

\begin{figure}[tbp]
\centering \includegraphics[width=15cm]{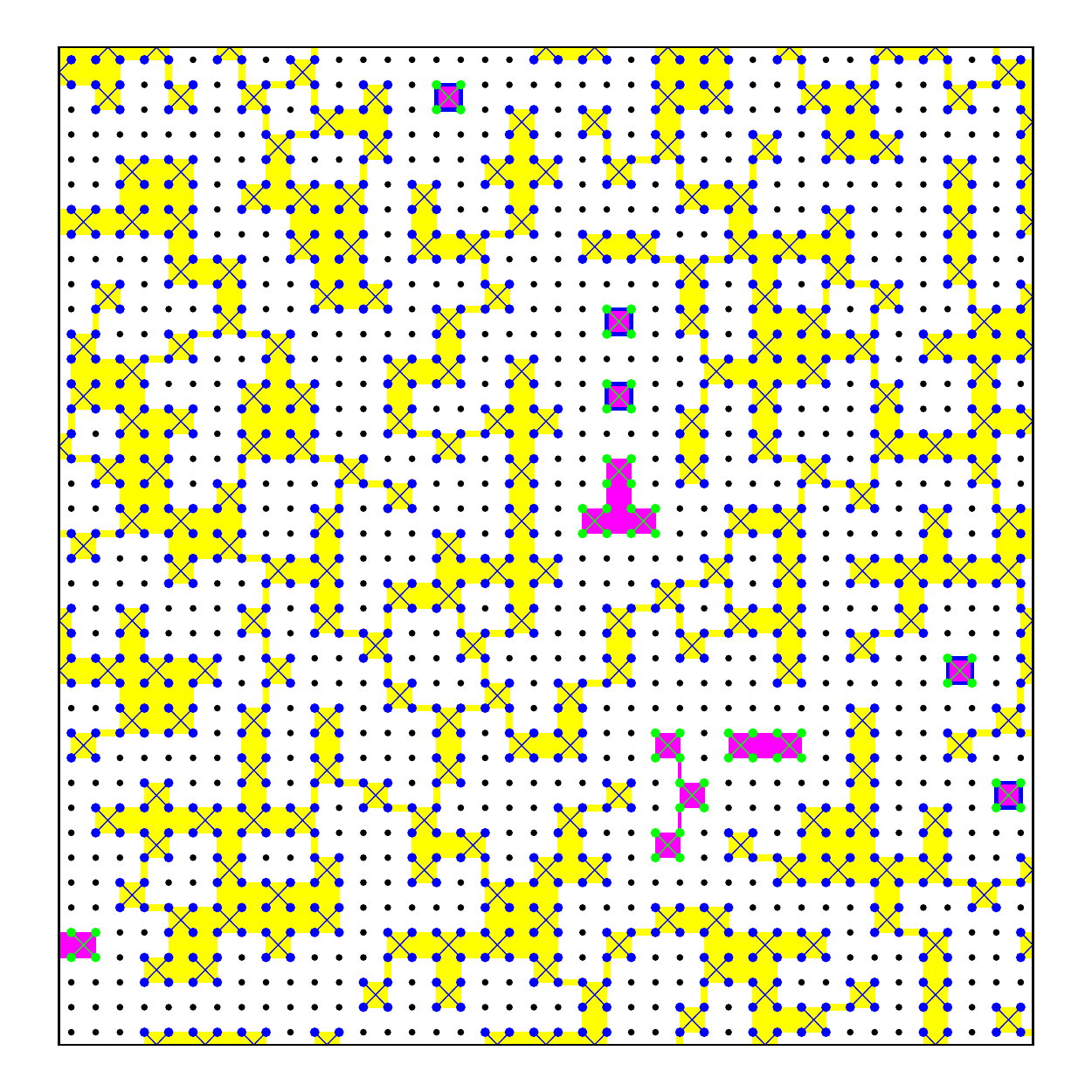}
\end{figure}
\noindent \textbf{Figure S15 (x = 0.15)}:
Plaquette doping of a $40\times40$ square CuO$_2$ lattice. Only the Cu sites are shown. The black dots are undoped AF Cu sites. The blue squares are isolated plaquettes (no neighboring plaquette). The yellow plaquette clusters are comprised of more than 4 plaquettes and contribute to the superconducting pairing because they are larger than the coherence length. The yellow overlay represents the metallic region comprised of planar Cu \dxxyy\ and planar O \psigma\ character. The magenta clusters are smaller than the Cooper pair coherence length and do not contribute to superconducting pairing. The magenta overlay represents metallic delocalization of the planar Cu \dxxyy\ and O \psigma\ orbitals. Fluctuating dumbbells are shown in every plaquette. There is no plaquette overlap at this doping. The 2D percolation threshold is approximately here.
\clearpage

\begin{figure}[tbp]
\centering \includegraphics[width=15cm]{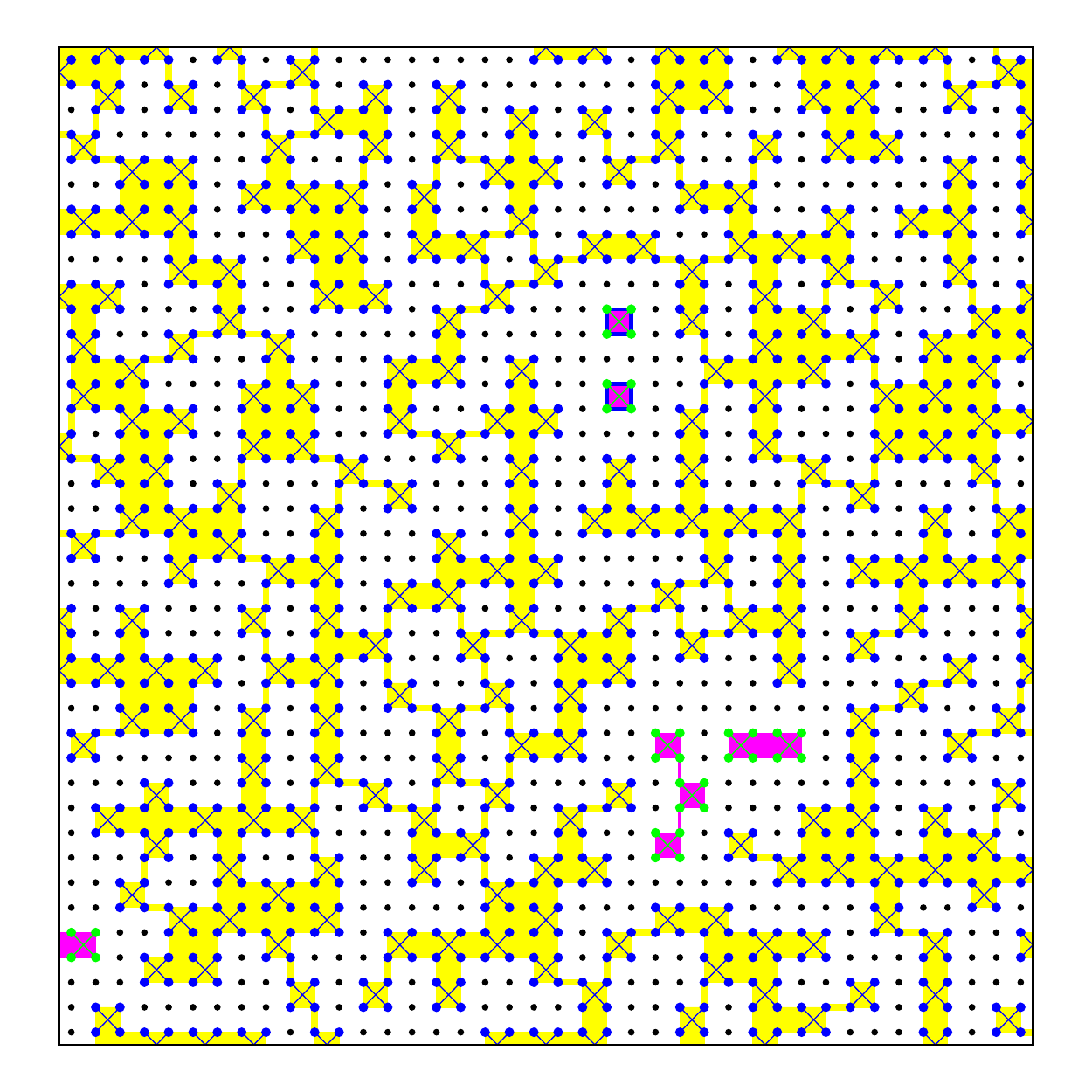}
\end{figure}
\noindent \textbf{Figure S16 (x = 0.16)}:
See the caption of Figure S15 for a definition of the symbols used here. 2D percolation occurs (see Figure \ref{pathway}), but the pathway is very tenuous. However, \tc\ is highest near this doping because the ratio of the surface metallic sites in the yellow region to total metallic sites (sum of yellow and magenta regions) is maximized.
\clearpage

\begin{figure}[tbp]
\centering \includegraphics[width=15cm]{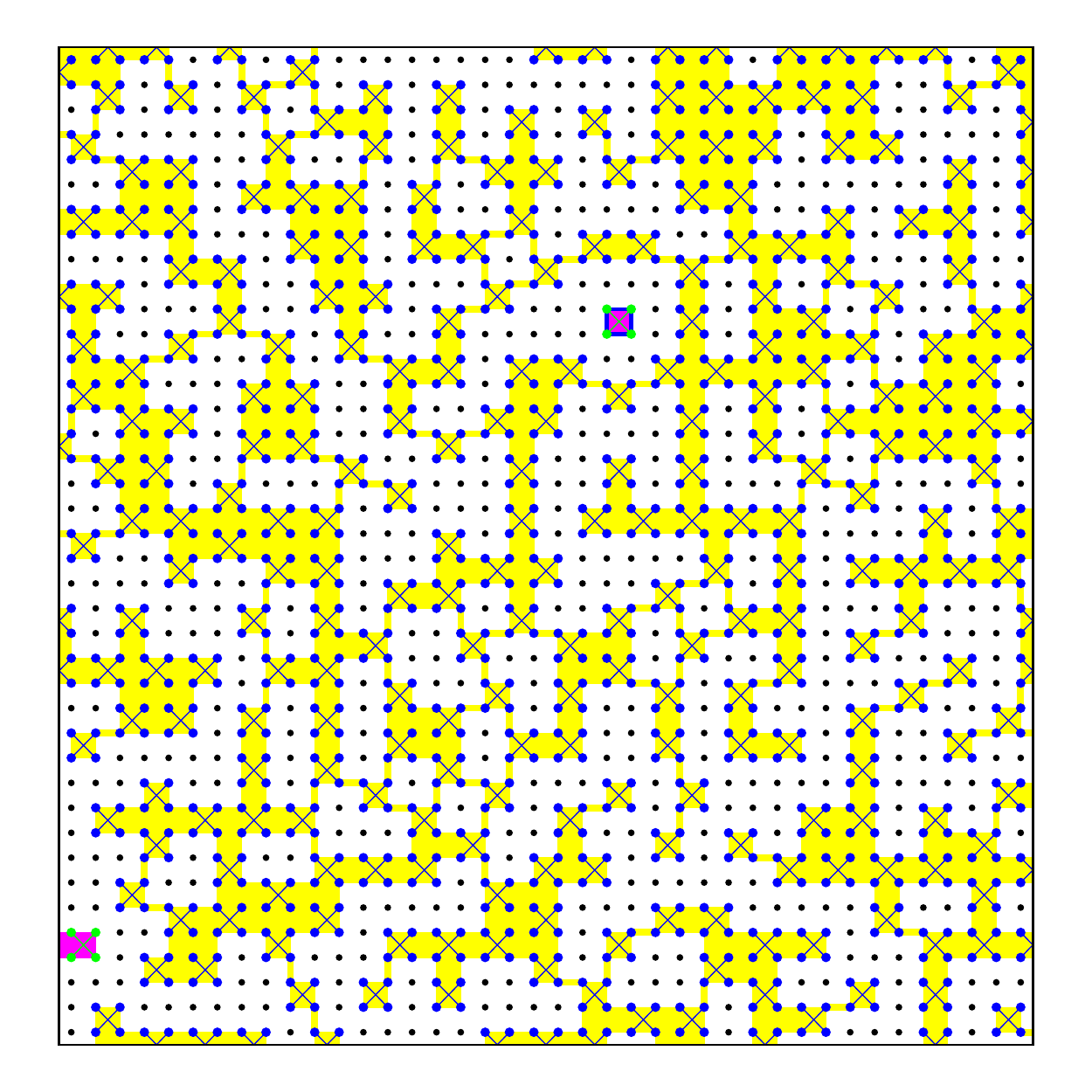}
\end{figure}
\noindent \textbf{Figure S17 (x = 0.17)}:
Plaquette doping of a $40\times40$ square CuO$_2$ lattice. Only the Cu sites are shown. The black dots are undoped AF Cu sites. The blue squares are isolated plaquettes (no neighboring plaquette). The yellow plaquette clusters are comprised of more than 4 plaquettes and contribute to the superconducting pairing because they are larger than the coherence length. The yellow overlay represents the metallic region comprised of planar Cu \dxxyy\ and planar O \psigma\ character. The magenta clusters are smaller than the Cooper pair coherence length and do not contribute to superconducting pairing. The magenta overlay represents metallic delocalization of the planar Cu \dxxyy\ and O \psigma\ orbitals. Fluctuating dumbbells are shown in every plaquette. There is no plaquette overlap at this doping. 2D percolation occurs here, but the pathway is very tenuous.
\clearpage

\begin{figure}[tbp]
\centering \includegraphics[width=15cm]{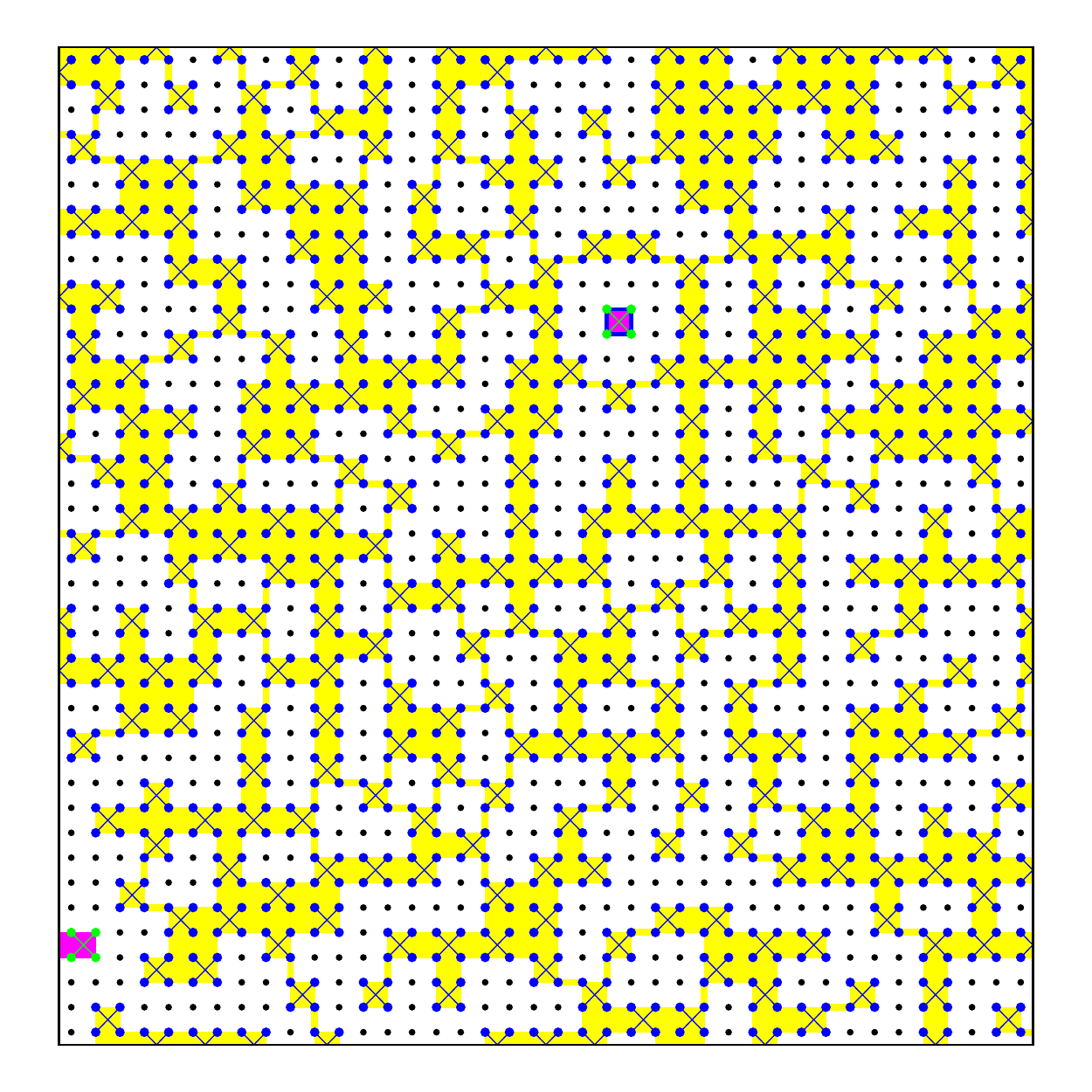}
\end{figure}
\noindent \textbf{Figure S18 (x = 0.18)}:
Plaquette doping of a $40\times40$ square CuO$_2$ lattice. Only the Cu sites are shown. The black dots are undoped AF Cu sites. The blue squares are isolated plaquettes (no neighboring plaquette). The yellow plaquette clusters are comprised of more than 4 plaquettes and contribute to the superconducting pairing because they are larger than the coherence length. The yellow overlay represents the metallic region comprised of planar Cu \dxxyy\ and planar O \psigma\ character. The magenta clusters are smaller than the Cooper pair coherence length and do not contribute to superconducting pairing. The magenta overlay represents metallic delocalization of the planar Cu \dxxyy\ and O \psigma\ orbitals. Fluctuating dumbbells are shown in every plaquette. There is no plaquette overlap at this doping. 2D percolation occurs here, but the pathway is very tenuous.
\clearpage

\begin{figure}[tbp]
\centering \includegraphics[width=15cm]{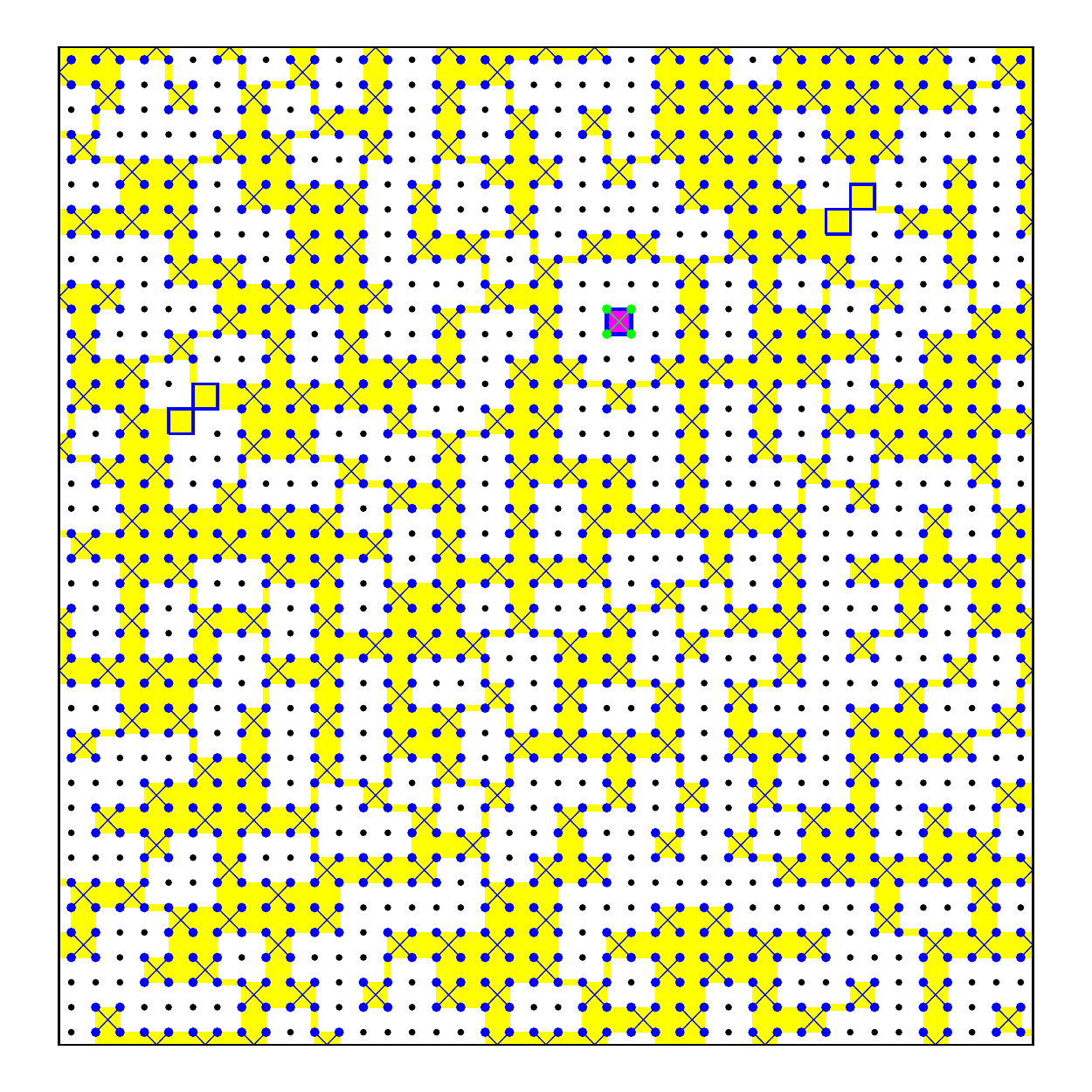}
\end{figure}
\noindent \textbf{Figure S19 (x = 0.19)}:
Plaquette doping of a $40\times40$ square CuO$_2$ lattice. Only the Cu sites are shown. The black dots are undoped AF Cu sites. One isolated plaquette (no neighboring plaquette) remains. The yellow plaquette clusters are comprised of more than 4 plaquettes and contribute to the superconducting pairing because they are larger than the coherence length. The yellow overlay represents the metallic region comprised of planar Cu \dxxyy\ and planar O \psigma\ character. The magenta clusters are smaller than the Cooper pair coherence length and do not contribute to superconducting pairing. The magenta overlay represents metallic delocalization of the planar Cu \dxxyy\ and O \psigma\ orbitals. Fluctuating dumbbells are inside non-overlapping plaquettes. Plaquette overlap begins (blue squares). In order to minimize their repulsion, the overlap occurs at the plaquette corners.
\clearpage

\begin{figure}[tbp]
\centering \includegraphics[width=15cm]{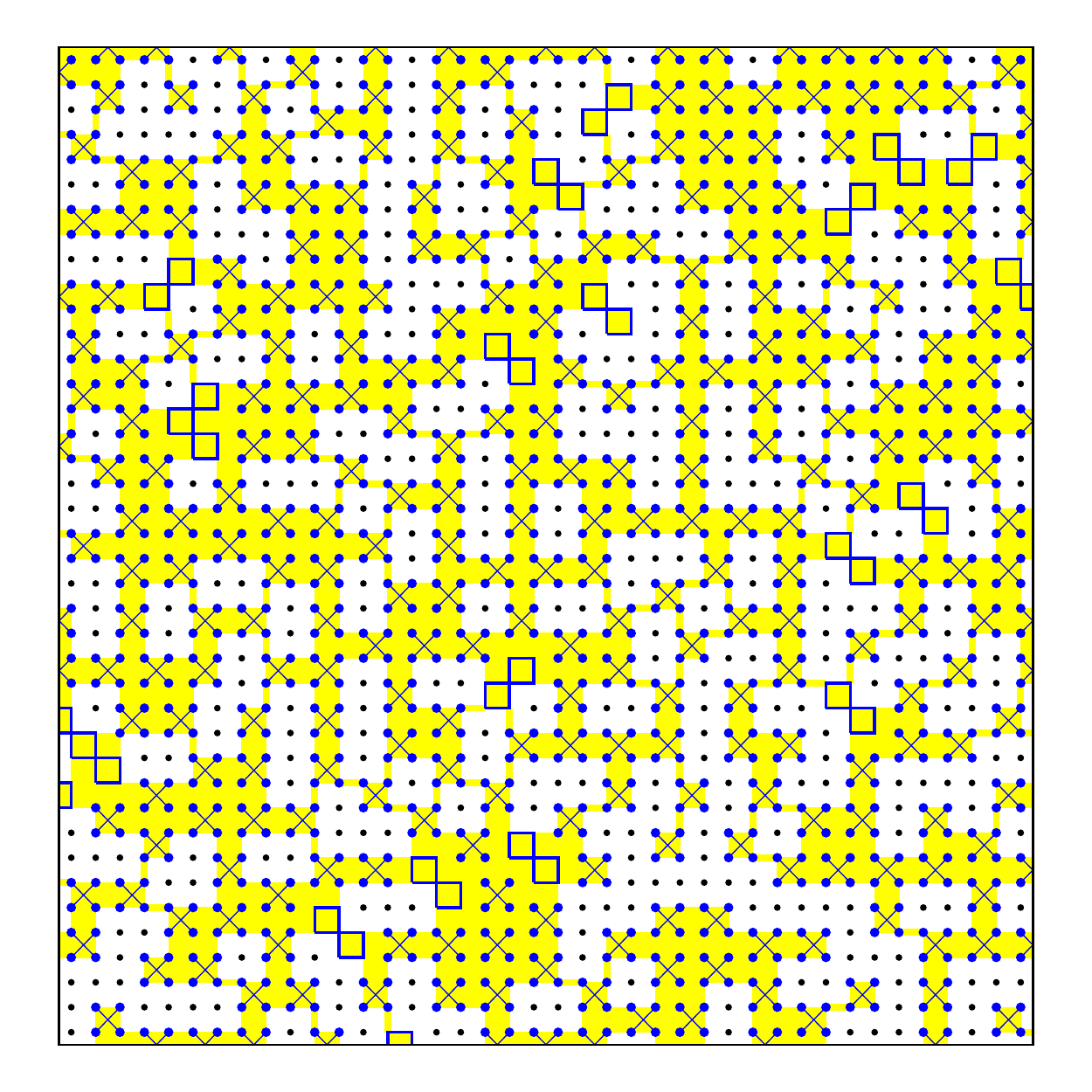}
\end{figure}
\noindent \textbf{Figure S20 (x = 0.20)}:
Plaquette doping of a $40\times40$ square CuO$_2$ lattice. Only the Cu sites are shown. The black dots are undoped AF Cu sites. The yellow plaquette clusters are comprised of more than 4 plaquettes and contribute to the superconducting pairing because they are larger than the coherence length. The yellow overlay represents the metallic region comprised of planar Cu \dxxyy\ and planar O \psigma\ character. Fluctuating dumbbells are inside non-overlapping plaquettes. Plaquette overlap is shown by blue squares. In order to minimize their repulsion, the overlap occurs at the plaquette corners only. There are no isolated plaquettes remaining in this lattice. The number of isolated plaquettes in the crystal is of measure zero. The pseudogap (arising from the isolated plaquettes) has vanished.\cite{Tahir-Kheli2011}
\clearpage

\begin{figure}[tbp]
\centering \includegraphics[width=15cm]{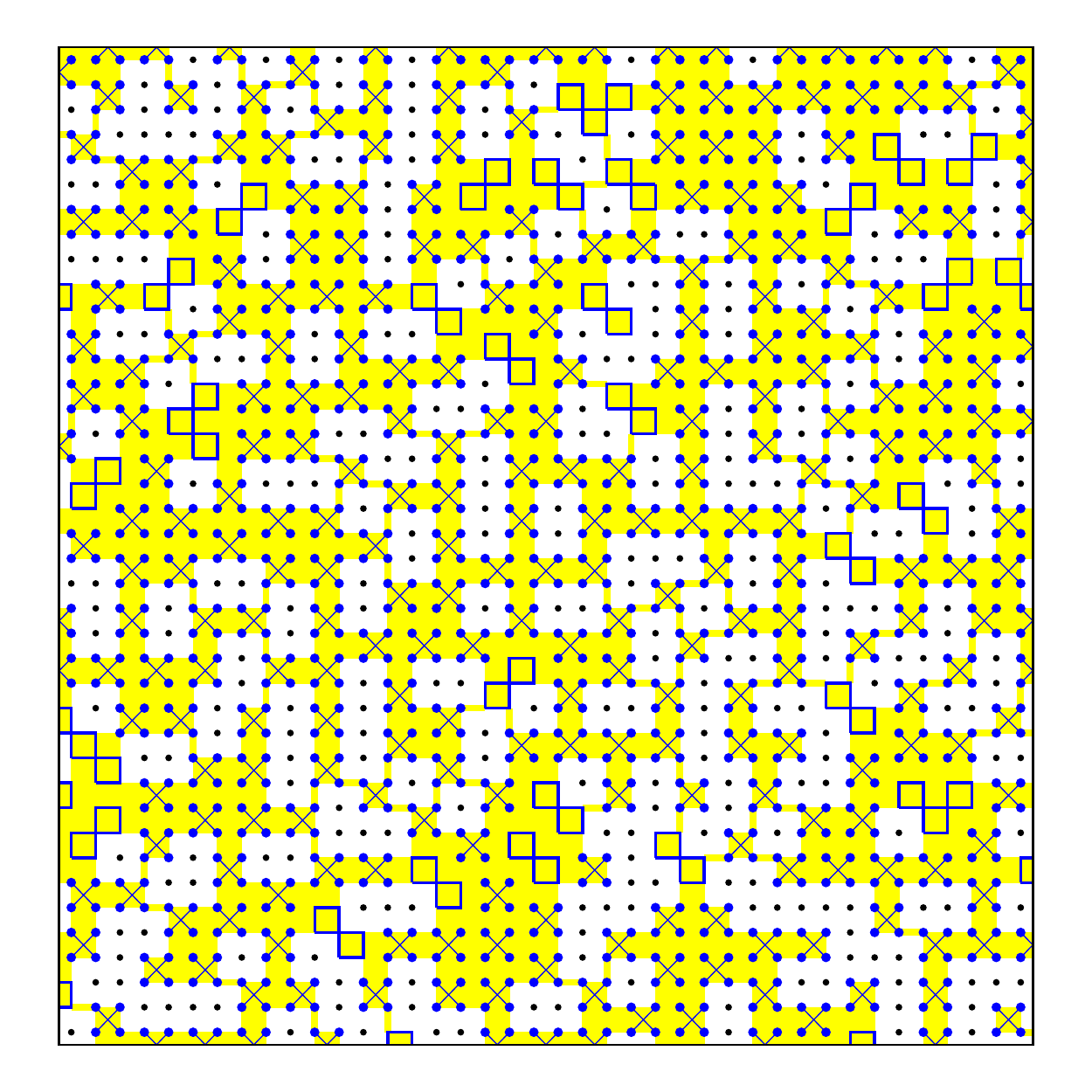}
\end{figure}
\noindent \textbf{Figure S21 (x = 0.21)}:
Plaquette doping of a $40\times40$ square CuO$_2$ lattice. Only the Cu sites are shown. The black dots are undoped AF Cu sites. The yellow plaquette clusters are comprised of more than 4 plaquettes and contribute to the superconducting pairing because they are larger than the coherence length. The yellow overlay represents the metallic region comprised of planar Cu \dxxyy\ and planar O \psigma\ character. Fluctuating dumbbells are inside non-overlapping plaquettes. Plaquette overlap is shown by blue squares. In order to minimize their repulsion, the overlap occurs at the plaquette corners only. There are no isolated plaquettes remaining in this lattice. The number of isolated plaquettes in the crystal is of measure zero.
\clearpage

\begin{figure}[tbp]
\centering \includegraphics[width=15cm]{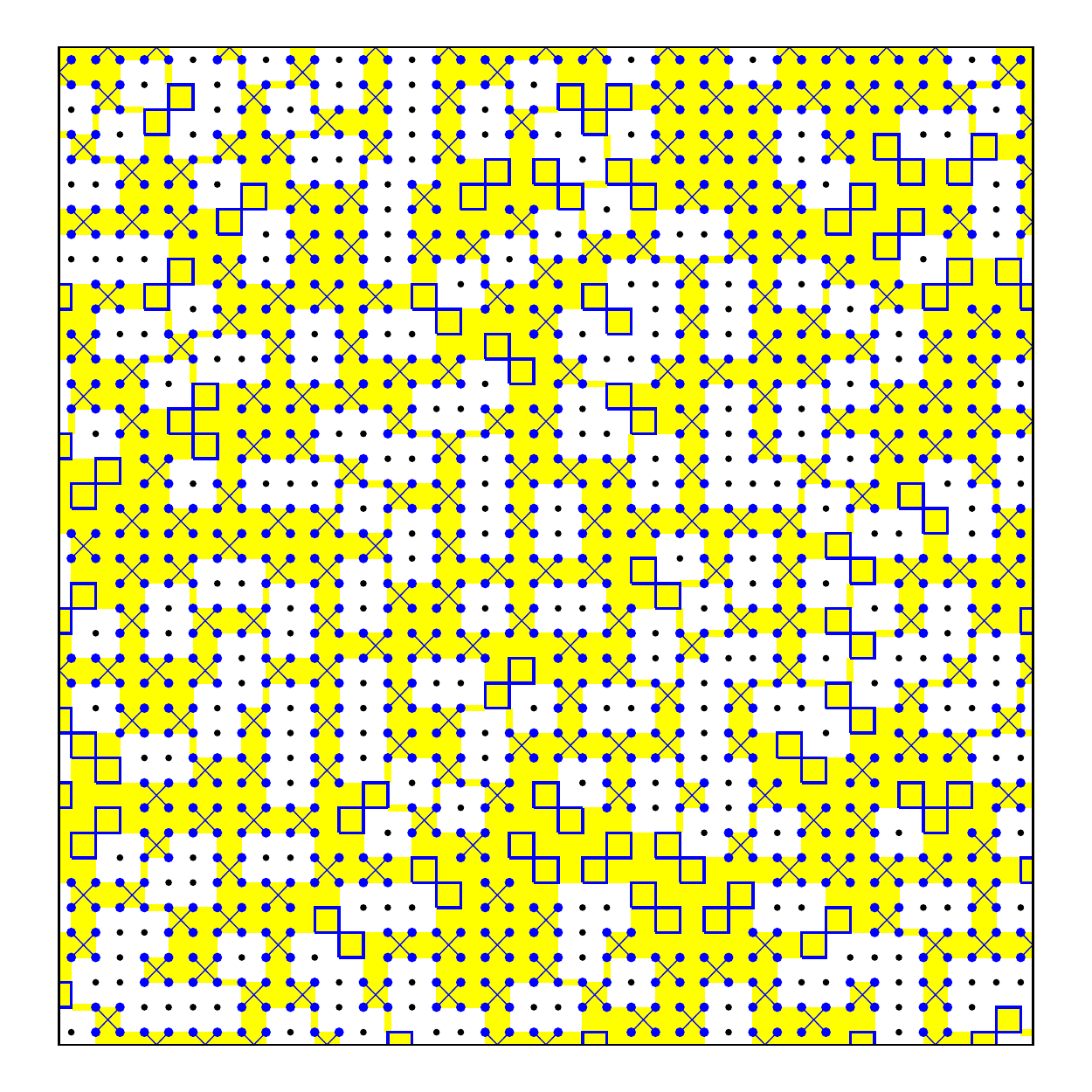}
\end{figure}
\noindent \textbf{Figure S22 (x = 0.22)}:
Plaquette doping of a $40\times40$ square CuO$_2$ lattice. Only the Cu sites are shown. The black dots are undoped AF Cu sites. The yellow plaquette clusters are comprised of more than 4 plaquettes and contribute to the superconducting pairing because they are larger than the coherence length. The yellow overlay represents the metallic region comprised of planar Cu \dxxyy\ and planar O \psigma\ character. Fluctuating dumbbells are inside non-overlapping plaquettes. Plaquette overlap is shown by blue squares. In order to minimize their repulsion, the overlap occurs at the plaquette corners only. There are no isolated plaquettes remaining in this lattice. The number of isolated plaquettes in the crystal is of measure zero.
\clearpage

\begin{figure}[tbp]
\centering \includegraphics[width=15cm]{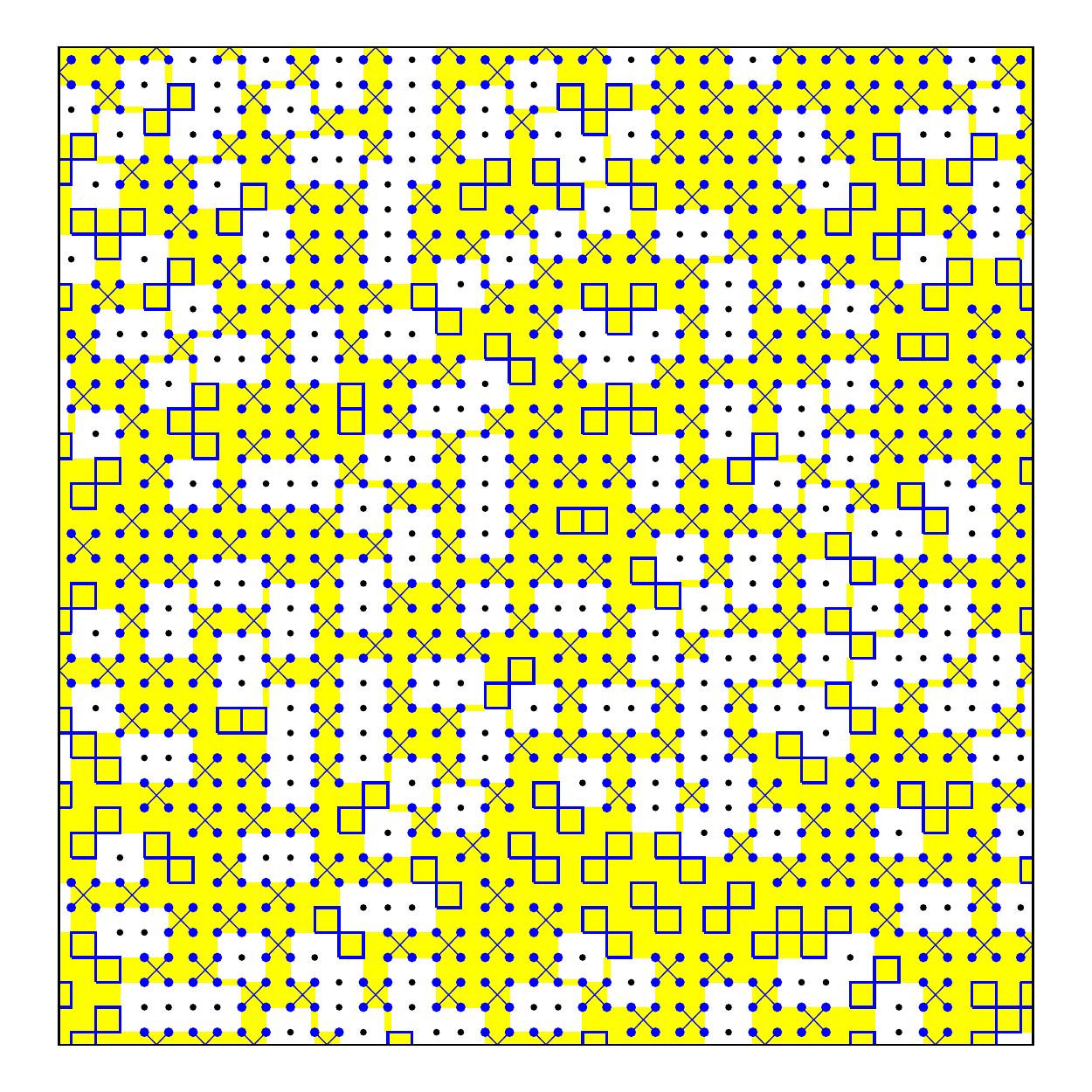}
\end{figure}
\noindent \textbf{Figure S23 (x = 0.23)}:
Plaquette doping of a $40\times40$ square CuO$_2$ lattice. Only the Cu sites are shown. The black dots are undoped AF Cu sites. The yellow plaquette clusters are comprised of more than 4 plaquettes and contribute to the superconducting pairing because they are larger than the coherence length. The yellow overlay represents the metallic region comprised of planar Cu \dxxyy\ and planar O \psigma\ character. Fluctuating dumbbells are inside non-overlapping plaquettes. Plaquette overlap is shown by blue squares. Corner overlap of plaquettes is no longer possible. Added plaquettes overlap existing plaquettes by sharing edges or overlapping two plaquettes at their corners. We believe the additional energy needed to overlap plaquette edges or two corners is the reason why YBa$\mathrm{_2}$Cu$\mathrm{_3}$O${\mathrm{_{7-\delta}}}$ cannot be doped beyond this point.
\clearpage

\begin{figure}[tbp]
\centering \includegraphics[width=15cm]{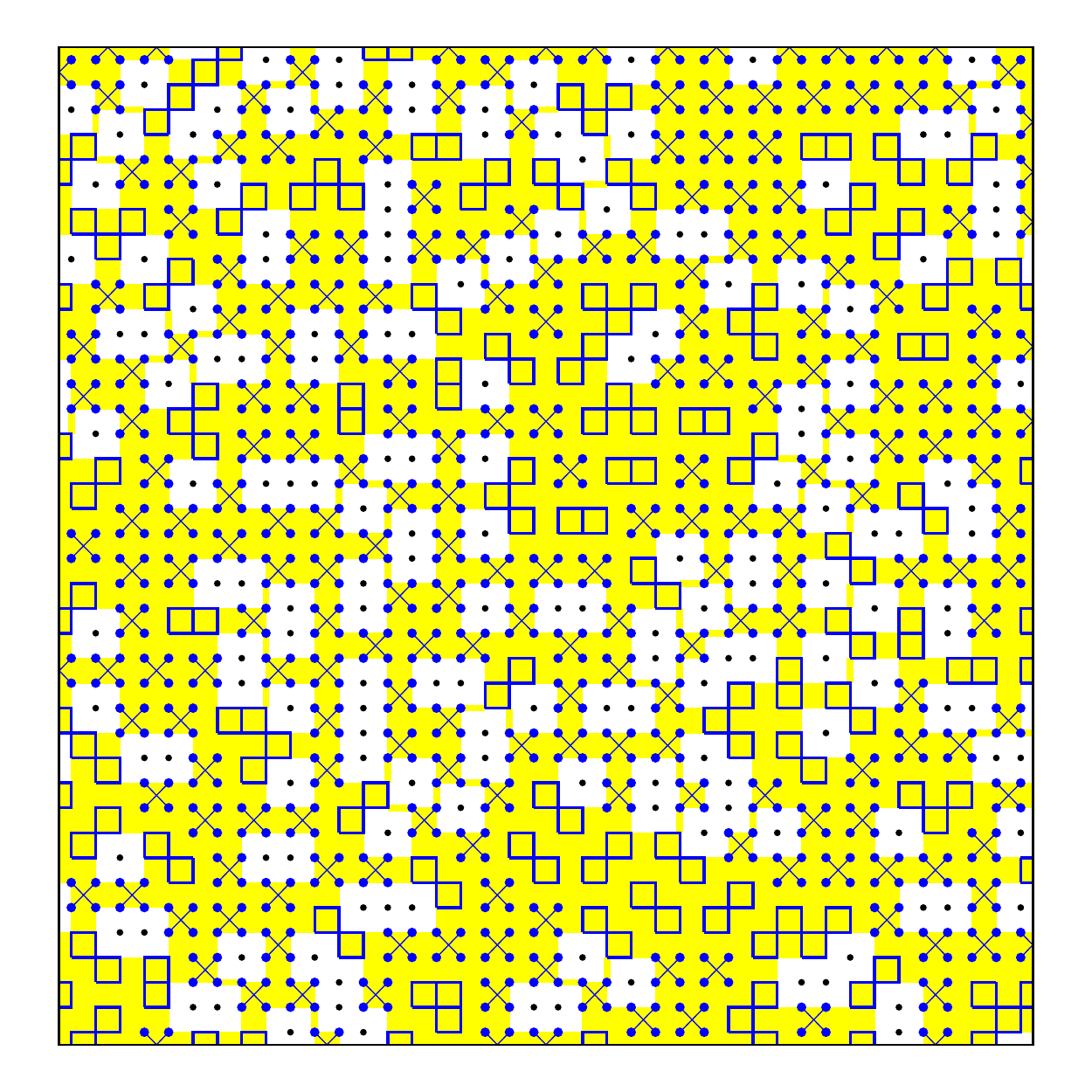}
\end{figure}
\noindent \textbf{Figure S24 (x = 0.24)}:
Plaquette doping of a $40\times40$ square CuO$_2$ lattice. Only the Cu sites are shown. The black dots are undoped AF Cu sites. The yellow plaquette clusters are comprised of more than 4 plaquettes and contribute to the superconducting pairing because they are larger than the coherence length. The yellow overlay represents the metallic region comprised of planar Cu \dxxyy\ and planar O \psigma\ character. Fluctuating dumbbells are inside non-overlapping plaquettes. Plaquette overlap is shown by blue squares. Corner overlap of plaquettes is no longer possible. Added plaquettes overlap existing plaquettes by sharing edges or overlapping two plaquettes at their corners.
\clearpage

\begin{figure}[tbp]
\centering \includegraphics[width=15cm]{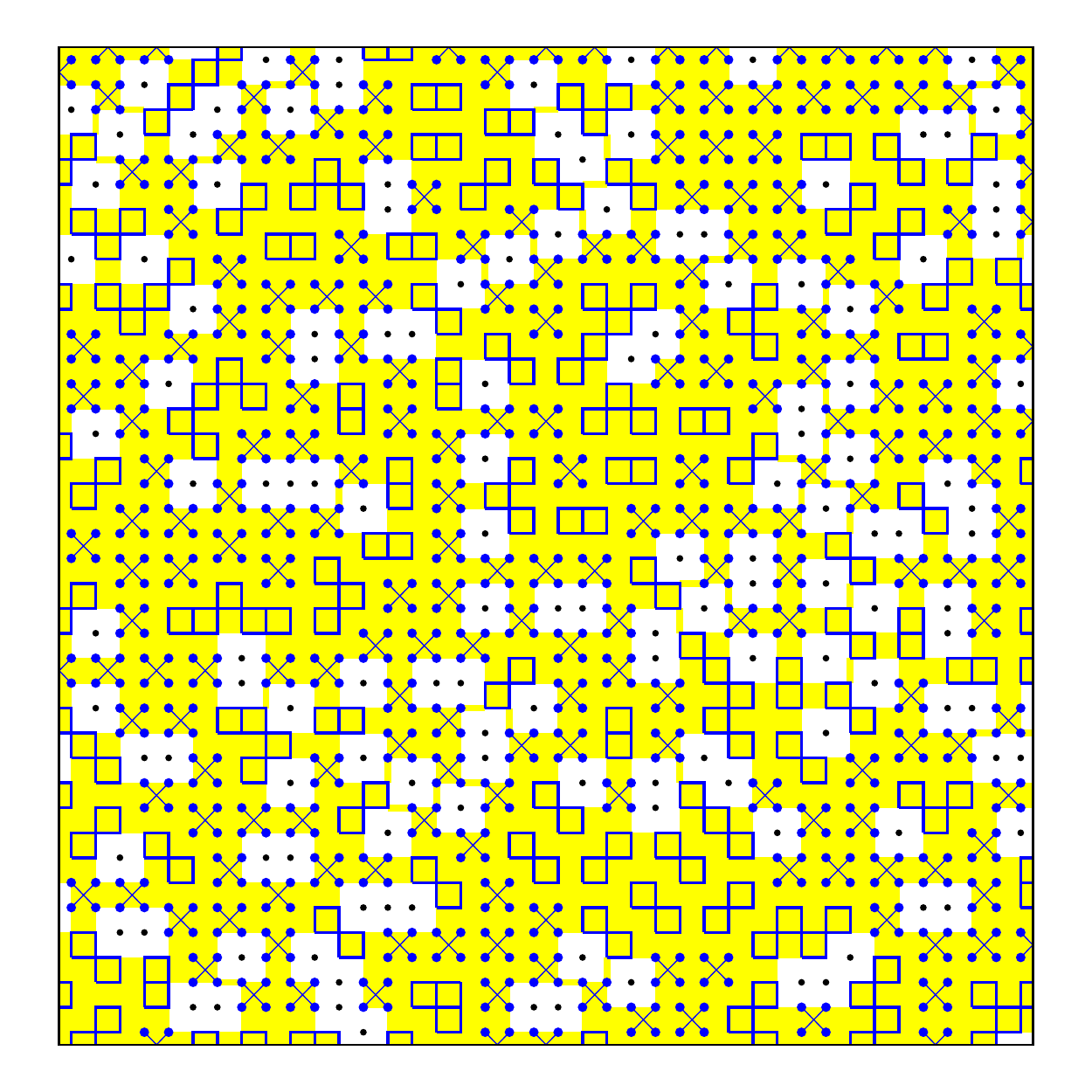}
\end{figure}
\noindent \textbf{Figure S25 (x = 0.25)}:
Plaquette doping of a $40\times40$ square CuO$_2$ lattice. Only the Cu sites are shown. The black dots are undoped AF Cu sites. The yellow plaquette clusters are comprised of more than 4 plaquettes and contribute to the superconducting pairing because they are larger than the coherence length. The yellow overlay represents the metallic region comprised of planar Cu \dxxyy\ and planar O \psigma\ character. Fluctuating dumbbells are inside non-overlapping plaquettes. Plaquette overlap is shown by blue squares. Corner overlap of plaquettes is no longer possible. Added plaquettes overlap existing plaquettes by sharing edges or overlapping two plaquettes at their corners.
\clearpage

\begin{figure}[tbp]
\centering \includegraphics[width=15cm]{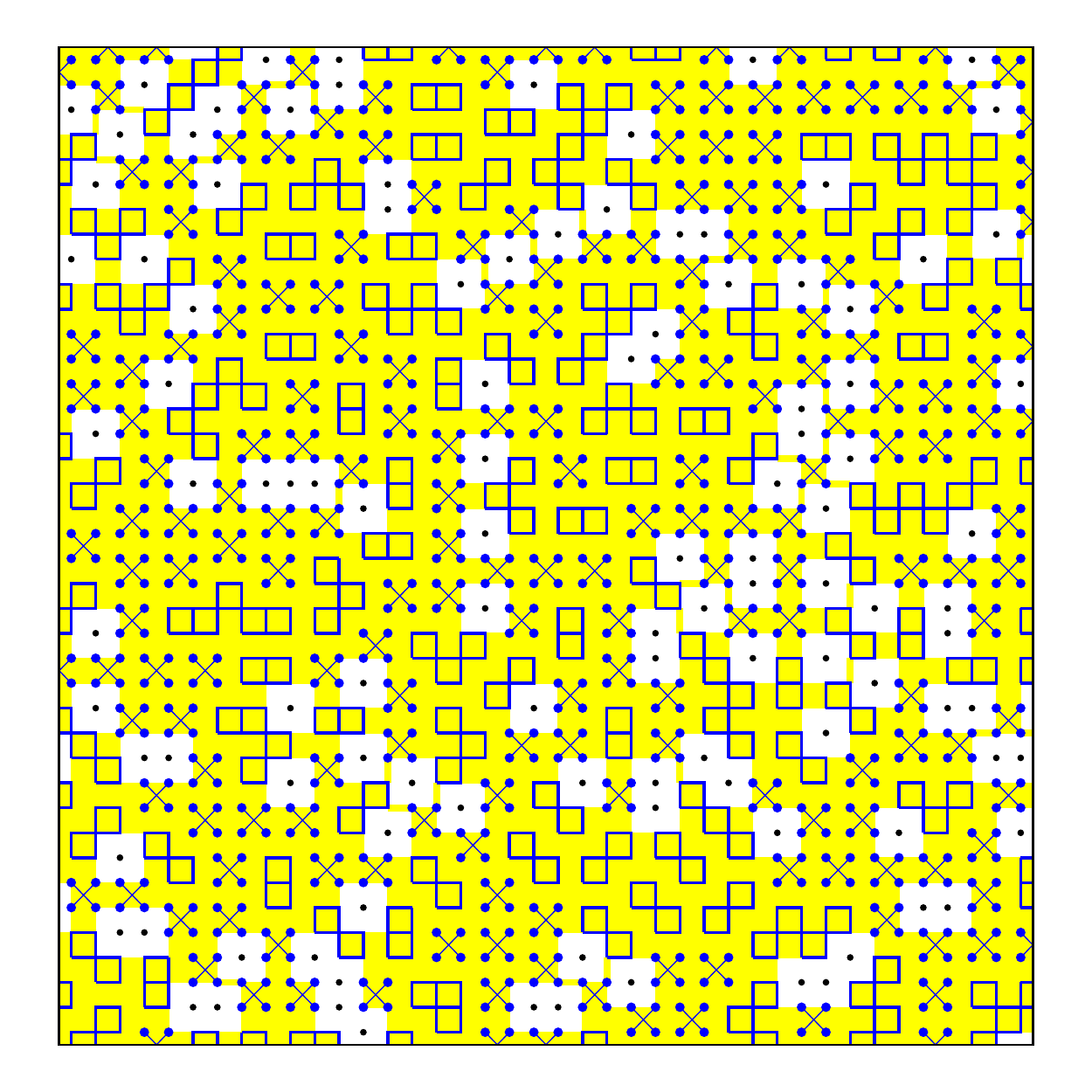}
\end{figure}
\noindent \textbf{Figure S26 (x = 0.26)}:
Plaquette doping of a $40\times40$ square CuO$_2$ lattice. Only the Cu sites are shown. The black dots are undoped AF Cu sites. The yellow plaquette clusters are comprised of more than 4 plaquettes and contribute to the superconducting pairing because they are larger than the coherence length. The yellow overlay represents the metallic region comprised of planar Cu \dxxyy\ and planar O \psigma\ character. Fluctuating dumbbells are inside non-overlapping plaquettes. Plaquette overlap is shown by blue squares. Corner overlap of plaquettes is no longer possible. Added plaquettes overlap existing plaquettes by sharing edges or overlapping two plaquettes at their corners.
\clearpage

\begin{figure}[tbp]
\centering \includegraphics[width=15cm]{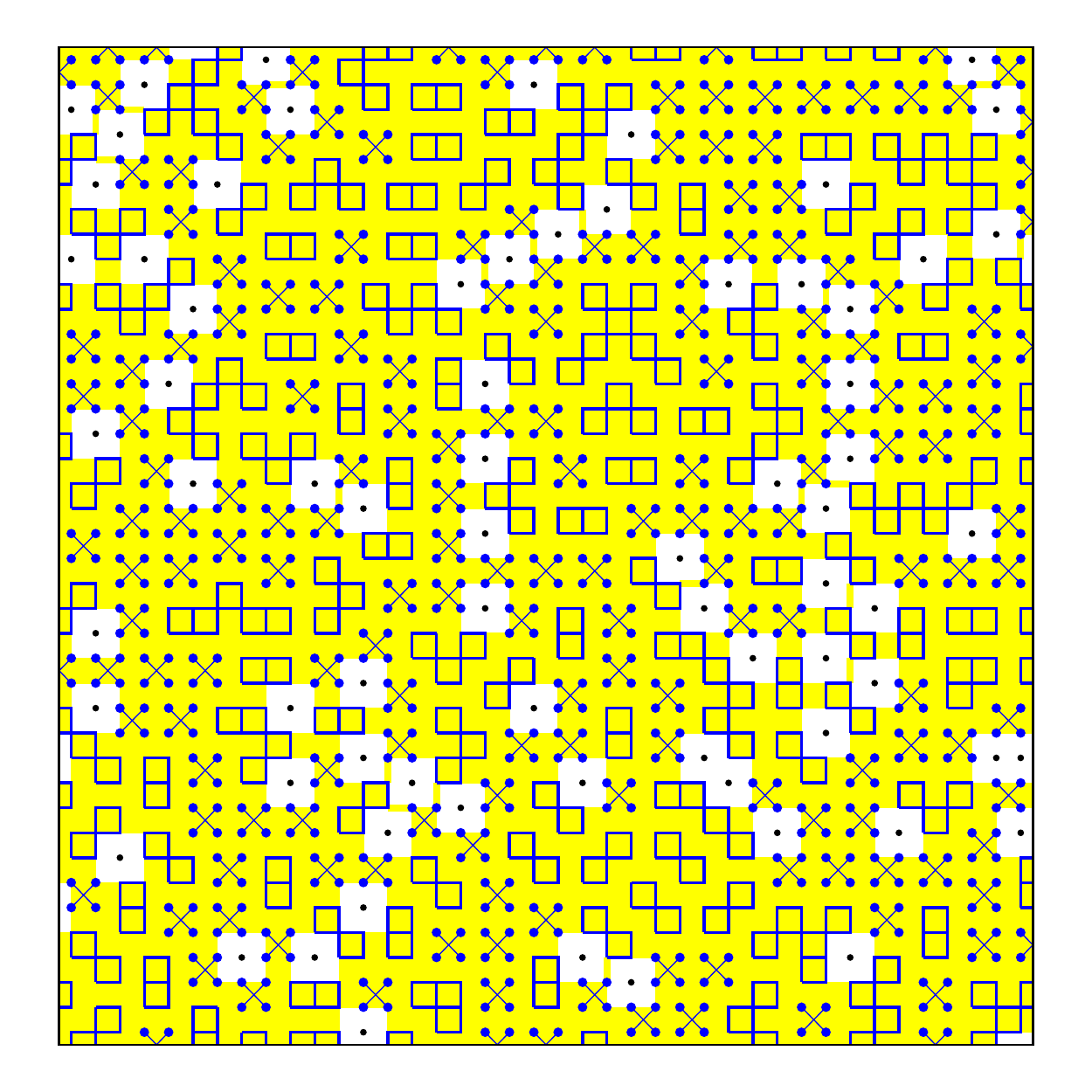}
\end{figure}
\noindent \textbf{Figure S27 (x = 0.27)}:
Plaquette doping of a $40\times40$ square CuO$_2$ lattice. Only the Cu sites are shown. The black dots are undoped AF Cu sites. The yellow plaquette clusters are comprised of more than 4 plaquettes and contribute to the superconducting pairing because they are larger than the coherence length. The yellow overlay represents the metallic region comprised of planar Cu \dxxyy\ and planar O \psigma\ character. Fluctuating dumbbells are inside non-overlapping plaquettes. Plaquette overlap is shown by blue squares. Corner overlap of plaquettes is no longer possible. Added plaquettes overlap existing plaquettes by sharing edges or overlapping two plaquettes at their corners.
\clearpage

\begin{figure}[tbp]
\centering \includegraphics[width=15cm]{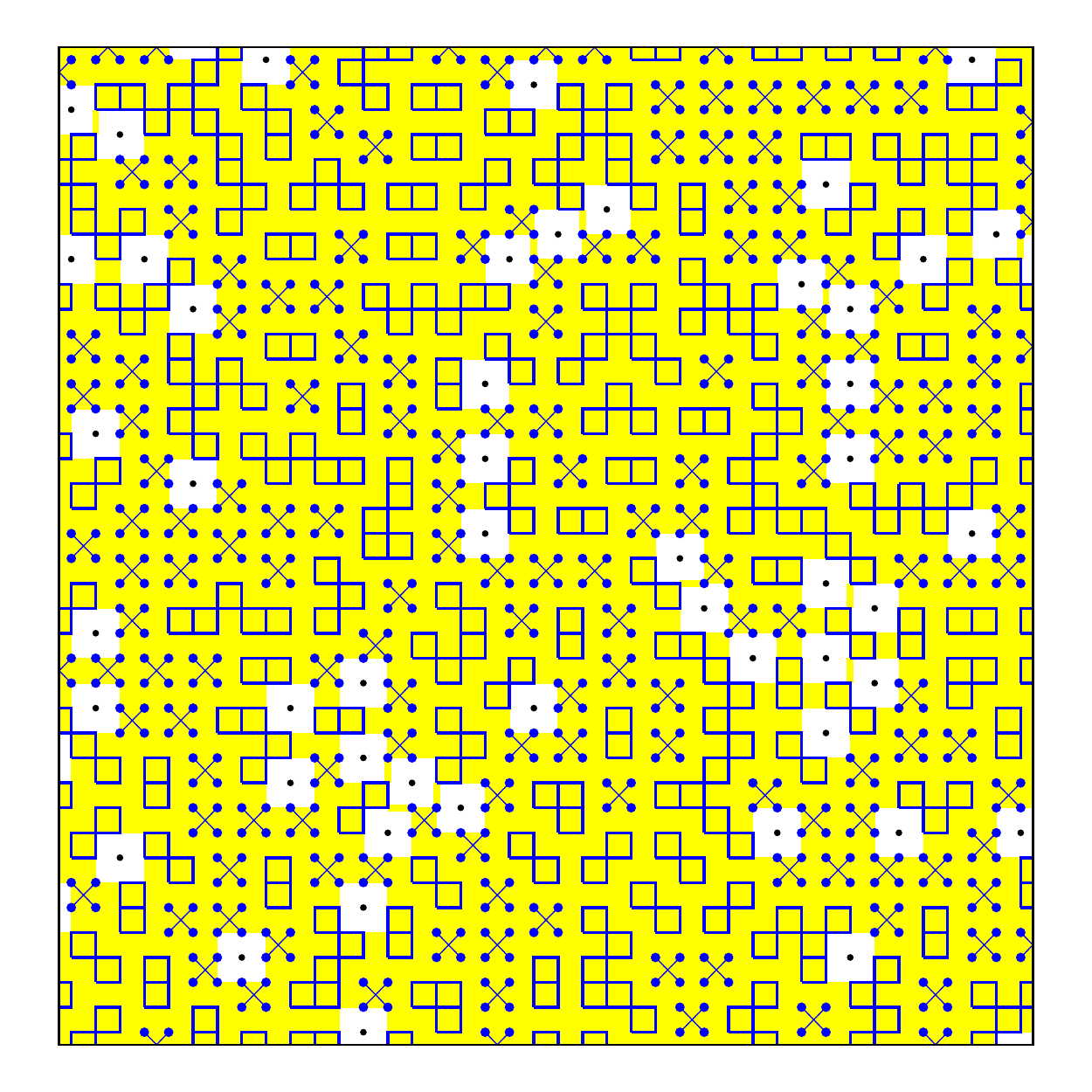}
\end{figure}
\noindent \textbf{Figure S28 (x = 0.28)}:
Plaquette doping of a $40\times40$ square CuO$_2$ lattice. Only the Cu sites are shown. The black dots are undoped AF Cu sites. The yellow plaquette clusters are comprised of more than 4 plaquettes and contribute to the superconducting pairing because they are larger than the coherence length. The yellow overlay represents the metallic region comprised of planar Cu \dxxyy\ and planar O \psigma\ character. Fluctuating dumbbells are inside non-overlapping plaquettes. Plaquette overlap is shown by blue squares. Edge overlap of plaquettes is no longer possible. Added plaquettes overlap existing plaquettes at three metallic sites. Only isolated localized spins remain (there are no adjacent black dots).
\clearpage

\begin{figure}[tbp]
\centering \includegraphics[width=15cm]{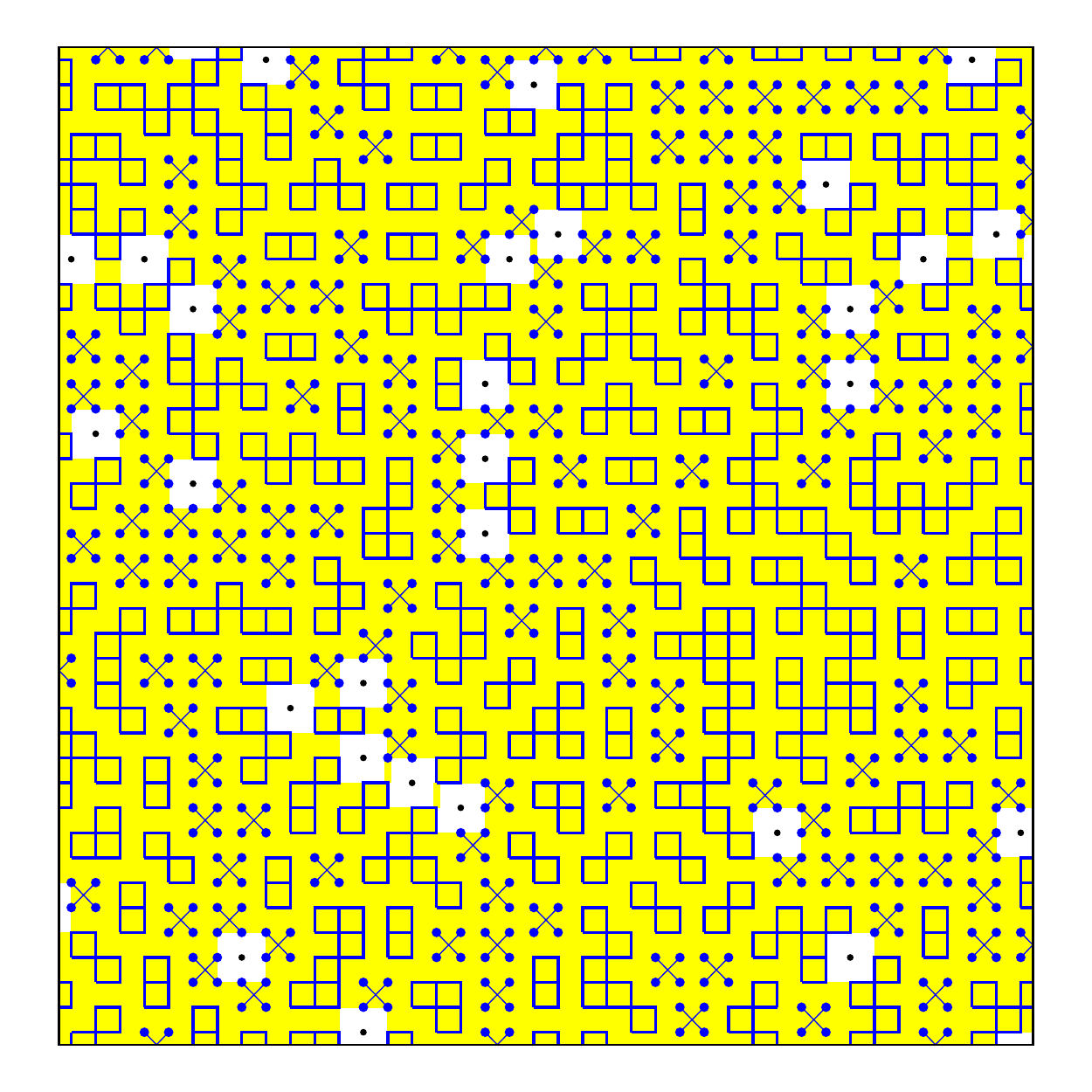}
\end{figure}
\noindent \textbf{Figure S29 (x = 0.29)}:
Plaquette doping of a $40\times40$ square CuO$_2$ lattice. Only the Cu sites are shown. The black dots are undoped AF Cu sites. The yellow plaquette clusters are comprised of more than 4 plaquettes and contribute to the superconducting pairing because they are larger than the coherence length. The yellow overlay represents the metallic region comprised of planar Cu \dxxyy\ and planar O \psigma\ character. Fluctuating dumbbells are inside non-overlapping plaquettes. Plaquette overlap is shown by blue squares. Edge overlap of plaquettes is no longer possible. Added plaquettes overlap existing plaquettes at three metallic sites. Only isolated localized spins remain (there are no adjacent black dots).
\clearpage

\begin{figure}[tbp]
\centering \includegraphics[width=15cm]{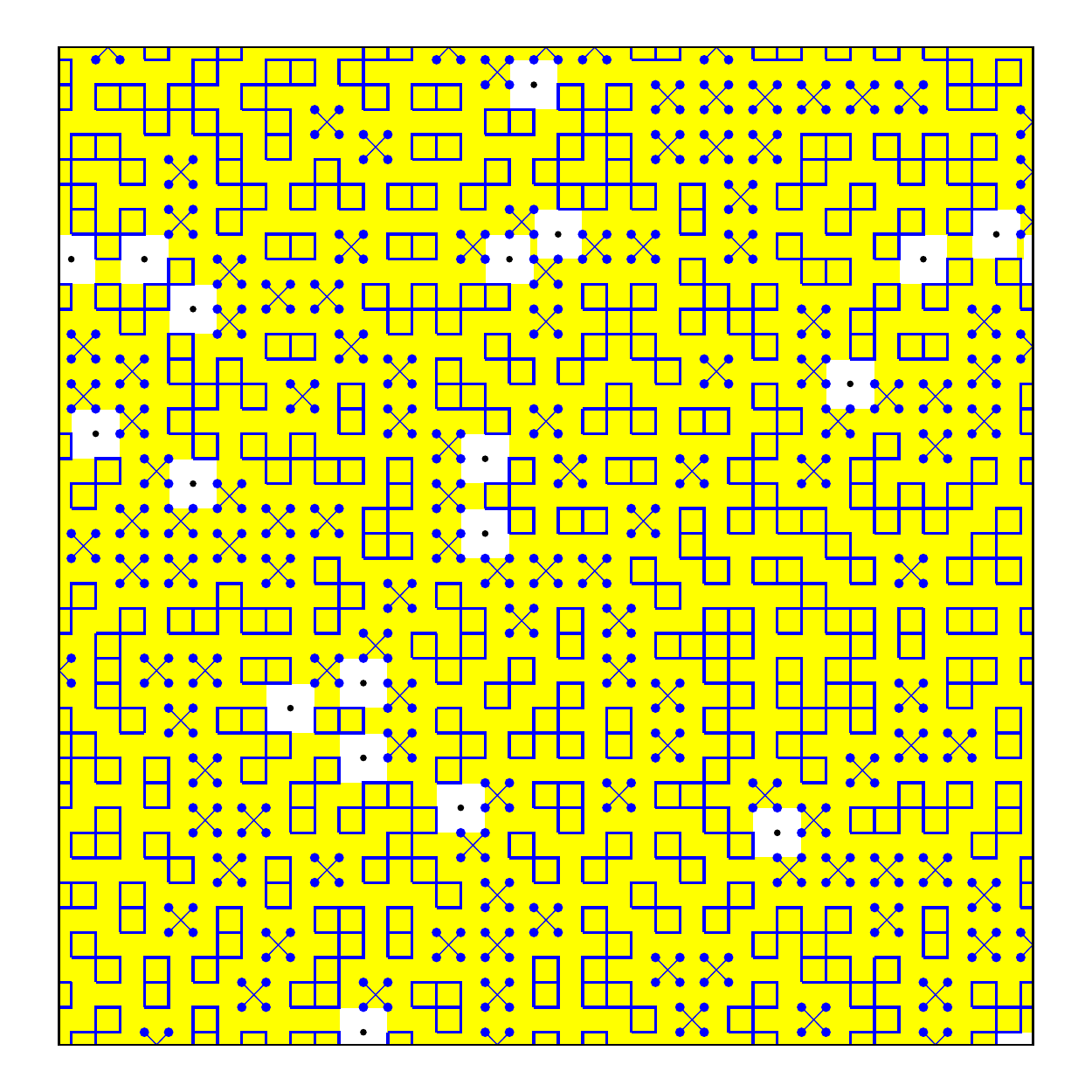}
\end{figure}
\noindent \textbf{Figure S30 (x = 0.30)}:
Plaquette doping of a $40\times40$ square CuO$_2$ lattice. Only the Cu sites are shown. The black dots are undoped AF Cu sites. The yellow plaquette clusters are comprised of more than 4 plaquettes and contribute to the superconducting pairing because they are larger than the coherence length. The yellow overlay represents the metallic region comprised of planar Cu \dxxyy\ and planar O \psigma\ character. Fluctuating dumbbells are inside non-overlapping plaquettes. Plaquette overlap is shown by blue squares. Edge overlap of plaquettes is no longer possible. Added plaquettes overlap existing plaquettes at three metallic sites. Only isolated localized spins remain (there are no adjacent black dots).
\clearpage

\begin{figure}[tbp]
\centering \includegraphics[width=15cm]{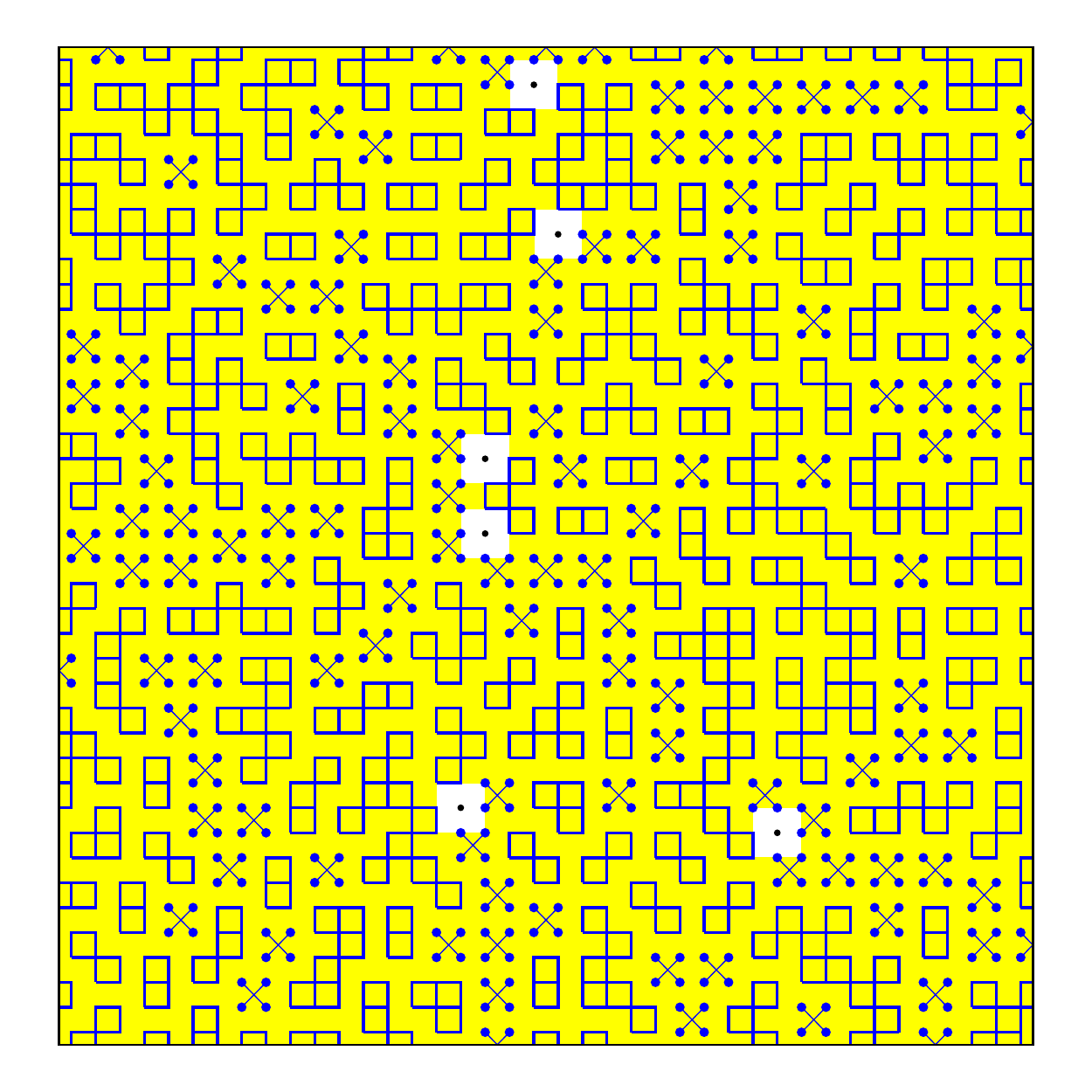}
\end{figure}
\noindent \textbf{Figure S31 (x = 0.31)}:
Plaquette doping of a $40\times40$ square CuO$_2$ lattice. Only the Cu sites are shown. The black dots are undoped AF Cu sites. The yellow plaquette clusters are comprised of more than 4 plaquettes and contribute to the superconducting pairing because they are larger than the coherence length. The yellow overlay represents the metallic region comprised of planar Cu \dxxyy\ and planar O \psigma\ character. Fluctuating dumbbells are inside non-overlapping plaquettes. Plaquette overlap is shown by blue squares. Edge overlap of plaquettes is no longer possible. Added plaquettes overlap existing plaquettes at three metallic sites. Only isolated localized spins remain (there are no adjacent black dots).
\clearpage

\begin{figure}[tbp]
\centering \includegraphics[width=15cm]{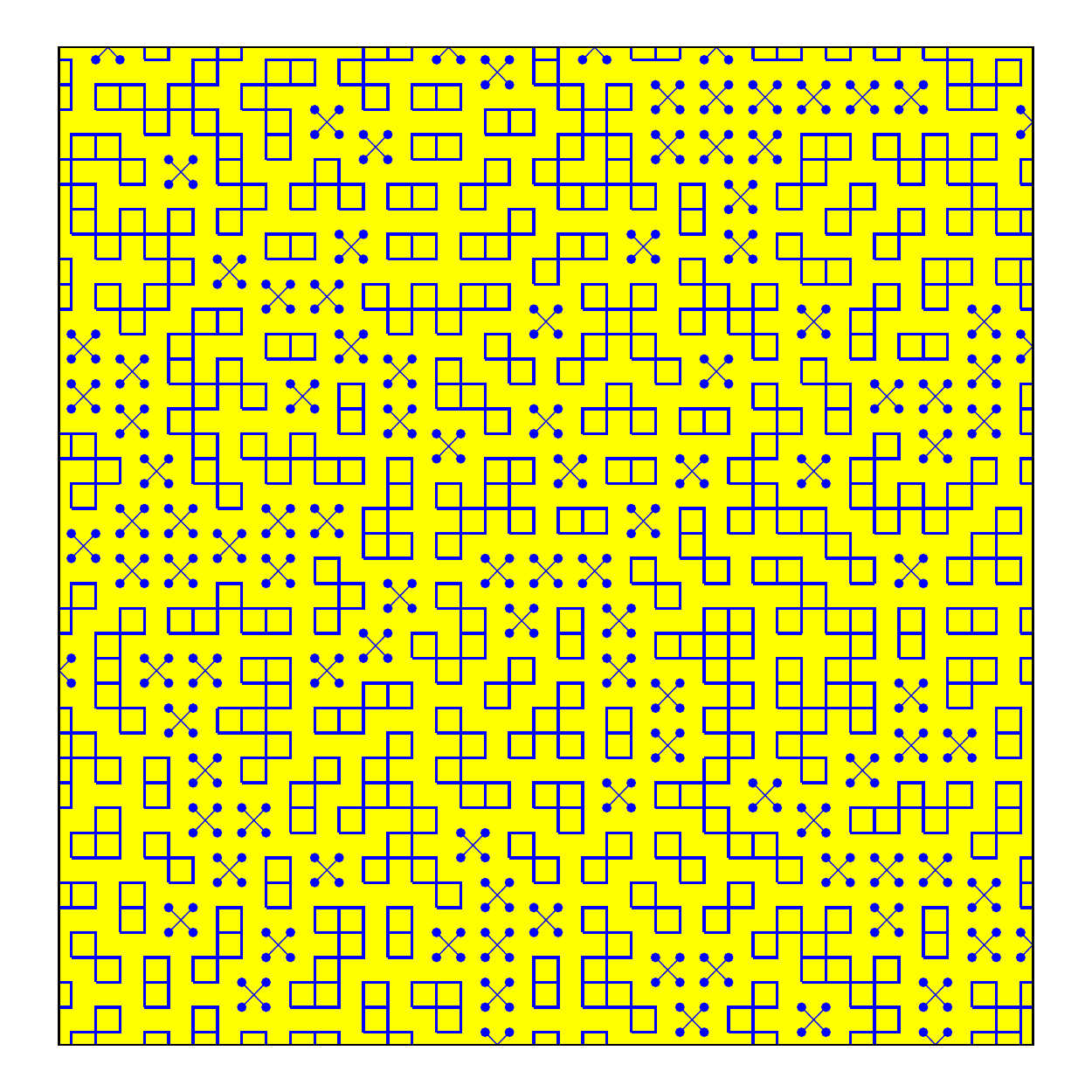}
\end{figure}
\noindent \textbf{Figure S32 (x = 0.32)}:
Plaquette doping of a $40\times40$ square CuO$_2$ lattice. Only the Cu sites are shown. The black dots are undoped AF Cu sites. The yellow plaquette clusters are comprised of more than 4 plaquettes and contribute to the superconducting pairing because they are larger than the coherence length. The yellow overlay represents the metallic region comprised of planar Cu \dxxyy\ and planar O \psigma\ character. Fluctuating dumbbells are inside non-overlapping plaquettes. Plaquette overlap is shown by blue squares. Edge overlap of plaquettes is no longer possible. Added plaquettes overlap existing plaquettes at three metallic sites. There are no remaining localized spins (black dots). The crystal is 100\% metallic. There is no metal-insulator interface to produce superconducting pairing, and therefore $T_c=0$. 

\clearpage

\end{document}